\documentclass[prd,preprint,tightenlines,floatfix,showpacs,preprintnumbers,nofootinbib,eqsecnum,superscriptaddress]{revtex4-1}
\pdfoutput=1 

\usepackage[usenames,dvipsnames]{color}

\usepackage{graphicx}

\usepackage{amsmath}        
\usepackage{amssymb}        
\usepackage{amsfonts}       
\usepackage{amsbsy}     

\usepackage{slashed}
\usepackage{hyperref}

\usepackage{soul}

\newcommand{\sina}{\sin\alpha}\newcommand{\cosa}{\cos\alpha}

\newcommand{\sa}{s_{\alpha}}
\newcommand{\ca}{c_{\alpha}}
\newcommand{\ztwo}{\mathbb{Z}_2}
\newcommand{\half}{\frac{1}{2}}
\newcommand{\hp}{{H^\pm}}
\newcommand{\wpm}{{W^\pm}}
\newcommand{\mhp}{{m_{H^\pm}}}

\newcommand{\M}{{\cal M}}

\newcommand{\ii}{\mathrm{i}}
\newcommand{\grp}[1]{\mathrm{#1}}
\newcommand{\hcal}{ \mathcal{H} }

\newcommand{\f}{\frac}

\newcommand*{\THDMC}{{\sc 2hdmc}}

\newcommand{\FA}{{\sc FeynArts}}
\newcommand{\FRU}{{\sc FeynRules}}
\newcommand{\FC}{{\sc FormCalc}}

\newcommand{\HiB}{{\sc HiggsBounds}}
\newcommand{\MadG}{{\sc MadGraph}}

\newcommand{\vev}[1]{ \left\langle {#1} \right\rangle }

\newcommand{\gev}{{\rm GeV}}

\newcommand{\reffig}[1]{{figure~\ref{#1}}}
\newcommand{\refeq}[1]{{eq.~(\ref{#1})}}
\newcommand{\eq}[1]{{(\ref{#1})}}

\newcommand{\be}{\begin{equation}}
\newcommand{\ee}{\end{equation}}
\newcommand{\bel}[1]{\be\label{#1}}
\newcommand{\eref}[1]{(\ref{#1})}
\newcommand{\Eref}[1]{eq.~(\ref{#1})}
\newcommand{\nn}{\nonumber}

\begin{document}

\begin{flushright}
LU-TP 13-43
\end{flushright}
\vspace{1cm}

\title{Higgs phenomenology in the Stealth Doublet Model}

\author{Rikard Enberg}
\email{Rikard.Enberg@physics.uu.se}
\affiliation{Department of Physics and Astronomy,
 Uppsala University, Box 516, SE--751 20 Uppsala, Sweden}

\author{Johan Rathsman}
\email{Johan.Rathsman@thep.lu.se}
\affiliation{Department of Astronomy
 and Theoretical Physics,
 Lund University, SE--223 62 Lund, Sweden}

\author{Glenn Wouda}
\email{Glenn.Wouda@physics.uu.se}
\affiliation{Department of Physics and Astronomy,
 Uppsala University, Box 516, SE--751 20 Uppsala, Sweden}


\begin{abstract}
We analyze a model for the Higgs sector with two scalar doublets and a $\ztwo$ symmetry that is manifest in the Yukawa sector but broken in the potential.
Thus, one of the doublets breaks the electroweak symmetry and has tree-level Yukawa couplings to fermions, whereas
the other doublet has no vacuum expectation value and no tree-level couplings to fermions. 
Since the $\ztwo$ parity is broken the two doublets can mix, which leads to a distinct and
novel phenomenology. This Stealth Doublet Model can be seen as a generalization of the Inert Doublet Model with a broken 
$\ztwo$ symmetry. We outline the model and present constraints from theory,
electroweak precision tests and collider searches, including the recent observation of a Higgs boson at the LHC. The charged scalar $\hp$ and the CP-odd scalar $A$ couple to fermions at one-loop level. We compute the decays of $\hp$ and $A$ and in particular the one-loop decays $A \to f \bar{f}$, $\hp \to f \bar{f}^\prime $,  $\hp \to W^\pm Z $ and $\hp \to W^\pm \gamma$. We also describe how to calculate and renormalize such processes in our model. We find that if one of $\hp$ or $A$ is the lightest scalar, $\hp\to W^\pm \gamma$ or $ A \to b \bar{b} $ are typically their respective dominating decay channels. Otherwise, the dominating decays of $\hp$ and $A$ are into a scalar and a vector. Due to the absence of tree-level fermion couplings for $\hp$ and $A$, we consider pair production and associated production with vector bosons and scalars at the LHC. If the parameter space of the model that favors $\hp \to W^\pm \gamma$ is realized in Nature, we estimate that there could be a considerable amount of such events in the present LHC data. 
\end{abstract}

\maketitle
\flushbottom

\section{Introduction}

The ATLAS \cite{ATLAShiggs} and CMS \cite{CMShiggs,Chatrchyan:2013lba} experiments at the Large Hadron Collider (LHC)
have after a long history of searches discovered a Higgs boson. By all accounts the properties of the  
observed particle agree within errors with what is expected of a Standard Model (SM) Higgs boson, but it 
will require much work to ascertain whether the SM Higgs doublet is all there is, or if
an extended Higgs sector exists. In earlier data there were some (not quite significant) 
hints of enhanced signal strengths in e.g.\ $H\to\gamma\gamma$, and moreover the results from ATLAS and CMS were not in complete agreement, but when all data from the first run of LHC are taken into account, the enhancement has disappeared and the two experiments agree, see e.g.\ \cite{Aad:2014eha,Khachatryan:2014ira} for the latest data on $H\to\gamma\gamma$. 
It is important to now probe and investigate the Higgs sector in detail to understand the observations
and what can be expected.

Much work has been dedicated to studying some standard scenarios for the electroweak symmetry breaking 
sector. Among these scenarios are the SM, the Minimal Supersymmetric Standard Model (MSSM), and
general CP-conserving two-Higgs doublet models (2HDMs). For the latter models one often imposes 
a, possibly softly broken, $\ztwo$ symmetry to prevent the occurrence of large flavor-changing neutral 
currents (FCNCs). General 2HDMs have been recently reviewed in Ref.~\cite{Branco:2011iw}.
Except for the SM, these models predict a set of additional Higgs bosons, each of which has 
characteristic production and decay channels for a given set of parameters.
 
In general CP-conserving models with two Higgs doublets there are two CP-even neutral Higgs bosons, 
$h$ and $H$, which have the same coupling structure to fermions and gauge bosons (up to mixing angles) 
as the SM Higgs. Their decay channels are the same as for the SM Higgs plus possible decays to 
lighter Higgs bosons. Of course their branching ratios can be very
different because of different coupling strengths and different decay channels being open.
There is additionally a CP-odd neutral
Higgs boson $A$, which mainly decays to the heaviest possible fermions, $A\to b\bar b$ or $t\bar
t$, or to a Higgs-vector boson pair, $A\to h Z$, $H^\pm W^\mp$. Finally, there is a charged Higgs boson $\hp$, which depending on its mass and couplings
decays mainly as $\hp\to\tau\nu$, $c s$ or $t b$, or as $\hp \to h W^\pm$ or $\hp \to A W^\pm$.

An alternative scenario is presented by the Inert Doublet Model (IDM)
\cite{Deshpande:1977rw,Ma:2006km,Barbieri:2006dq}, where there is a SM-like Higgs boson, but in addition
there is another doublet that is odd under a discrete $\ztwo$ symmetry. Making all other SM particles even 
under this symmetry and demanding that the Lagrangian is $\ztwo$ symmetric, the scalars from the other 
doublet become fermiophobic, i.e. do not couple to fermions. Thus, if the $\ztwo$ symmetry is exact the lightest
scalar from this doublet is stable, providing a possible dark matter candidate (see e.g.~\cite{Goudelis:2013uca,Krawczyk:2013jta} for constraints on the IDM from dark matter).
This makes for a very different phenomenology, so that if 
an alternative scenario such as the IDM or some other non-standard model is realized in Nature, the common searches
may prove inadequate.

The Stealth Doublet Model (SDM) studied in this paper was recently proposed in~Ref.~\cite{Enberg:2013ara}. 
It can be seen as a generalization of the IDM, but with the $\ztwo$ symmetry  broken in the scalar potential. 
This means that, in general, there is no stable scalar particle, but instead there are now two particles, $h$ and $H$,
that can play the role of the Higgs boson observed at LHC. In \cite{Enberg:2013ara} we showed that this model can describe the observations 
of ATLAS and CMS very well. In this paper we will study the model in more detail, and we will in particular study
some of the properties of the charged scalar $\hp$ and the CP-odd scalar~$A$.

As in the IDM, the $\hp$ and $A$ have no tree-level couplings to fermions, and must therefore be
produced and decay in different channels than in the standard
scenarios. However, contrary to the IDM, because of the broken $\ztwo$ symmetry, couplings to fermions
are now generated at the one-loop level. The usual decay channels of the $\hp$ and $A$ bosons 
into fermions are therefore loop suppressed in our model. Consequently, model-dependent constraints 
do not always apply, and $\hp$ and $A$ can be lighter in our model than in standard scenarios. For
example, the main decay of the charged Higgs is typically $\hp\to W^\pm\gamma$, provided that $\hp$ is the lightest scalar. 
Another example is that the production of the CP-odd Higgs $A$ through gluon-gluon fusion is strongly suppressed, but still the main 
decay channel is typically into $b \bar{b} $ as in the standard scenarios. 

Fermiophobic models have been discussed
previously~\cite{Diaz:1994pk,Akeroyd:1995hg,Akeroyd:1998ui,Barroso:1999bf,Brucher:1999tx,Gabrielli:2011bj,Gabrielli:2012yz},
for the case where the lightest CP-even Higgs boson is fermiophobic. Such a Higgs boson has an
increased branching ratio for $h\to\gamma\gamma$ but is not produced in $gg\to h$. In our model,
instead, the lightest CP-even Higgs boson has the same types of interactions as in standard 2HDMs, but the $H^\pm$ and $A$ are fermiophobic. Fermiophobic charged Higgs bosons have recently been discussed in~\cite{Celis:2013ixa}~and~\cite{Ilisie:2014hea}. 

As already mentioned,
a $\ztwo$ symmetry is usually imposed on 2HDMs in order to not run into dangerous FCNCs. 
One possibility is to arrange the symmetry such that only one of the
doublets couples to fermions. This is known as a Type-I Yukawa sector, and our model is an example
of such a Yukawa sector. It is worth pointing out that the model can not be obtained by simply taking the 
$\tan \beta \to 0$ or $\tan \beta \to \infty$ limit of a Type-I 2HDM with a broken 
$\ztwo$ symmetry, similarly as the IDM can not be obtained from a Type-I 2HDM with an exact $\ztwo$ symmetry~\cite{Barbieri:2006dq}. 
An additional motivation for considering Type-I models is that recent work
in string theory \cite{Ambroso:2008kb} seems to imply that they are generic in heterotic string
theories, where selection rules forbid additional Higgs doublets from coupling to fermions. 
Type-I models by definition have an exact $\ztwo$ symmetry in the Yukawa sector. 
As a consequence, if the symmetry is only broken  in the Higgs potential, then no
dangerous FCNCs are generated at tree-level. This also applies to our model, where new sources of FCNCs only appear at the two-loop level.

Furthermore, it is possible to avoid FCNC by imposing \textit{alignment} in the Yukawa sector \cite{Pich:2009sp}. In the Aligned 2HDM (A2HDM), the Yukawa couplings are governed by the three parameters $\tan \beta^{U,D,L}$ in place of the $\tan \beta$ parameter of the previously mentioned $\ztwo$-symmetrical 2HDMs. We note that our model is very similar to the fermiophobic limit of the A2HDM, see Section \ref{sec:yukawa}.\footnote{We also note that in Ref.\ \cite{Ilisie:2014hea}, which appeared some time after the first arXiv version of this paper, our calculations of the decay widths of fermiophobic $H^\pm$ presented in Section \ref{sect:decays} are reproduced in the A2HDM with compatible results.}  For recent analyses of the A2HDM we refer to \cite{Celis:2013ixa} and \cite{Ilisie:2014hea}. 

The organization of this paper is as follows: in section \ref{sect:model} we discuss the definition 
of the model and derive masses as well as define the free parameters of the model. We then consider constraints on the model from theoretical considerations and electroweak precision tests (EWPT) in Section~\ref{sect:constraints}. The recently observed Higgs boson at the LHC is discussed in the 
context of our model in Section~\ref{sect:collconstraints}. Decays of the scalar particles are 
discussed in Section~\ref{sect:decays}. Finally, we briefly discuss the collider phenomenology of the 
charged scalar and the CP-odd scalar in Section~\ref{SDMatColliders}. 
Some more technical matters are relegated to the appendices.

\section{The Stealth Doublet Model}\label{sect:model}

In this paper we construct and study a model with two scalar doublets where 
 only one of the doublets 
couples to fermions at tree-level. This is achieved by  imposing a $\ztwo$ 
symmetry in the Yukawa sector, which, however, is broken in the potential.
 In this section we will 
first analyze the scalar potential of the model. We will then derive the scalar mass 
eigenstates, and consider the free parameters and the constraints on them.
 We will finally consider the structure of the Yukawa couplings in section \ref{sec:yukawa}.

We will in the following refer to the model as the Stealth Doublet Model (SDM). The model has previously been presented in \cite{Enberg:2013ara} and in the conference proceedings \cite{Wouda:2010zz}.

\subsection{The scalar potential}
We introduce two $\grp{SU(2)}_L$-doublet, hypercharge $Y=1$, complex scalar fields $\Phi_{1,2}$, which may
be written in terms of their component fields as
\bel{eq:doublets}
\Phi_{1,2} = \binom{\varphi^+_{1,2}}{\varphi_{1,2}},
\ee
or in components $[\Phi_{1,2}]^+ = \varphi^+_{1,2}$ and $[\Phi_{1,2}]^0 = \varphi_{1,2}$. We then consider the most general gauge invariant and renormalizable scalar potential,
\begin{align}
\mathcal{V} \left[\Phi_1 , \Phi_2 \right]   &= M_{11}^2\Phi_1^\dagger\Phi_1+M_{22}^2\Phi_2^\dagger\Phi_2
-[M_{12}^2\Phi_1^\dagger\Phi_2+{\rm h.c.}]\nonumber\\
& +\half\Lambda_1(\Phi_1^\dagger\Phi_1)^2
+\half\Lambda_2(\Phi_2^\dagger\Phi_2)^2
+\Lambda_3(\Phi_1^\dagger\Phi_1)(\Phi_2^\dagger\Phi_2)
+\Lambda_4(\Phi_1^\dagger\Phi_2)(\Phi_2^\dagger\Phi_1)
\nonumber\\
& +\left\{\half\Lambda_5(\Phi_1^\dagger\Phi_2)^2
+\big[\Lambda_6(\Phi_1^\dagger\Phi_1)
+\Lambda_7(\Phi_2^\dagger\Phi_2)\big]
\Phi_1^\dagger\Phi_2+{\rm h.c.}\right\}\,,
\label{eq:Vpotential}
\end{align}
where all parameters are real except $\Lambda_{5,6,7}$ and $M_{12}^2$, which may be complex. In this
paper we are only concerned with CP-conserving models and will from now on assume all couplings to
be real. 

A priori there is no physical difference between the two fields $\Phi_1$ and $\Phi_2$ in the scalar potential
\protect\eq{eq:Vpotential}, since they have the same quantum numbers and transformation properties.
We will now consider the effect on the scalar potential \protect\eq{eq:Vpotential} 
of global $\grp{U(2)}$ transformations of the two doublets, 
$\Phi_a \to U_{ab}\Phi_b$ with $U\in \grp{U(2)}$. The potential is in general not invariant under such transformations, 
but since there is no difference between the doublets, any linear combination of them can be the physical fields. 

It is therefore convenient to define a \textit{basis} for the doublets in terms
of their vacuum expectation values (vevs) as
\begin{align}
 \vev{\Phi_1} &= \frac{1}{\sqrt{2}} \binom{0}{v_1} \\
 \vev{\Phi_2} &= \frac{1}{\sqrt{2}} \binom{0}{v_2 \, e^{\ii \xi}},
\end{align}
where $v^2=v_1^2 + v_2^2 \approx (246\, \gev)^2$ is the total vev, and where $\xi$ is a possible phase 
that could allow spontaneous CP~breaking, which we therefore set to zero. 
A particular choice of vevs $v_1$ and $v_2$ of the two doublets then corresponds to a choice of a particular basis,
and the $\grp{U(2)}$ transformations may be seen as 
changes of basis for the doublets, where the total vev is rotated between the doublets. Once again, the physics related to the scalar potential, such as the mass spectrum of the scalars, is not affected by basis transformations. 
(See \cite{Ginzburg:2004vp,Davidson:2005cw,Haber:2006ue,Branco:2011iw} for clear discussions of basis changes in 2HDMs.)

One particular example of $\grp{U(2)}$ transformations is the transformations belonging to the discrete $\ztwo$ subgroup,
\begin{align}
 \Phi_1 &\to  \Phi_1\\
 \Phi_2 &\to -\Phi_2.
\end{align}
The potential is in general not invariant under such transformations. The non-invariant terms are
the dimension-two operator $\Phi_1^\dagger\Phi_2+$h.c.\ with coupling $M_{12}^2$ and the
dimension-four operators $(\Phi_1^\dagger\Phi_1)(\Phi_1^\dagger\Phi_2)+$h.c.\ and
$(\Phi_2^\dagger\Phi_2)(\Phi_1^\dagger\Phi_2)+$h.c.\ with couplings $\Lambda_6$ and $\Lambda_7$. 

The $\ztwo$ symmetry is often imposed to remove these symmetry breaking terms. It is also imposed, with various schemes
for assignments of $\ztwo$ charges to fermions, in order to avoid large flavor-changing neutral
currents (FCNC)~\cite{Glashow:1976nt,Paschos:1976ay}, by arranging the Yukawa couplings such that 
each fermion only couples to one doublet. If the symmetry is broken, large FCNC may potentially occur, 
but in our model we will only encounter new sources of FCNC at the two-loop level (see section \ref{sec:yukawa} below).

If the fields $\Phi_1$ and $\Phi_2$ would only occur in the scalar potential (and in the kinetic terms), there would, as already mentioned, be no difference between them. However, 
once the fields are coupled to fermions and a specific structure for the Yukawa couplings is introduced, they are no longer equivalent and a particular basis is singled out as the physical one.

In our model, only one of the doublets, which we take to be $\Phi_1$, couples to fermions, and we will from now on therefore work 
in what is known as the \textit{Higgs basis},
which is precisely the basis where only $\Phi_1$ has
a vev (see Section \ref{sec:yukawa}). The vacuum 
expectation values (vevs) of the doublets are then
\begin{align}
 \vev{\Phi_1} &= \frac{1}{\sqrt{2}} \binom{0}{v} \\
 \vev{\Phi_2} &= \binom{0}{0},
\end{align}
where $v \approx 246\, \gev$. 

The minimization conditions for electroweak symmetry breaking in the Higgs basis become
\begin{align}
 m_{11}^2 &= \, -\f{1}{2} v^2 \lambda_1, \label{eq:minim11}\\
 m_{12}^2 &= \quad \f{1}{2} v^2 \lambda_6, \label{eq:minim12} 
\end{align}
giving no constraint on $m_{22}^2$, which is therefore a free parameter in this basis and in our model. From 
now on we will use lowercase letters to specify that we are working in the Higgs basis.

\subsection{Physical states and mass relations}
\label{sec:states}
We choose $\Phi_1$ to be the doublet that gets a vev, with $\ztwo$ parity $+1$, and $\Phi_2$ to
be the one with zero vev and $\ztwo$ parity $-1$. In a CP-conserving 2HDM, there are two CP-even
neutral states $h,H$, one CP-odd neutral state $A$, and two charged states $H^\pm$. We may then write the
doublets in the Higgs basis as
\begin{align}
 \Phi_1 &= \f{1}{\sqrt{2}} \binom{\sqrt{2} G^+}{v+\phi_1+ \ii G^0} \\
 \Phi_2 &= \f{1}{\sqrt{2}} \binom{\sqrt{2} H^+}{\phi_2 + \ii A} ,
\end{align}
where $G^\pm$ and $G^0$ are the Goldstone bosons and $\phi_{1,2}$ are the neutral CP-even interaction
eigenstates. The doublet $\Phi_2$ is fermiophobic, i.e., the states $H^\pm$, $A$, and $\phi_2$ do
not interact with fermions at tree-level. From now on, we will call
the mass eigenstates in our model ``scalars'', not Higgs bosons, in accordance with the usual IDM
nomenclature \cite{Barbieri:2006dq}.

The masses for the $A$ and
$\hp$ can be found directly from the potential,
\begin{align}
&m_A^2 \:  = m_{22}^2 + \half v^2 (\lambda_3+\lambda_4-\lambda_5) = m_\hp^2 - \half v^2 (\lambda_5-\lambda_4)  \label{eq:Amass}\\
&m_\hp^2 = m_{22}^2 + \half v^2 \lambda_3. \label{eq:Hpmass}
\end{align}
The mass matrix for the CP-even states has non-diagonal elements, and we
may find the physical mass eigenstates by diagonalizing this matrix. 
Taking the minimization conditions (\ref{eq:minim11}, \ref{eq:minim12}) into account, we have
\be \label{eq:massmatrix}
\M^2 = \begin{pmatrix}
\lambda_1 v^2 & \lambda_6 v^2 \\
\lambda_6 v^2 & \: m_{22}^2 + \lambda_{345} v^2 \\
\end{pmatrix}
= \begin{pmatrix}
\lambda_1 v^2 & \lambda_6 v^2 \\
\lambda_6 v^2 & \: m_A^2 + \lambda_5 v^2 \\
\end{pmatrix},
\ee
where $\lambda_{345}=\lambda_3+\lambda_4+\lambda_5$. The matrix $\M^2$ may be diagonalized by an
orthogonal matrix $V$, defined by a rotation angle $\alpha$, as
\be
\begin{pmatrix}
 m_H^2 & 0\\
0 & m_h^2\\
\end{pmatrix}
= V^T \M^2 V.
\ee
The physical CP-even states are then given by (with $\alpha$ defined so that $m_H > m_h$)
\begin{align}
 \begin{pmatrix}
  H\\h
 \end{pmatrix}
= V^T
 \begin{pmatrix}
  \phi_1\\
  \phi_2\\
 \end{pmatrix}
=
\begin{pmatrix}
 \quad \cosa & \sina \\
 -\sina & \cosa \\
 \end{pmatrix}
 \begin{pmatrix}
  \phi_1\\
  \phi_2\\
 \end{pmatrix},  \quad \text{where } \, -\frac{\pi}{2} \leq \alpha \leq \frac{\pi}{2} \,.
\end{align}
The physical CP-even scalar masses can be expressed as 
\begin{align}\label{eq:Hhmasses}
 m_h^2 & = c_\alpha^2 m_A^2 + s_\alpha^2 v^2 \lambda_1 +c_\alpha^2 v^2 \lambda_5 - 2 s_\alpha c_\alpha v^2 \lambda_6 \\
 m_H^2 & = s_\alpha^2 m_A^2 + c_\alpha^2 v^2 \lambda_1 + s_\alpha^2 v^2 \lambda_5 + 2 s_\alpha c_\alpha v^2 \lambda_6, \label{eq:Hhmasses2} 
\end{align}
where we defined the abbreviations $\sa\equiv\sina, \, \ca\equiv\cosa$.
Finally, we have the following explicit expressions for the potential parameters $\lambda_{1,3,4,5}$ in terms of the masses, the mixing angle $\alpha$, and the couplings $\lambda_6$ and $m_{22}^2$,
\begin{align}
\lambda_1 v^2 &= \frac{m_H^2 + m_h^2}{2} + \f{\left(m_H^2-m_h^2\right)}{2\cos 2 \alpha}  -  v^2 \lambda _6 \tan 2 \alpha
 \label{eq:lambda1} \\
\lambda_3 v^2 &= {2 \left(m_{H^\pm}^2-m_{22}^2\right) }
 \label{eq:lambda3} \\
\lambda_4 v^2 &= \frac{m_H^2 + m_h^2}{2} - \f{\left(m_H^2-m_h^2\right)}{2\cos 2 \alpha} + v^2 \lambda _6 \tan 2 \alpha +  m_A^2 - 2 m_{H^\pm}^2
 \label{eq:lambda4} \\
\lambda_5 v^2 &= \frac{m_H^2 + m_h^2}{2} - \f{\left(m_H^2-m_h^2\right)}{2\cos 2 \alpha} + v^2 \lambda _6 \tan 2 \alpha -  m_A^2,
 \label{eq:lambda5}
\end{align}
allowing us to use the masses of the scalars as parameters of the model. The mixing angle $\alpha$ is given by
\begin{align}
 \tan 2\alpha = \frac{2v^2\lambda_6}{ v^2(\lambda_1-\lambda_5)- m_A^2}  , \label{eq:t2a}
\end{align}
or, in terms of the masses and $\lambda_6$ only,
\be
\sin 2\alpha = \frac{2v^2\lambda_6}{m_H^2-m_h^2}. \label{eq:s2a}
\ee

Note that the mass relations eqs.\ \eref{eq:Amass}, \eref{eq:Hpmass}, \eref{eq:Hhmasses} and \eref{eq:Hhmasses2} are invariant under $\sin\alpha\to-\sin\alpha$. Equivalently, from  eqs.\ (\ref{eq:lambda1}--\ref{eq:lambda5}), the parameters $\lambda_1$, $\lambda_3$, $\lambda_4$ and $\lambda_5$ are also invariant. This is easily seen, since as we have $-\tfrac{\pi}{2} \leq \alpha \leq \tfrac{\pi}{2}$, the parameter $\sin\alpha$ can take any value $-1 \leq \sin\alpha \leq 1$, and  $\cos\alpha$ is always non-negative. This implies that under $\sin\alpha\to-\sin\alpha$, we have $\sin 2\alpha \to -\sin 2\alpha$ and $\lambda_6 \to -\lambda_6$.

Eqs.\ (\ref{eq:lambda1}--\ref{eq:lambda5}) are not valid in the case of maximal mixing, $\alpha = \pm \tfrac{\pi}{4}$. In this case one instead obtains
\begin{align}
\lambda_1 v^2 &= \frac{m_H^2 + m_h^2}{2}
\\
\lambda_3 v^2 &= {2 \left(m_{H^\pm}^2-m_{22}^2\right) }
\\
\lambda_4 v^2 &= \frac{m_H^2 + m_h^2}{2} +  m_A^2 - 2 m_{H^\pm}^2
\\
\lambda_5 v^2 &= \frac{m_H^2 + m_h^2}{2} -  m_A^2.
\end{align}

Eqs. \eq{eq:massmatrix} and \eq{eq:s2a} show that when the $\ztwo$ symmetry is exact ($\lambda_6=0$), the
mass matrix is diagonal and there will be no mixing between $h$ and $H$. This is the case in the
Inert Doublet Model; in fact all our results reduce to the IDM in the limit $\lambda_6\to 0$, $\lambda_7\to 0$ and $\sina\to 1$ or $-1$.\footnote{Note that in this case the relation $m_H>m_h$ is not valid, since no rotation is performed to diagonalize the mass matrix $\M^2$.} In this sense, our model is a
generalization of the IDM.

The scalar-scalar couplings depend on the potential parameters and are straightforward to obtain from the potential. 
The scalar-gauge boson couplings are obtained from the covariant derivatives and depend on the mixing angle only.
The relevant three-particle couplings are listed in Appendix \ref{sect:couplings}.

\subsection{Yukawa sector}
\label{sec:yukawa}
Now we are in a position to specify the Yukawa couplings of the model.
The most general Yukawa Lagrangian in the Higgs basis reads~\cite{Haber:2006ue}
\begin{equation}\begin{split}
-\mathcal{L}_{\:\text{Yukawa}}\, =  &\:\kappa^L_0 \bar{L}_L \Phi_1 E_R + \kappa^U_0 \bar{Q}_L (-\ii \sigma_2 \Phi_1^*)\, U_R +\kappa^D_0 \bar{Q}_L \Phi_1 D_R   \\
& + \rho^L_0 \bar{L}_L \Phi_2 E_R + \rho^U_0 \bar{Q}_L (-\ii \sigma_2  \Phi_2^*)\, U_R +\rho^D_0 \bar{Q}_L \Phi_2 D_R
\end{split}
\label{Lyuk}
\end{equation}
and is written in terms of the electroweak interaction eigenstates. In order to obtain the fermion mass
eigenstates, the matrices $\kappa^F_0, \rho_0^F$ ($F=U,D,L)$ are transformed by a biunitary transformation
that diagonalizes $\kappa^F_0$ using the unitary matrices
$V_L^F, V_R^F$ according to
\be
\kappa^F = V_L^F \kappa^F_0 V_R^F = \frac{\sqrt{2}}{v}M^F,\quad \rho^F = V_L^F \rho^F_0 V_R^F,
\ee
where $M^F$ is the diagonal mass matrix for fermions $F$, e.g.\ $\left[M^L\right]_{22} = m_\mu$ etc. 

The $\rho^F$ matrices are in general non-diagonal and will generate FCNC. 
However, in our model we demand the $\ztwo$ symmetry to only be broken in the potential part of the Lagrangian. Since the $\ztwo$ symmetry must be exact in $\mathcal{L}_{\:\text{Yukawa}}$, we impose $\rho^F =0$ at tree-level. As a result, $ \Phi_2 $ has no tree-level couplings to fermions, and therefore large FCNC are avoided.  The fermions will acquire mass through Yukawa couplings with the Higgs doublet $ \Phi_1 $ only. The Yukawa Lagrangian in unitary gauge then reads
\be
-\mathcal{L}_{\:\text{Yukawa}} = \: \frac{m_f}{v}\, \bar{\Psi}_f \Psi_f \,\phi_1 \: = \frac{m_f}{v}\, \bar{\Psi}_f \Psi_f \left( \, H \,\cos \alpha-  h \,\sin\alpha \, \right) ,
\ee
for all fermions $f$. As will be shown in sections \ref{hpff} and \ref{Aff} the soft breaking terms $m^2_{12}
\Phi_1^\dagger \Phi_2 + \text{ h.c.} $ will generate couplings between $\Phi_2$ and fermions, i.e.\
$\rho^F \neq 0$ at one-loop level. Furthermore, we will show in section \ref{hpff} that the $\rho^F$ matrices are diagonal and UV-finite at one-loop level. At higher orders in perturbation theory, $\rho^F $ will develop off-diagonal elements and
introduce additional sources of FCNC\footnote{In our model, just as in the SM, we will have, e.g., $h b \bar{s}$ couplings generated by a loop with two $W^\pm$~bosons with off-diagonal CKM matrix elements.}. Finally we also note that the couplings of fermions to $A$ and $\hp$ are  governed by $\rho^F$; more specifically we have terms of the form 
$\ii\bar{F}\rho^F\gamma_5 F A$ and $\bar{U}\left[V_{\rm CKM}\rho^D(1+\gamma_5) - \rho^U V_{\rm CKM}(1-\gamma_5) \right] D H^+$. 

It is interesting to compare our model with the A2HDM, where the Yukawa matrices are imposed to be aligned in the general basis ($\tan \beta \neq 0$)~\cite{Pich:2009sp}. This condition makes $\rho^F_0$ proportional to $\kappa^F_0$ and they can be diagonalized simultaneously, without invoking a $\ztwo$-symmetry. In this sense, our model can be seen as the fermiophobic limit of the A2HDM, where the alignment parameters are set to zero~\cite{Ilisie:2014hea}. 
It should be noted that, due to the lack of a $\ztwo$-symmetry in the A2HDM, the alignment of the Yukawa couplings in this model are in the general case not protected with respect to higher-order corrections. In other words, the alignment condition is in general not stable under  renormalisation group evolution (RGE) at the one-loop level as emphasized by Ferreira et al.~\cite{Ferreira:2010xe}. However, the special case of setting $\rho^F=0$ is stable at one-loop. Thus the structure of the Yukawa sector of the SDM is stable under RGE at this level.

Before ending this section, we want to emphasize that the physical basis, i.e.\ the fermionic structure, in the SDM and the A2HDM is not related to a particular value of $\tan \beta = v_2/v_1$. There are no observables that depend on $\tan \beta$, i.e., the relation between the physical Yukawa couplings $\rho^F$ and $\kappa^F$ is unchanged even if $\tan \beta$ is modified \cite{Pich:2009sp,Haber:2006ue}. Therefore $\tan\beta$ should be regarded as an auxiliary parameter. As a matter of principle one can of course work in an arbitrary basis, with a related value of $\tan\beta$. However, it is convenient to work in a specified basis and in this article, we choose to work in the previously introduced Higgs~basis.

\subsection{Parameters of the model}
\label{sec:params}
We consider models with CP conservation by imposing only real parameters and thus the scalar potential has ten free parameters.
The minimization conditions (\ref{eq:minim11}, \ref{eq:minim12}) remove $m_{11}^2$ and $m_{12}^2$,
leaving us with the eight parameters $\lambda_1$--$\lambda_7$ and $m_{22}^2$. We may use the relations
(\ref{eq:lambda1}--\ref{eq:lambda5}) to relate $\lambda_1$, $\lambda_3$, $\lambda_4$, and
$\lambda_5$ to the four physical scalar masses $m_h, \ m_H, \ m_A$ and $m_{H^\pm}$. The parameter $\lambda_6$ can be used to specify
the amount of $\ztwo$ breaking, but considering~eqs.~(\ref{eq:t2a},~\ref{eq:s2a}) we choose to
instead use the mixing angle $\alpha$ for this purpose, since in a general 2HDM $\sin(\alpha - \beta )$ 
is invariant under basis changes. 

Of the remaining $\lambda$-parameters, we note that
$\lambda_2$ only enters indirectly through the stability and tree-level unitarity constraints etc. to be discussed below,
 as its only direct effect is to set the strength of the self-interaction of the
$\Phi_2$ field, whereas, as we will see in more detail later, $\lambda_3$ and $\lambda_7$ 
govern couplings between the two doublets such as $g_{hH^+H^-}$.   
 Finally, we can relate $\lambda_3$ and $m_{22}^2$
using \refeq{eq:Hpmass}. We choose $\lambda_3$ as input parameter, as this parameter enters the
coupling between the CP-even states and pairs of charged scalars, see sections~\ref{sect:collconstraints}, \ref{sect:cpeven},
 and Appendix~\ref{sect:couplings} for more details.

The eight parameters of the model that we will use are then
\[ m_h, \ m_H, \ m_A, \ m_{H^\pm},  \ \sina , \ \lambda_2, \ \lambda_3, \ \lambda_7 . \]

To simplify our analysis we will often make the following assumptions. To start with, we choose $\lambda_2=\lambda_1$ and $\lambda_7=\lambda_6$. Sometimes we will also be using a set of representative values for 
$\lambda_3$,  chosen as $\lambda_3 = 0$, $2m_{\hp}^2/v^2$ and $4m_{\hp}^2/v^2$, 
corresponding to $m_{22}^2 = m^2_\hp$, 0  and $ -m^2_\hp$, respectively. 
In Sections \ref{sect:collconstraints}, \ref{sect:decays} and \ref{SDMatColliders}, we
will vary $\lambda_2$, $\lambda_3$ and $\lambda_7$, within theoretically allowed regions, to deduce their impact
on the signal strengths for $h \rightarrow \gamma \gamma$ and $H \rightarrow \gamma \gamma$, and the decays of $H$.

We must also consider bounds on the parameters from the requirement that the potential is bounded
from below \cite{Deshpande:1977rw,Sher:1988mj}.
Stability of the potential gives rise to a number of constraints on the parameters in the quartic
part of the potential. The simplest constraints are
\begin{align}\label{eq:boundedness}
 \lambda_1 > 0 , \qquad
 \lambda_2 > 0 , \qquad
 \lambda_3 > -\sqrt{\lambda_1 \lambda_2}, \qquad
 \lambda_3 + \lambda_4 - \lambda_5 > -\sqrt{\lambda_1 \lambda_2},
\end{align}
where the last equation applies for $\lambda_6\ne 0$ or $\lambda_7\ne 0$.
There are also additional constraints that we do not list here, which can be found in references \cite{Deshpande:1977rw,Sher:1988mj,Kaffas:2006nt}. 
In addition, one can
also constrain the parameters by requiring 
perturbativity of the various four-Higgs couplings and tree-level unitarity as we will return to below in  section \ref{sect:constraints}.

\section{Constraints on the SDM}\label{sect:constraints}

\begin{figure}[t]
\begin{center}
\begin{tabular}{cc}
\includegraphics[width=0.48\textwidth]{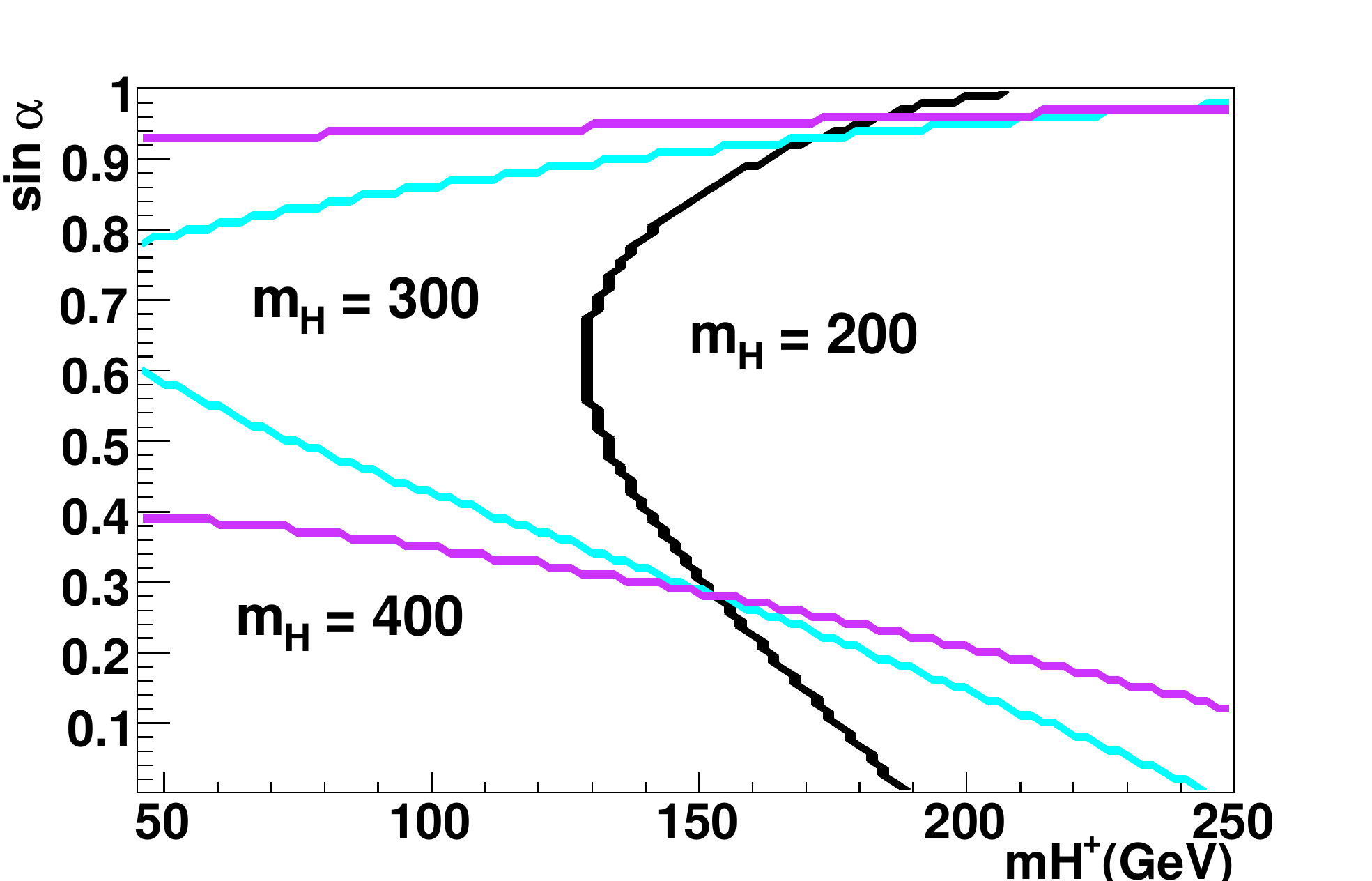}  &  \includegraphics[width=0.48\textwidth]{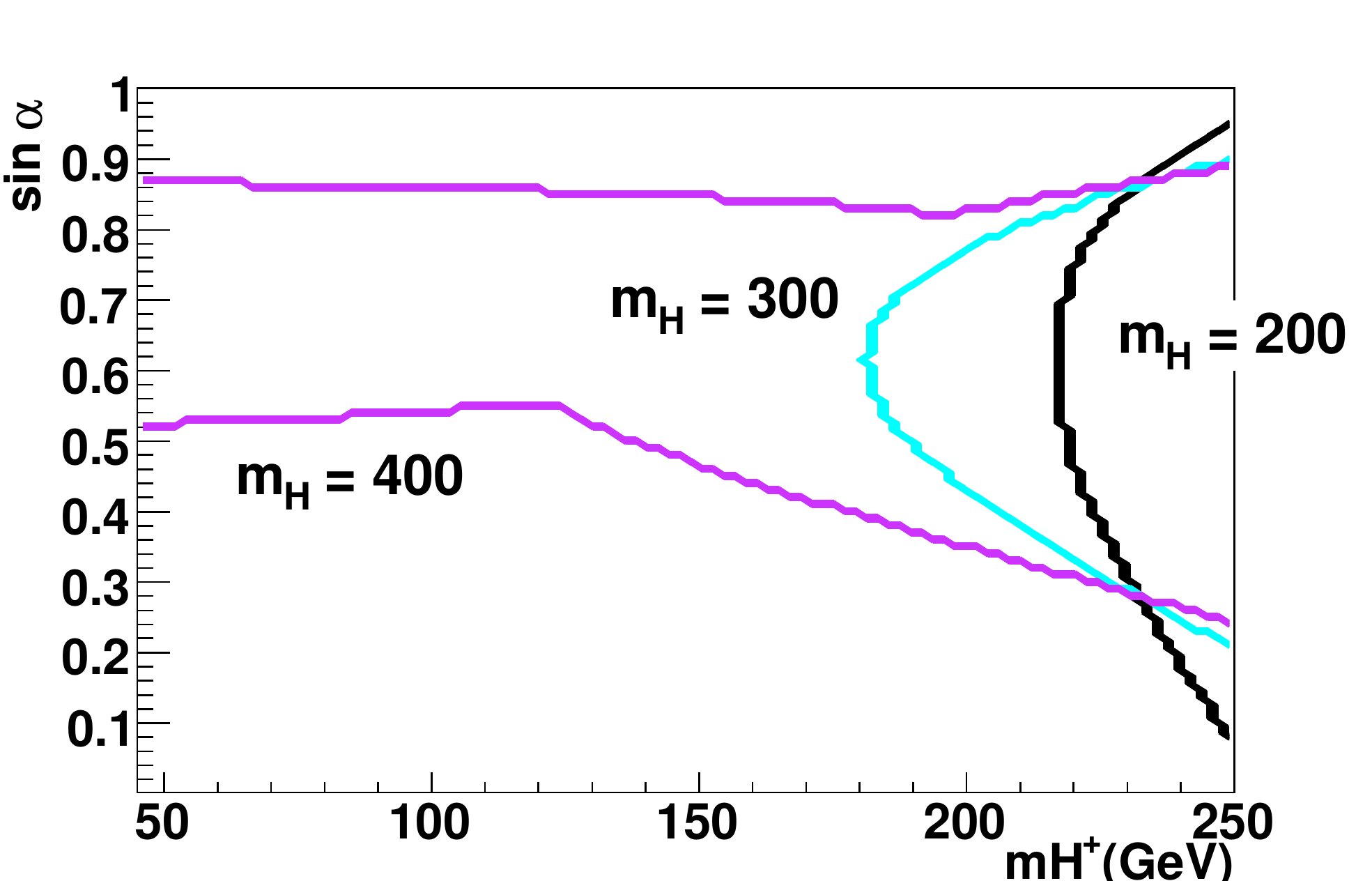} \\
(a) & (b)\\
$m_h = 125$  GeV, $m_A = m_{H^\pm}, \, \lambda_3 = 0$.    & $m_h = 125$ GeV, $m_A = m_{H^\pm}, \,  \lambda_3 = 1$.  
\end{tabular}
\end{center}
\caption{Contours displaying allowed regions in parameter space (to the left of/above/below the contour lines), taking into account the theoretical constraints
of stability, tree-level unitarity and perturbativity. The black contour displays the allowed region for $m_H = 200$ GeV, cyan $m_H = 300$~GeV and magenta $m_H = 400$ GeV. Here, we have used $\lambda_2 = \lambda_1$
and $\lambda_7 = \lambda_6$, which makes the allowed regions depend only on $|\sin \alpha |$. }
\label{fig:teorcons}
\end{figure}

Apart from the constraints discussed above, namely that we require electroweak symmetry breaking with a vacuum
bounded from below, we impose several other theoretical and
experimental constraints on the model. All of the constraints discussed in this section are
included in our numerical work by using the two-Higgs doublet model calculator
\THDMC~\cite{Eriksson:2009ws,Eriksson:2010zzb}, where we have implemented our model as a special
case.

The electroweak vacuum selected by the symmetry breaking mechanism must be stable, which requires
that the potential should be bounded from below for any values of the fields. We also impose the requirements
that tree-level scattering of scalars and longitudinal $W$ and $Z$ bosons must be unitary at 
high energies (the eigenvalues $L_i$ of the $S$-matrix elements fulfill $|L_i| \leq 16 \pi$)
\cite{Huffel:1980sk,Maalampi:1991fb,Kanemura:1993hm,Akeroyd:2000wc,Ginzburg:2005dt}, and 
that the quartic scalar couplings are perturbative $|C_{{h_i}{h_j}{h_k}{h_\ell}}| \leq 4 \pi$. 
We will collectively call these constraints  ``theoretical constraints''. Two examples of the 
allowed regions in the parameter space of the model are shown in \reffig{fig:teorcons}. 
For simplicity we choose $\lambda_2 = \lambda_1$ and $\lambda_7 = \lambda_6$, which makes the allowed regions depend only on $|\sina|$.

In general, one could also consider constraints from renormalization group evolution of Yukawa 
couplings and masses in a similar way as in \cite{Bijnens:2011gd,Goudelis:2013uca}. Furthermore, 
one could consider constraints on metastable vacua as in \cite{Barroso:2012mj,Barroso:2013awa}. However, this is beyond the scope of this study.

Any model with new particles that couple to gauge bosons can potentially lead to large contributions
to the gauge boson self-energies. Such corrections are constrained by experimental measurements,
and can be parametrized by the oblique Peskin--Takeuchi $S$, $T$, and $U$ parameters~\cite{Peskin:1991sw},
which are defined in terms of contributions to the vacuum polarizations of the electroweak gauge
bosons. In particular, the $T$ parameter is proportional to the deviation from the SM value of the $\rho$ parameter
$\rho=m_W^2/(m_Z^2\cos^2\theta_W)$. We do not list the explicit expressions here, which are lengthy
and involve all scalars. It should be noted that $S$, $T$ and $U$ do not depend explicitly on the parameters in \Eref{eq:Vpotential} 
but only implicitly through the scalar masses of the model, equations (\ref{eq:Amass}), 
(\ref{eq:Hpmass}) and (\ref{eq:Hhmasses}). Additionally, the mixing angle $\alpha$ only enters 
as $s_{\alpha}^{2}$ and $c_{\alpha}^{2}$, so $S$, $T$ and $U$ do not depend on the sign of $\sa$.

We use \THDMC{} to compute the oblique parameters $S,T$ and $U$ and 
require the obtained values of $S$ and $T$ to fall within the 90\% C.L.\ ellipse of
figure~10.7 in \cite{PhysRevD.86.010001}. This ellipse is given by values of constant
$\mathcal{E}_{\text{ST}}\, (S,T)$, where, approximately,
\be
\mathcal{E}_{\text{ST}}\, (S,T) =
\left( \frac{\tilde{S} \cos \theta + \tilde{T} \sin \theta }{0.224} \right)^2 + \left( \frac{\tilde{T} \cos\theta  - \tilde{S}\sin \theta }{0.068} \right)^2 ,
\label{SSTT}
\ee
with $\theta = 0.753$, $\tilde{S} = S - 0.051$ and $\tilde{T} = T - 0.077$. In other words, figure~10.7 in
\cite{PhysRevD.86.010001} shows the  $\mathcal{E}_{\text{ST}}\, (S,T) = 1$ ellipse.
We use the reference value $m_H^\text{ref} = 125$~GeV, which is
to be compared with the values $115.5 < m_H^\text{ref} < 127$~GeV used in
\cite{PhysRevD.86.010001}, where $U$ was fixed at $U=0$, the expected result for models
without anomalous gauge couplings. We find that for parameter points in our model with allowed $S$ and 
$T$
values, we have $ 0 \lesssim U \lesssim 0.02 $. 

\begin{figure}[t]
\begin{centering}
\begin{tabular}{cc}
\includegraphics[width=0.48\textwidth]{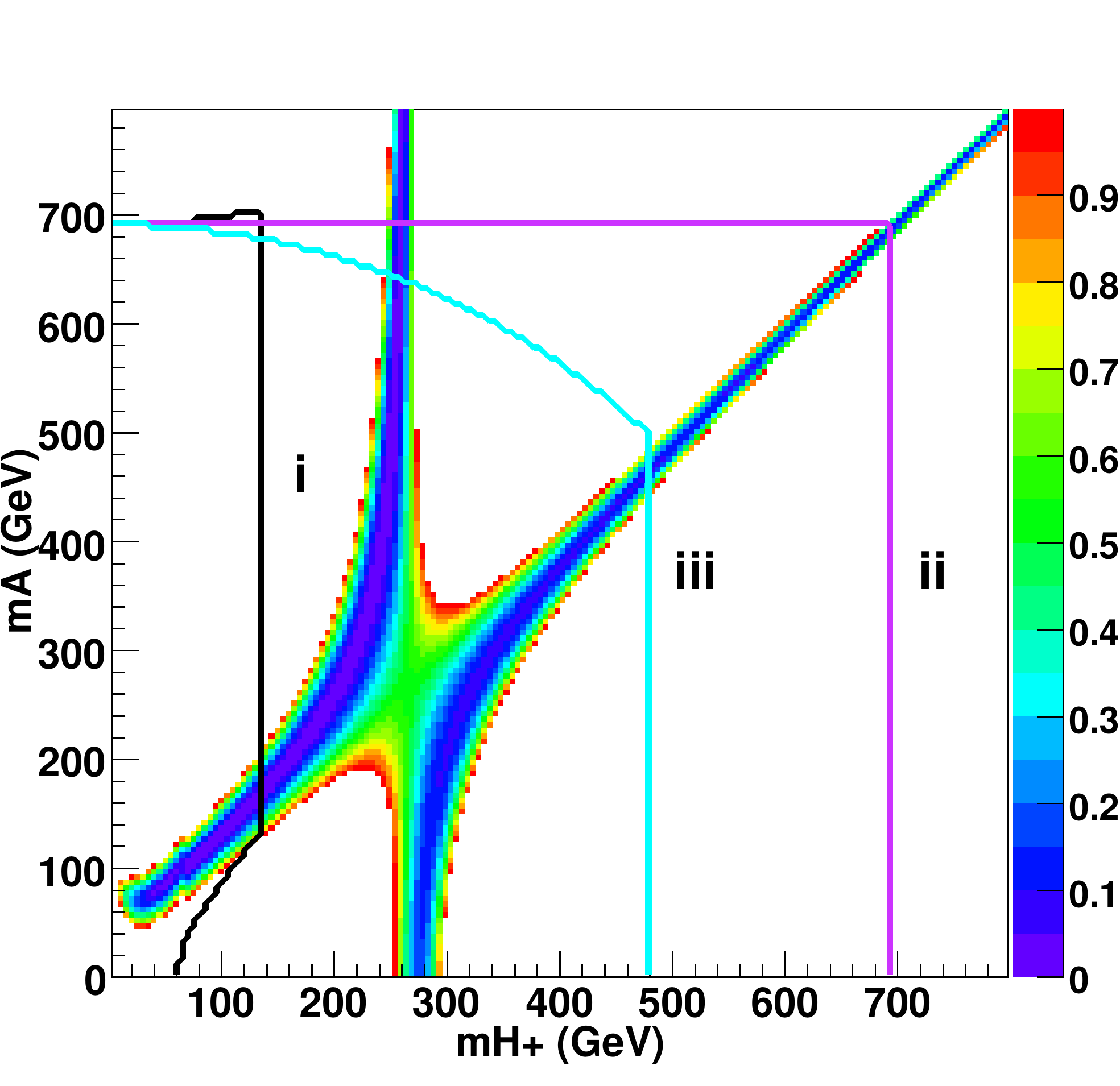}  & $\:$   \includegraphics[width=0.48\textwidth]{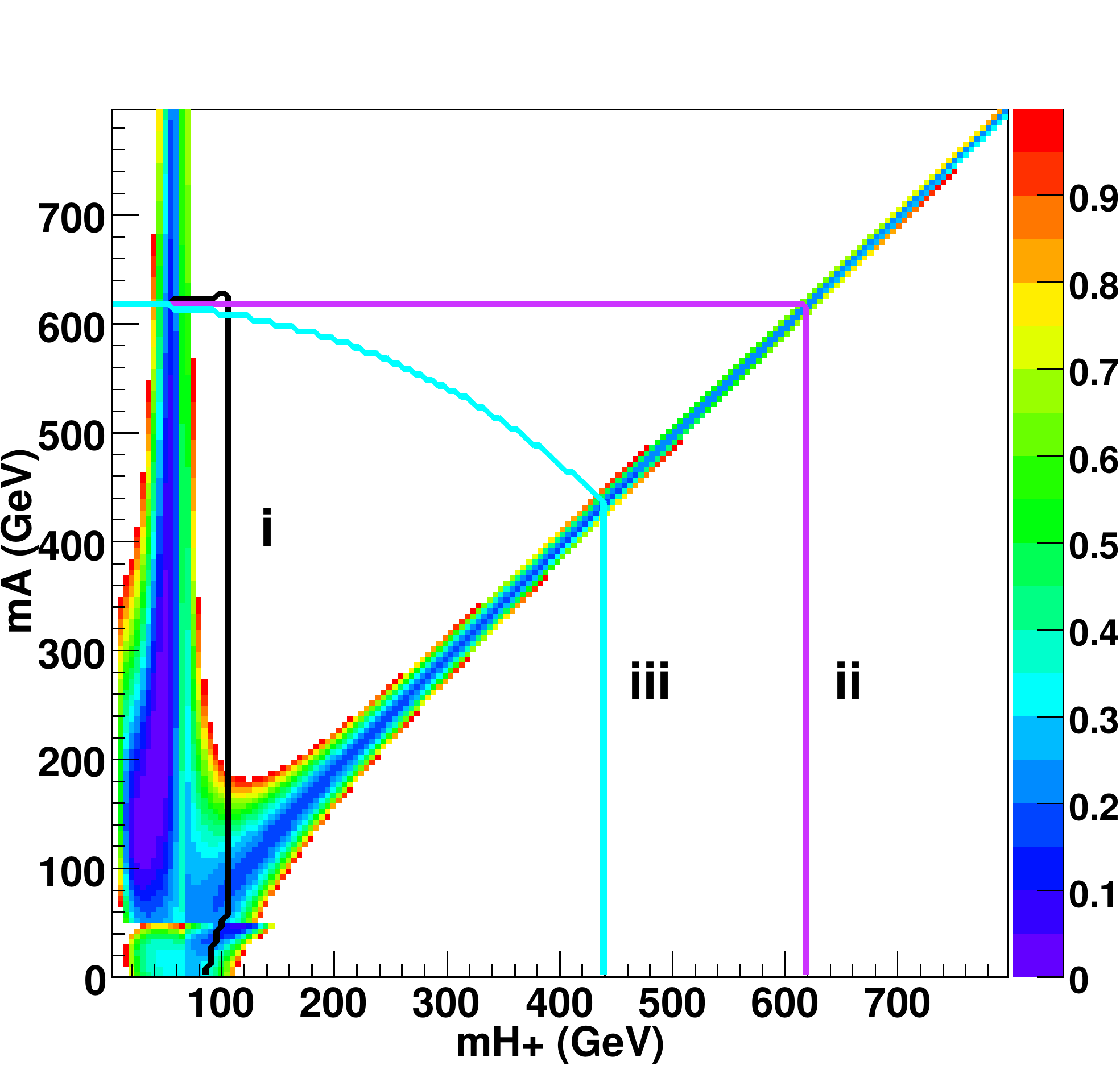} \\
(a) & $\quad$  (b)\\
$m_h = 125 $ GeV, $\, m_H = 300$ GeV,  $s_\alpha = 0.9 $.   & $\quad$  $m_h = 75$ GeV, $m_{H} = 125$ GeV,  $s_\alpha = 0.1 $. 
\end{tabular}
\caption{Some examples of allowed regions in parameter space taking into account theoretical
constraints and experimental $S$ and $T$ values. The $x$-axis shows the charged scalar mass
$m_{H^\pm}$ and the $y$-axis the CP odd scalar mass $m_A$. The $z$-axis displays the value of
$\mathcal{E}_{\text{ST}}\, (S,T) $ if it fulfills $ \mathcal{E}_{\text{ST}} \leq 1.0 $, see
\Eref{SSTT}. The regions to the left of the lines in the figure are the allowed by theoretical 
constraints for the different values of $\lambda_3$ indicated: black (i) $\lambda_3 = 0$, 
magenta (ii) $\lambda_3 = 2 m_\hp^2 /v^2$ and cyan (iii) $\lambda_3 = 4 m_\hp^2 /v^2$. 
Here, we have also used $\lambda_2 = \lambda_1$ and $\lambda_7 = \lambda_6$. }
\label{fig:constraints}
\end{centering}
\end{figure}

In \reffig{fig:constraints} we show some examples
of regions satisfying the experimental constraints on the $S$ and $T$ parameters as well as the theoretical constraints discussed above.
We note that in our model, there are two candidates for the new observed Higgs boson,~$\mathcal{H}$, with
mass $m_{\mathcal{H}} \approx 125$~GeV:\ either the lightest CP-even scalar $h$, or the
heaviest~$H$. We will in the following refer to the scenario $m_h = 125$ GeV  as ``Case~1'' and to $m_H
= 125$ GeV as ``Case~2''. In section \ref{sect:collconstraints} we will see that in
order to accommodate the experimentally observed signal strengths, $|\sin \alpha|$ must be close to
unity in Case~1 and small in Case~2. Motivated by these relationships between $m_{h,H}$ and $\sina$,
we present the
constraints in the ($m_\hp,m_A$)-plane from theory and $S$ and $T$ parameters, using $\sina =
0.9$ for $m_{h} = 125$~GeV, and $\sina=0.1$ for $m_{H} = 125$ GeV
in \reffig{fig:constraints}.

We also present the boundaries for different values
of $\lambda_3$ (corresponding to the three values $m^2_{22} = 0$ and $m^2_{22} = \pm m^2_\hp$, according to \Eref{eq:lambda3}), shown as the
regions inside the black, magenta and cyan lines in \reffig{fig:constraints}. First of all, we see
that in order to satisfy the theoretical constraints, the scalar masses can typically not exceed
$\sim$ 700 GeV. Secondly, as noted in \cite{Mahmoudi:2009zx} for 2HDMs, in order to have a small contribution to the $S$ and $T$ parameters,
the $\hp$ and $A$ masses must satisfy an approximate custodial symmetry (the two branches in the figure). If we define~\cite{Mahmoudi:2009zx}
\begin{equation}
M^2 \equiv m_h^2 \cos^2 \alpha + m_H^2 \sin^2 \alpha,
\label{eq:ms2}
\end{equation}
then there is an approximate custodial symmetry if either $m_A \approx m_\hp + \,50$ GeV when $m^2_\hp \lesssim M^2  $, or $m_A \approx m_\hp$ when $m^2_\hp \gtrsim M^2  $,  or $ 0 \lesssim m_A \lesssim 700 $ GeV when $ m^2_\hp \approx M^2 $.

When presenting the results in \reffig{fig:constraints} we
use $\lambda_2 = \lambda_1$ and $\lambda_7 = \lambda_6$ for simplicity, but the results 
are not sensitive to the precise values chosen. It is always possible to find parameters such 
that $m_\hp,m_A$ up to $\sim 700$ GeV are allowed.

In models with charged scalars $\hp$, any Feynman diagram that contains a $W^\pm$ 
also occurs with a $\hp$. In particular, this 
will affect low energy observables such as decay widths of $B$-mesons. By considering 
the effects of $\hp$ and $A$ on low energy observables, one can indirectly constrain e.g.\ 
$\mhp$ for a given set of couplings $C_{\hp f \bar{f}^\prime}$, or in other words 
$\rho^F$. For a discussion 
of the impact of constraints from meson decays on $\hp$ in general 2HDMs we refer 
to e.g.\ Ref.~\cite{Mahmoudi:2009zx}. In our model, we will assume that the sizes 
of the loop-induced couplings between $\hp$ and fermions are well below current 
limits from such flavor observables. In other words, indirect constraints from 
flavor observables do not apply to the $\hp$ and $A$ of our model. The only direct, model 
independent, constraint prior to LHC that applies to our $\hp$ is the measurement 
of $\Gamma_Z$, which gives the limit $m_{\hp}>39.6$~GeV~\cite{Abdallah:2003wd}.

\section{The SDM and the observed Higgs boson at the LHC}
\label{sect:collconstraints}

In this section, we include collider constraints in our analysis of the SDM parameter space. This is implemented through the
\THDMC{} interface to \HiB{} (version 4.1.3)~\cite{Bechtle:2008jh,Bechtle:2011sb}, which includes Higgs searches at LEP, the Tevatron and the LHC. Limits on $m_{\hp}$ and $m_A$ are not tested with \HiB{}, since \THDMC{} only calculates tree-level branching ratios for the charged scalar $\hp$ and the CP-odd scalar~$A$, see section \ref{sect:hpdecays} and further below. We will refer to the recently discovered Higgs boson as $\hcal$ and the SM Higgs boson as $H_{\text{SM}}$. 

We here mainly consider the $\gamma \gamma$-channel, which was the most significant channel in the discovery of $\hcal$. Studies of the impact of the $\gamma \gamma$-signal on the IDM has been studied in~e.g.~\cite{Arhrib:2012ia,Swiezewska:2012eh,Goudelis:2013uca,Krawczyk:2013jta}. In ref.~\cite{Chiang:2013ixa} constraints on general 2HDMs with a softly broken $\ztwo$ symmetry and $\tan \beta \neq 0$ are studied in the light of the new LHC data.

The ATLAS experiment previously observed a small excess in the signal strength $ \gamma \gamma$ compared to the SM, which was in slight disagreement with the CMS measurement. With the higher statistics of the most recent data, this excess is no longer present and the two experiments are compatible.

The signal strength $ \mu_{\hcal \, \gamma\gamma}$ is defined as
\begin{equation}
 \mu_{\hcal \, \gamma\gamma}  = \frac{\sum_k \sigma_{k} ( pp \to \hcal + X_k ) \, \times \, \text{BR}(\hcal \to \gamma \gamma)  }{  \sum_k \sigma_{k} ( pp \to H_{\text{SM}} + X_k ) \, \times \, \text{BR}(H_{\text{SM}} \to \gamma \gamma) } \: \: ,
\end{equation}
where $ \hcal = h,H $ in our model, and $\sigma_k$ are the gluon-fusion and vector boson fusion (VBF) hadronic cross sections. The signal strength for other channels, such as $\mu_{\hcal \, ZZ}$, are defined in an analogous way.

At the time of writing, ATLAS reports for the $\hcal \to \gamma\gamma$ channel the signal strength
$\mu_{\hcal \gamma \gamma} = 1.17\pm 0.27$
at a mass of $m_\hcal = 125.4 \pm 0.4$ GeV \cite{Aad:2014eha} whereas CMS reports $\mu_{\hcal \gamma \gamma} = 1.14^{+0.26}_{-0.23}$ at a mass of $ m_\hcal=124.70\pm 0.34$ GeV \cite{Khachatryan:2014ira}.
In the $\hcal\to ZZ \to 4 \ell$ channel, ATLAS measures the signal strength $ \mu_{\hcal ZZ} = 1.44^{+0.40}_{-0.33}$ at the mass $ m_\hcal = 125.36$ GeV  \cite{Aad:2014eva} and the CMS experiment obtains the signal strength $ \mu_{\hcal ZZ} = 0.93^{+0.29}_{-0.25}$ at $m_\hcal = 125.6\pm  0.45 $ GeV  \cite{Chatrchyan:2013mxa}. 
We also note that CMS reports a combined best fit value for all decay channels of 
$ \mu_{\hcal} = 1.00^{+0.14}_{-0.13}$ with a best-fit mass  of 
$m_\hcal = 125.03^{+0.29}_{-0.31}$~GeV~\cite{CMS-PAS-HIG-14-009}.

In the following, we will use the weighted averages of the ATLAS and CMS signal strengths. We use symmetric errors, choosing in the case of asymmetric errors the smaller of the two in order to be conservative and reject a larger portion of parameter space. This gives $ \mu_{\hcal \gamma\gamma} = 1.15 \pm 0.35$ and $\mu_{\hcal ZZ} = 1.12 \pm 0.41 $.

In our model, where $\hcal = h,H$, the signal strength $\mu_{\hcal \, \gamma\gamma}$ becomes 
\begin{equation}
\mu_{h\gamma\gamma} =  \sin^2 \alpha \, \frac{ \text{BR}(h \to \gamma \gamma)}{\text{BR}(H_{\text{SM}} \to \gamma \gamma)} \:, \quad \mu_{H\gamma\gamma} =   \cos^2\alpha \, \frac{ \text{BR}(H \to \gamma \gamma)}{\text{BR}(H_{\text{SM}} \to \gamma \gamma)} \:,
\end{equation}
at leading order see~\eref{eq:pphH}. This is because the $h$ couples as $\sina$ both to quarks in the
$gg$-fusion process and to vector boson pairs  in VBF, whereas $H$ couples as $\cosa$. 

\begin{figure}[bt]
\begin{center}
\begin{tabular}{cc}
\includegraphics[width=0.25\textwidth]{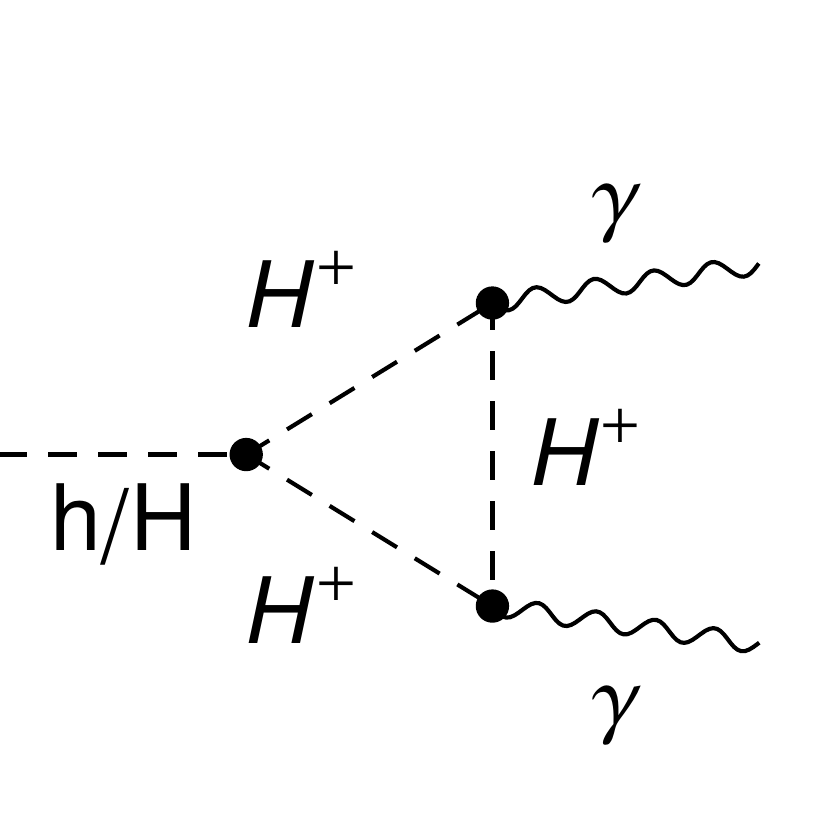}  &    \includegraphics[width=0.25\textwidth]{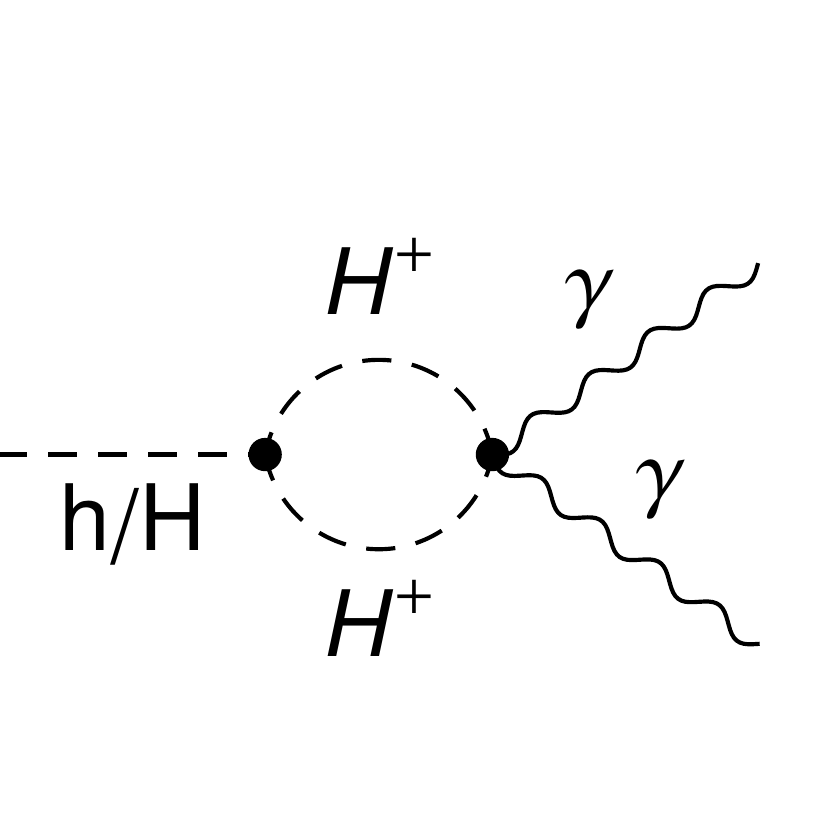}
\end{tabular}
\end{center}
\caption{ The two additional Feynman diagrams for the process $\hcal \to \gamma \gamma$ in 2HDMs, $\hcal = h, H$. }
\label{hgagadiag}
\end{figure}

The matrix element for $\hcal \to \gamma \gamma $ at lowest order in 2HDMs, and in particular in our model, has contributions from two additional Feynman diagrams compared to the SM, with a pair of charged scalars in the loop, as shown in \reffig{hgagadiag}. These two diagrams contain the couplings between $\hcal$ and $H^+H^-$
\begin{equation}
g_{hH^+H^-} = - \ii v \left( -  \lambda_3\sina + \lambda_7 \cosa  \right)  ,\quad g_{HH^+H^-} = - \ii v \left(  \lambda_3 \cosa + \lambda_7 \sina  \right)  \, .
\label{hphmhco}
\end{equation}
The inclusion of the charged scalars in the loop can enhance the $\Gamma_{\hcal \to \gamma \gamma } $ and BR$(\hcal \to \gamma \gamma ) $ compared to the SM and therefore also $\mu_{\hcal \gamma \gamma}$. 

\begin{figure}[bt]
\begin{center}
\includegraphics[width=0.6\textwidth]{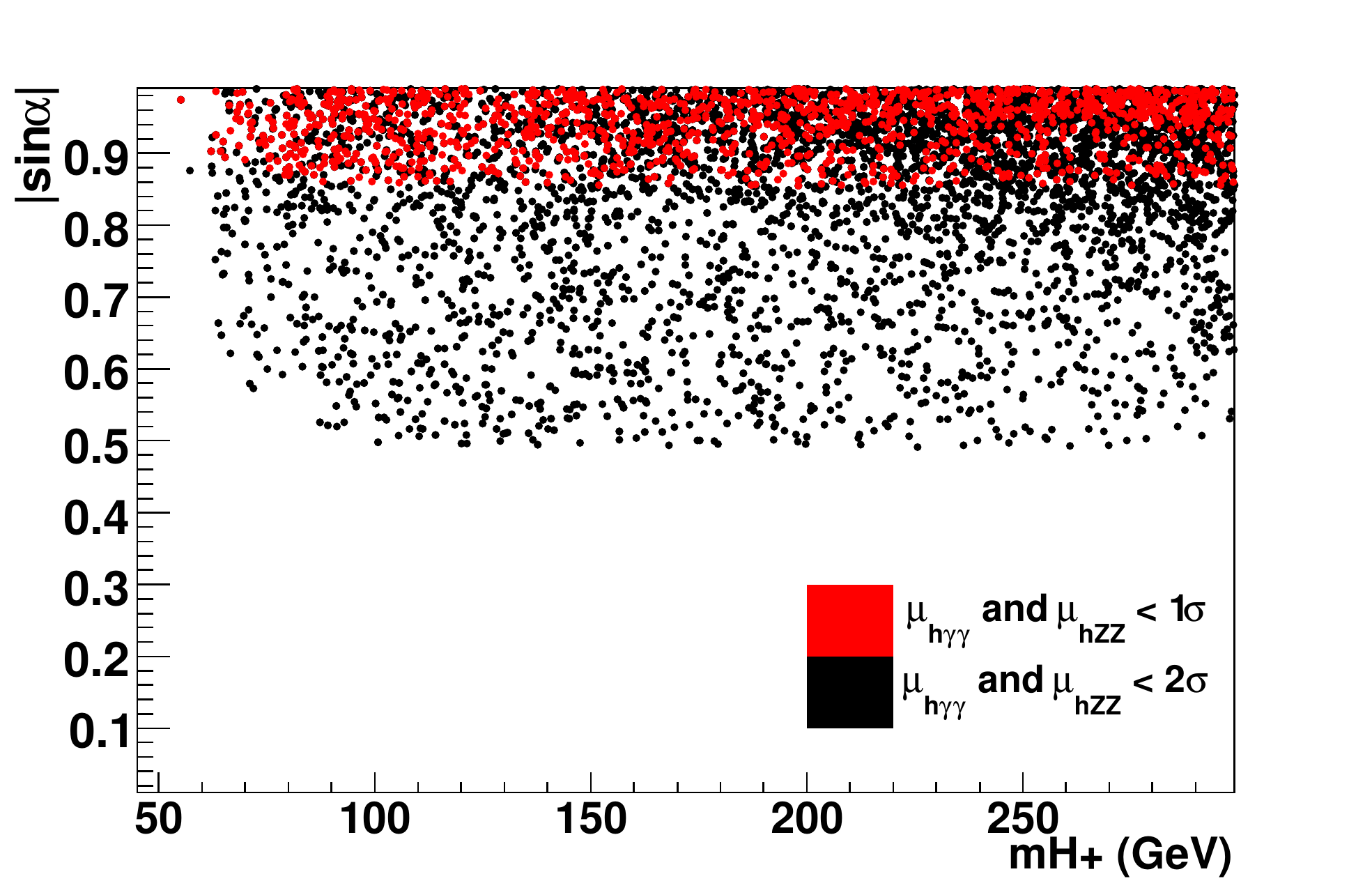}
\end{center}
\caption{Points in Case 1, with $m_h = 125$ GeV and $m_H = 300$ GeV, that satisfy all constraints from theory, collider searches with the use of \HiB{} version 4.1.3. The red (black) points have both the predicted $\mu_{h \gamma\gamma}$ and $\mu_{h ZZ} $ within $1\sigma$ ($2\sigma$) from their experimental values given in the text. The scan is described in the text. }
\label{fig:collider3}
\end{figure}

\begin{figure}[bt]
\begin{center}
\begin{tabular}{cc}
\includegraphics[width=0.48\textwidth]{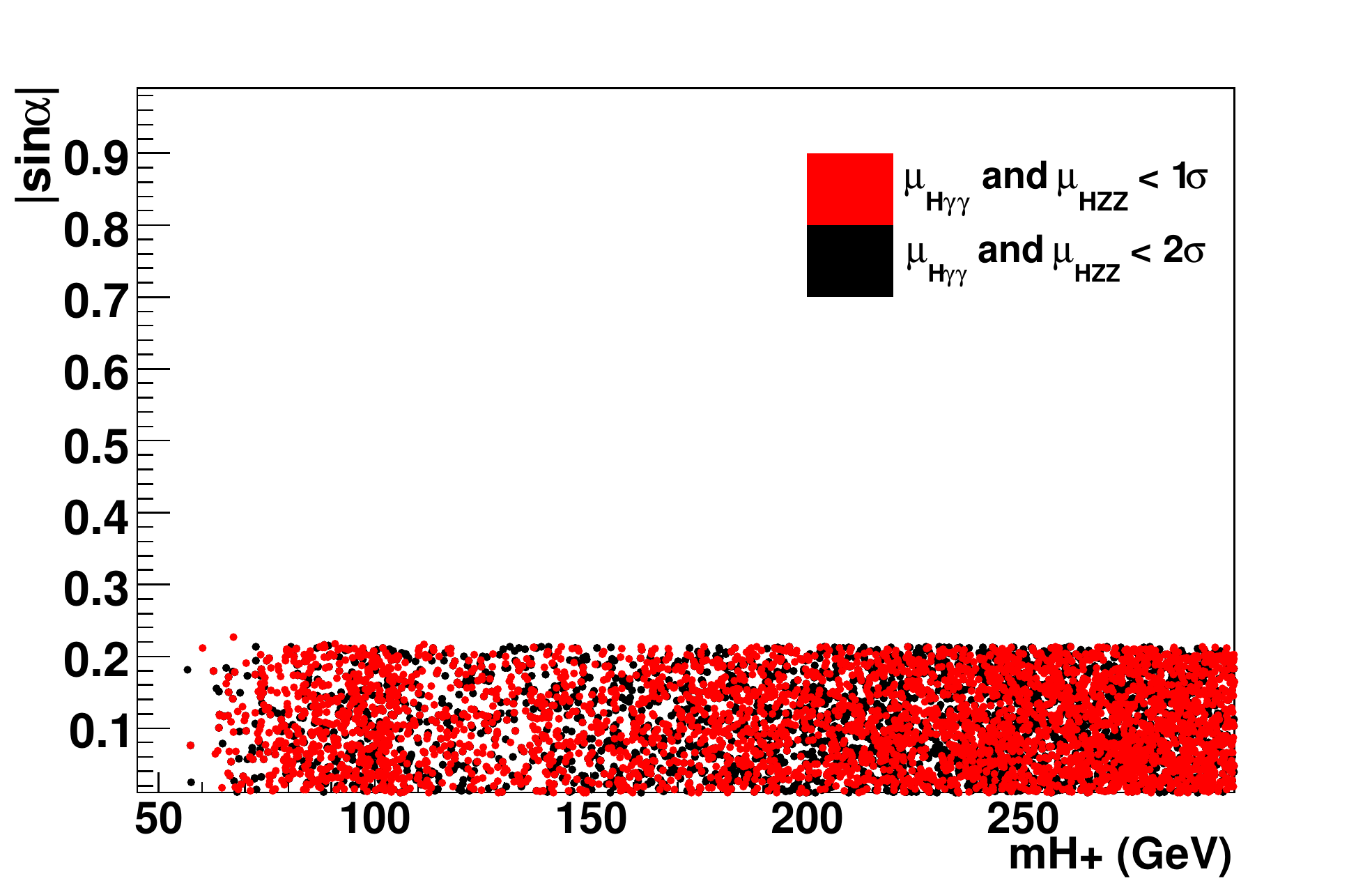} & \includegraphics[width=0.48\textwidth]{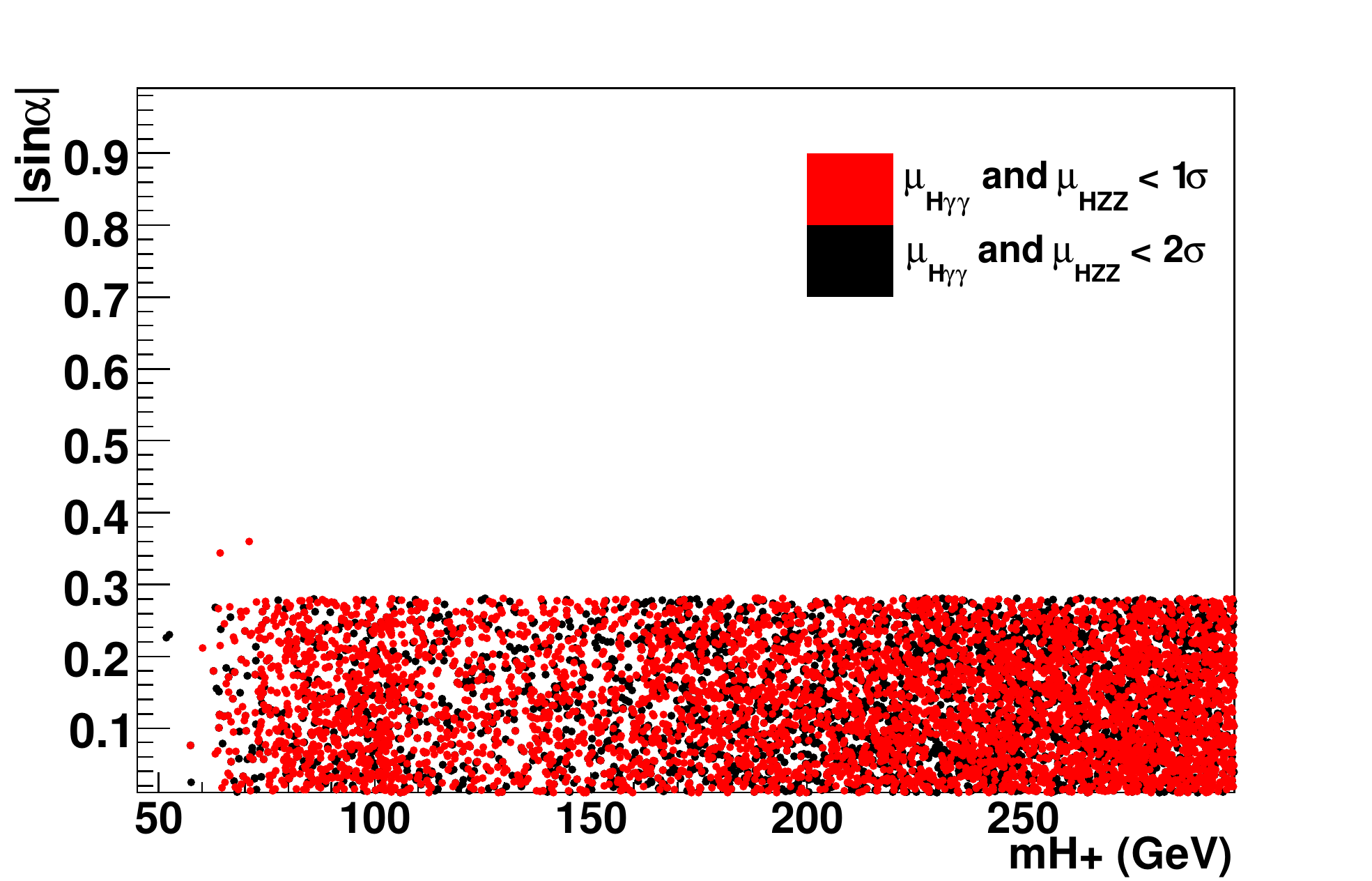} \\
(a) & (b) \\
$m_h = 75$ GeV and $m_H = 125$ GeV. & $m_h = 95$ GeV and $m_H = 125$ GeV. 
\end{tabular}
\end{center}
\caption{Points in Case 2 that satisfy all constraints from theory, collider searches with the use of \HiB{} version 4.1.3. The red (black) points have both the predicted $\mu_{h \gamma\gamma}$ and $\mu_{h ZZ} $ within $1\sigma$ ($2\sigma$) from their experimental values given in the text. The scan is described in the text.}
\label{fig:collider4}
\end{figure}

In order to deduce the regions of parameter space in our model that are compatible 
with the experimentally observed $\gamma \gamma$ and $ZZ$ signal strengths and that
satisfy constraints from EWPT, theory and limits from previous collider experiments 
(through \HiB), we scan in the ($\mhp,\sina$)-plane over the $\lambda_2$, $\lambda_3$ and $\lambda_7$ parameters. 

The scan proceeds by sampling uniformly from the following intervals: 
\begin{equation}
\begin{split}
&m_{\hp} \in [45,300] \, \text{GeV}\, ,\quad | \sin \alpha | \in [0,1] \,, \\  
&\lambda_2 \in [0,4\pi]\, , \quad \lambda_3 \in [-\sqrt{\lambda_1 \lambda_2}\,,4\pi]\, , \quad \lambda_7 \in [-4\pi,4\pi].
\end{split}
\label{Lrandomscan}
\end{equation}
We need only consider $ | \sin \alpha |$ since the allowed region is independent on the sign of $\sina$.

In Case~1, $m_A$ is taken as $m_A = m_\hp + 50$ GeV in order to fulfill the constraints from EWPT. In Case~1 we also use $m_H = 300$ GeV as a representative value. In Case~2 we use $m_h = 75$ or 95~GeV with $m_A = m_\hp$ to fulfill EWPT constraints. The allowed points that satisfy all the constraints are shown in figures \ref{fig:collider3} and \ref{fig:collider4} for Case~1 and Case~2 respectively, showing points within $1\sigma$ and $2\sigma$ of the experimental measurement.

We find an allowed region for Case~1 compatible with observed signal strengths,
such that $ | \sina | \gtrsim 0.85 $ at $1\sigma$ or $ | \sina | \gtrsim 0.5 $ at $2\sigma$. 
For Case~2, with $m_h = 75$~GeV or $m_h = 95$~GeV, the preferred regions at both $1\sigma$ and $2\sigma$ are $|\sina | \lesssim  0.2$ or $|\sina | \lesssim  0.3$ respectively, both with $m_h \lesssim m_\hp $, see \reffig{fig:collider4}.

We also note that there are allowed regions with $m_\hp < m_\hcal / 2$, where the coupling $g_{hH^+ H^-}$ is small enough to make BR$(\hcal \to H^+ H^-)$ negligible. For $m_{\hp} \lesssim 80$ GeV, one might think that the LEP constraints on $m_\hp$ are violated \cite{Abdallah:2003wd,Heister:2002ev,Abbiendi:2013hk}. However, the majority of the allowed points in the scan have BR$(\hp \to W^\pm \gamma) > 99$\% and are therefore not excluded by the LEP constraints. We refer the reader to sections \ref{Hp-results} and \ref{HpwgammaZ} for details concerning the $\hp$ decays in our model. 

Because of the smallness of $\Gamma_{H^\pm}$ and $\Gamma_A$ we have not considered 
the off-shell decay channels $\hcal \to H^{+(*)} H^{-*}$ or $\hcal \to A^{(*)}A^*$ (see sections \ref{Hp-results} and \ref{A-results}).

\begin{figure}[bt]
\begin{center}
\includegraphics[width=0.6\textwidth]{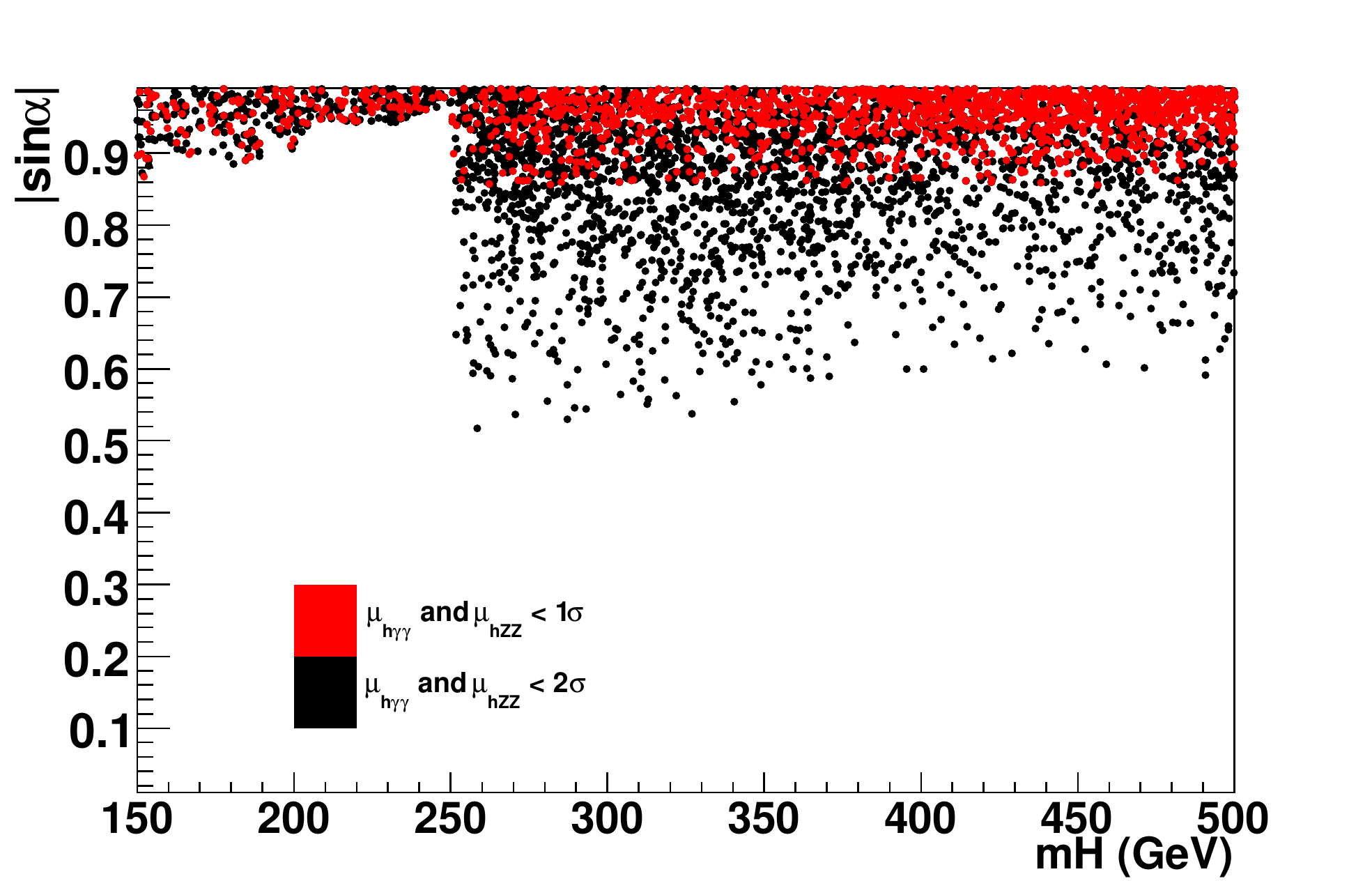}
\end{center}
\caption{Similar to Figs.\ \ref{fig:collider3} but scanning over $m_H$, showing
points with $m_h = 125$~GeV and $\mhp = m_H$, $m_A=\mhp+50$~GeV that satisfy all constraints from theory and collider searches. }
\label{fig:mHscan}
\end{figure}

The heavier scalar $H$ in Case~1 is also constrained by the LHC data. In Fig.\ \ref{fig:mHscan}  we present the allowed points in the $(m_H, \sina )$-plane for $m_\hp = m_H$ and $m_A = m_\hp + 50$~GeV. When $m_H < 2 m_h$, the $H$ has the same decay modes as the SM Higgs boson and the SM Higgs searches apply directly. When $m_H \gtrsim 2 m_h$, the decay channel $H\to hh$ opens up, which has the effect of suppressing the branching ratios of the SM channels, thus allowing more of parameter space. This boundary in $m_H$ is clearly seen in Fig.\ \ref{fig:mHscan}.

\section{Decays of the Scalars in the SDM}\label{sect:decays}
In this section we present the decay branching ratios and widths for the scalars 
in our model. We first briefly discuss the decays of the CP-even bosons $h$ and $H$ 
followed by a longer discussion of the decays of the charged scalar $\hp$. 
Some of the discussion regarding technical details of the $\hp$ decays is relegated 
to Appendices \ref{appendix-renormalization} and \ref{verticesandselfE}. 
We then finish this section by briefly discussing the decays of the $A$ bosons, 
which are computed analogously to the $\hp$ decays.

\subsection{Decays of the non SM-like CP even scalar $h$ or $H$ }\label{sect:cpeven}
In this section, we will focus on Case~1 and Case~2, which were discussed in 
section \ref{sect:collconstraints}. For the calculations of the branching ratios of $h$ and $H$, we use \THDMC{}.   

We first consider Case~1, where $m_h = 125$ GeV. The decay modes of $h$ must be 
SM-like in order to reproduce the recent LHC results. This constrains the masses 
of the charged scalar $\hp$ and the CP-odd $A$ to be large enough to prohibit 
e.g.\ $h \to  H^+ H^-$ and $h \to  A A$, unless the couplings are small as discussed in section \ref{sect:collconstraints}. 
The heavier $H$~boson can decay 
into $hh$, $ H^\pm W^\mp $, $H^+H^-$, $AA$ and $AZ$ if any of these channels are open. 
In this case they will be potential production channels for charged scalars and 
CP-odd scalars, see section~\ref{sec:HpProd}.

\begin{figure}[t]
\begin{center}
\begin{tabular}{cc}
\includegraphics[width=0.48\textwidth]{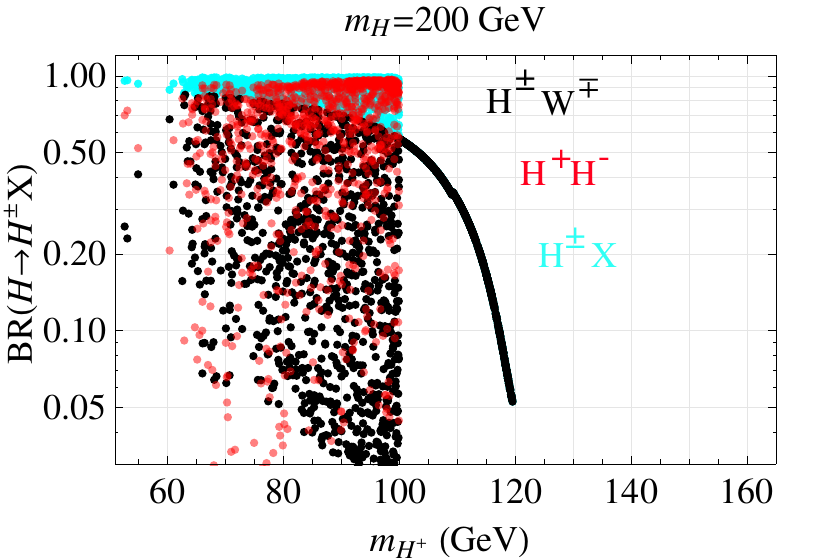}  &   \includegraphics[width=0.48\textwidth]{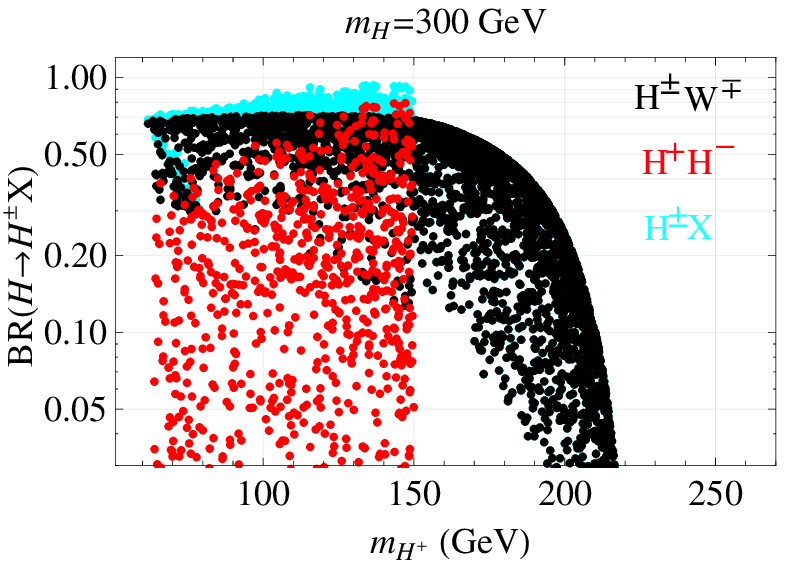} \\
(a) & (b) 
\end{tabular}
\end{center}
\caption{The branching ratios of the $H$ boson as a function of $m_\hp$ when scanning over $\lambda_3$ 
and $\lambda_7$ (see the text for details) for $m_H = 200 $ GeV (left) and  $m_H = 300 $ GeV (right): 
BR($H \to H^+H^-$) is shown as red points, BR($H \to H^\pm W^\mp$) as black points, and the cyan points shows the sum BR($H \to \hp X$). }    
\label{fig:Hdecay}
\end{figure}

In order to investigate these decays in more detail, we scan the parameters $\lambda_2,\lambda_3$ and $\lambda_7$ as in \eref{Lrandomscan} with $\sina = 0.9$, $m_h = 125$~GeV and $m_A = m_\hp + 50 $~GeV. We impose the theoretical constraints and demand the points to fulfil 0.8 $ < \mu_{h\gamma\gamma}  < $ 1.5 and 0.71 $ < \mu_{hZZ} < $ 1.53 as before (the points shown in red in figures~\ref{fig:collider3},~\ref{fig:collider4}~and~\ref{fig:mHscan}). In the scan, it is possible to obtain $\Gamma_H \gtrsim m_H$ through the $Hhh$, $HH^+H^-$ and $HAA$ couplings, which depend on the scanned $\lambda_3$ and $\lambda_7$ parameters. This means that the partial widths $\Gamma_{H\to hh}$, $\Gamma_{H\to H^+H^-}$ and $\Gamma_{H\to AA}$ can become very large. In order to have well defined particle properties, e.g. narrow resonances, we demand the width of $H$ to fulfil $\Gamma_H < 0.1 m_H$ as an additional constraint. The results are summarized in \reffig{fig:Hdecay}a and \reffig{fig:Hdecay}b for $m_H = 200$ and 300~GeV, respectively. 

In the case $m_H = 200$~GeV, the kinematically open non SM-like decays are $H\to H^+H^-$ and $H\to H^\pm W^\mp$.  From \reffig{fig:Hdecay}a we see that for $m_H = 200$ GeV the decay $H \to H^+H^-$ can dominate completely whereas $H \to \hp W^\mp$ is substantial for $m_\hp \lesssim 120 $~GeV. We also note that the branching ratio for $H \to H^+H^-$ grows all the way up to the threshold. This is due to the constraint on $\Gamma_H$ which puts limits on the magnitude of the $HH^+H^-$ coupling for $m_\hp < m_H/2$. When $m_\hp$ goes to $m_H/2$, larger values for the coupling is allowed and therefore also larger BR($H \to H^+H^-$) is possible. Without the constraint $\Gamma_H < 0.1m_H$ it is possible to obtain BR$(H\to H^+H^-) \approx 1$ as $m_\hp$ goes to $m_H/2$. This is because in this case the only decay that is open and depends on $\lambda_3$ and $\lambda_7$ is $H\to H^+ H^-$.\footnote{The decays $H\to \gamma \gamma$ and $H \to Z \gamma$ are open and depends on $\lambda_3$ and $\lambda_7$ but are loop-suppressed.} 

Turning to the case $m_H = 300$ GeV, the decay $H\to hh$ is now open. Furthermore, the decays of $H$ into $H^+H^-$ and $H^\pm W^\mp$ are open for $m_\hp \lesssim 150$~GeV and $m_\hp \lesssim 220$~GeV respectively. Finally, the $AA$ and $AZ$ channels are open for $m_\hp \lesssim 100$~GeV and $m_\hp \lesssim 160$~GeV respectively. From the results shown in \reffig{fig:Hdecay}b we see that the branching ratio of the $H$ scalar into a pair of charged scalars $H^+ H^-$ can be as large as 80\% and  $H \to \hp W^\mp$ can be up to 70\%. 
Looking at the sum of the two, we see that the branching ratio for  $H \to \hp X$ 
is substantial for $m_\hp \lesssim 150 $ GeV. Without the constraint on $\Gamma_H$ is possible to enhance BR($H \to H^+H^-$) further. However, the $H \to H^+H^-$ has to compete against the $H \to hh$ and $H \to AA$ modes. For $m_\hp \lesssim 100$~GeV, BR($H \to H^+H^-$) can reach 80\%. For larger $m_\hp$, BR($H \to H^+H^-) = 0.95$ is possible. 

In Case~2, where $m_H=125$~GeV, the possible decays modes of $H$ are the same as in Case~1. However, 
in order to accommodate the recent LHC results, the signal strengths must be very SM-like 
and this puts limits on $\mhp$ and $m_A$. 
The branching ratios of $h$ in Case~2 should then also be SM-like since no other decay channels are open.

\subsection{Decays of the charged scalar $\hp$}
\label{sect:hpdecays}
We now turn to the decay of the charged scalar $\hp$. The main issue here is that 
below the $\hp \to W^\pm S$ threshold, where $S$ is the lightest of the neutral 
scalars, it is  not known \textit{a~priori} which is the largest of the partial 
decay widths: $\hp \to f\bar{f}^\prime $, $\hp \to W^\pm Z/\gamma$ (which proceeds 
at one-loop level at lowest order) or $\hp \to W^{\pm *} S^* \to 4$ or 6 fermions 
(which are tree level processes, suppressed by massive propagators and 
multi-particle phase-space).

All loop calculations of the $H^\pm$ and $A$ decays in this paper have been 
performed by implementing the model in the \FA{} \cite{FeynArts} and \FC{} 
\cite{FormCalc,LoopTools} packages with the help of the \FRU{} package 
\cite{Christensen:2008py}.\footnote{The \FRU{} model can be obtained from 
the authors.} The calculations have been performed in Feynman--'t~Hooft gauge,
i.e.\  $R_\xi$ gauge with $\xi=1$,
and renormalization conditions and counterterms have been implemented in \FC{} 
directly as this is not included in models generated using \FRU. Details of the 
calculations are given in the rest of this section, and details of the 
renormalization and the chosen on-shell renormalization scheme are given in 
Appendix~\ref{appendix-renormalization}.

\begin{figure}[b]
\begin{center}
\begin{tabular}{cc}
\includegraphics[scale=0.8]{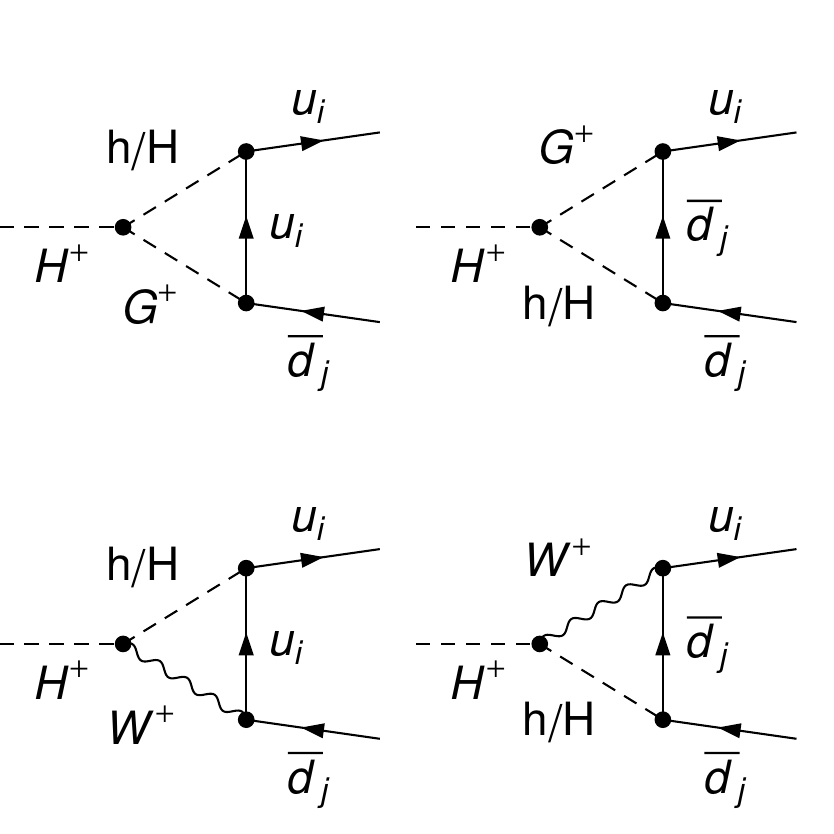} & \includegraphics[scale=0.8]{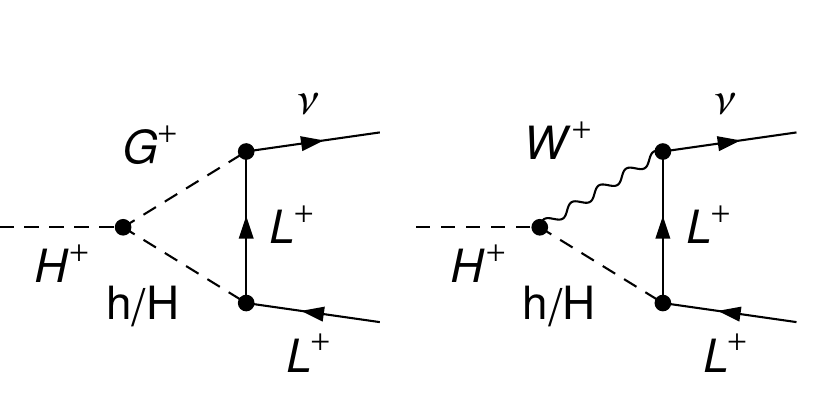} \\
(a) & (b)
\end{tabular}
\end{center}
\caption{Feynman diagrams in $R_\xi$ gauge for the effective vertex for (a): $H^+ \to
u_i\,\bar{d}_j $ and (b): $H^+ \to L^+\,\nu $. Here, $u_i$ and $d_i$ denote up- and down-type quarks of family $i$.
$L^+$ denotes a positively charged lepton; $e^+,\,\mu^+,\,\tau^+$ and $\nu$ the corresponding neutrino. Diagrams that contain propagators
denoted by $h/H$ are to be counted as two diagrams:\ one with a $h$ boson running in the loop and one with
a $H$ boson instead. The effective vertices for $A u_i\bar{u}_i $ and $A L^+L^- $ are described at
one-loop order by the same set of diagrams as in (a) and (b) but with the replacements $
H^+ \to A$, $W^+  \to  Z$, $ G^+ \to G^0 $, $\bar{d}_i
\rightarrow \bar{u}_i$ and $ \nu \to L^- $.}
\label{fig:HpLnuHpUD1}
\end{figure}

\subsubsection{$H^\pm \to f\bar{f}^\prime$}
\label{Hptwoferm}\label{hpff}
Due to the assigned $\ztwo$ parities of the $ \Phi_{1,2} $ fields and the fermions, the charged
scalar, which resides solely in $\Phi_2$, does not couple to fermions
at tree level. Since the CP-even mass eigenstates are a mixture of the neutral and real
components from $\Phi_1 $ and $\Phi_2 $ it is possible for the charged scalar to interact with
fermions through the terms $ m^2_{12} \Phi_{1}^\dagger \Phi_{2} + \text{h.c.} $ in the scalar
potential. Because of the mixing, the amplitudes for all such diagrams will be proportional to $ \sin 2\alpha \propto |m_{12}|$ (see eqs.\ (\ref{eq:minim12}) and (\ref{eq:s2a})).

There are several different ways for the charged scalar to couple to two fermions. We start by considering 
the effective vertex generated by the Feynman diagrams shown in
\reffig{fig:HpLnuHpUD1}, and given in eq.\ (\ref{eq:vertex}) in appendix \ref{verticesandselfE}. Since the coupling $C_{H^\pm f\bar{f}^\prime} \sim  \rho^F $ is absent
at tree level and no counterterm is obtained by performing field and coupling expansions in
$\mathcal{L}_{\:\text{Yukawa}}$, the loop-generated coupling is UV~finite. This has also been
verified explicitly using the \FA{} and \FC{} implementation.

\begin{figure}[b]
\begin{center}
\includegraphics[scale=1.3]{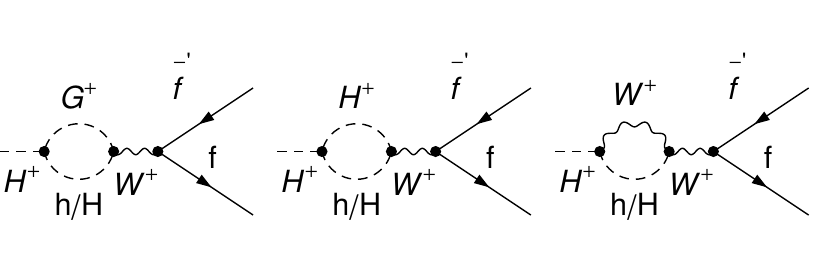}
\end{center}
\caption{$H^+W^-$ mixing contribution to $H^+ \to
f\bar{f}^\prime$.  The same set of diagrams exists for the $AZ$ mixing contribution
to $A \to f \bar{f}$ with the replacements $H^+ \to A$, $W^+  \to  Z$, $
G^+ \to G^0 $ and $\bar{f}^\prime \rightarrow \bar{f} $. There is also the possibility to draw
diagrams where the $A$~boson mixes with a $h/H$ boson which in turn go into a pair of fermions, but all
such diagrams vanish due to CP conservation in the scalar sector. Diagrams that contain propagators
denoted by $h/H$ are to be counted as two diagrams:\ one with a $h$ boson running in the loop and one with
a $H$ boson instead.}
\label{fig:HpmixFF2}
\end{figure}
\begin{figure}[b]
\begin{center}
\begin{tabular}{cccc}
\includegraphics[scale=0.4]{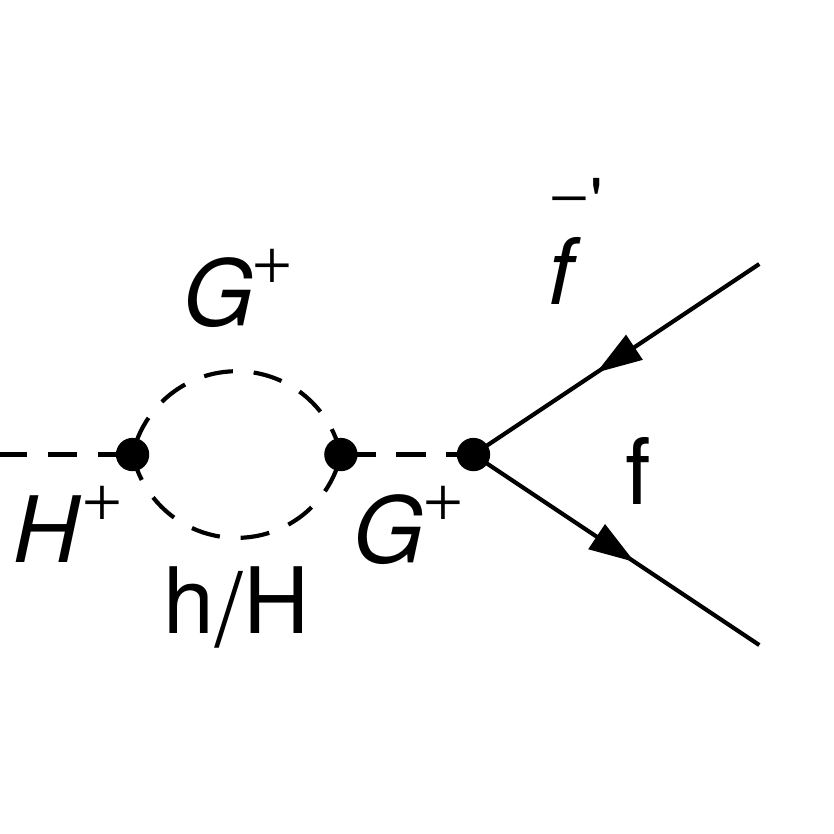} & \includegraphics[scale=0.4]{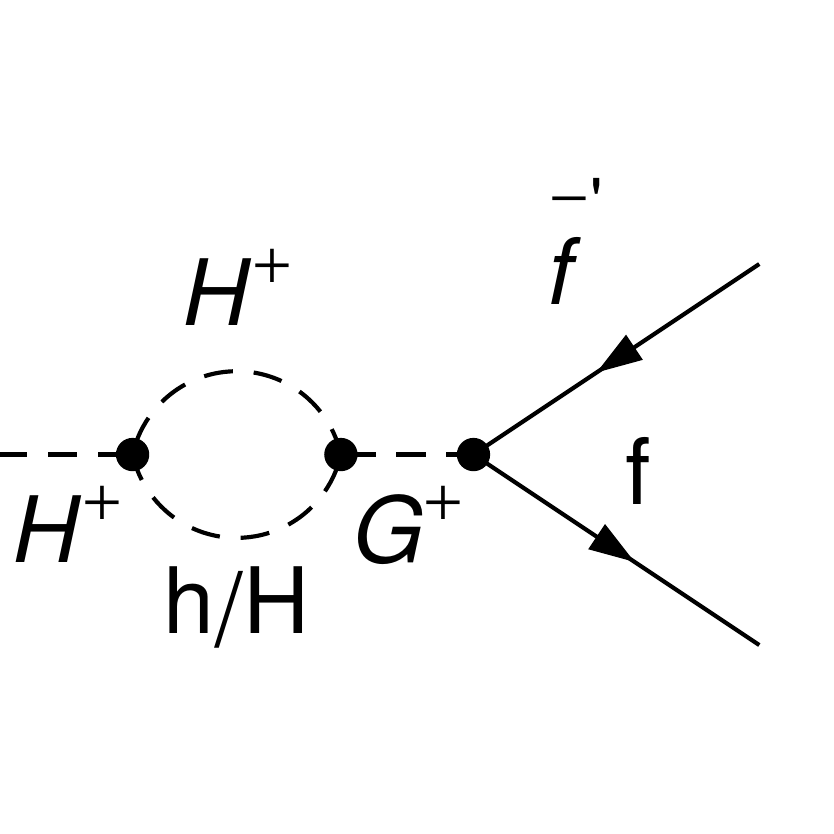} &\includegraphics[scale=0.4]{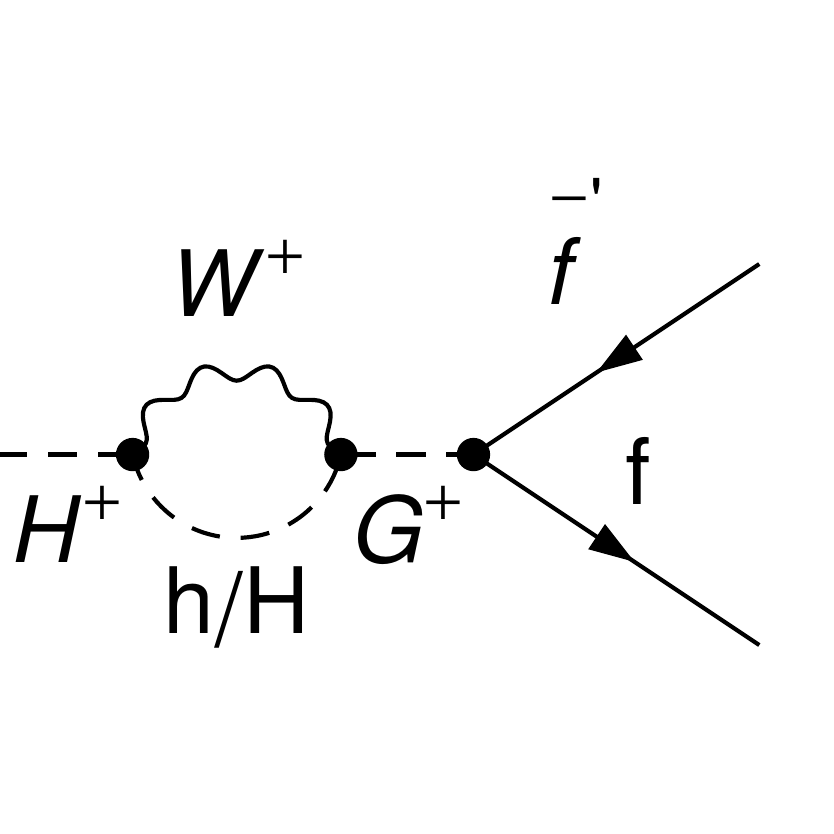} & 
\includegraphics[scale=0.4]{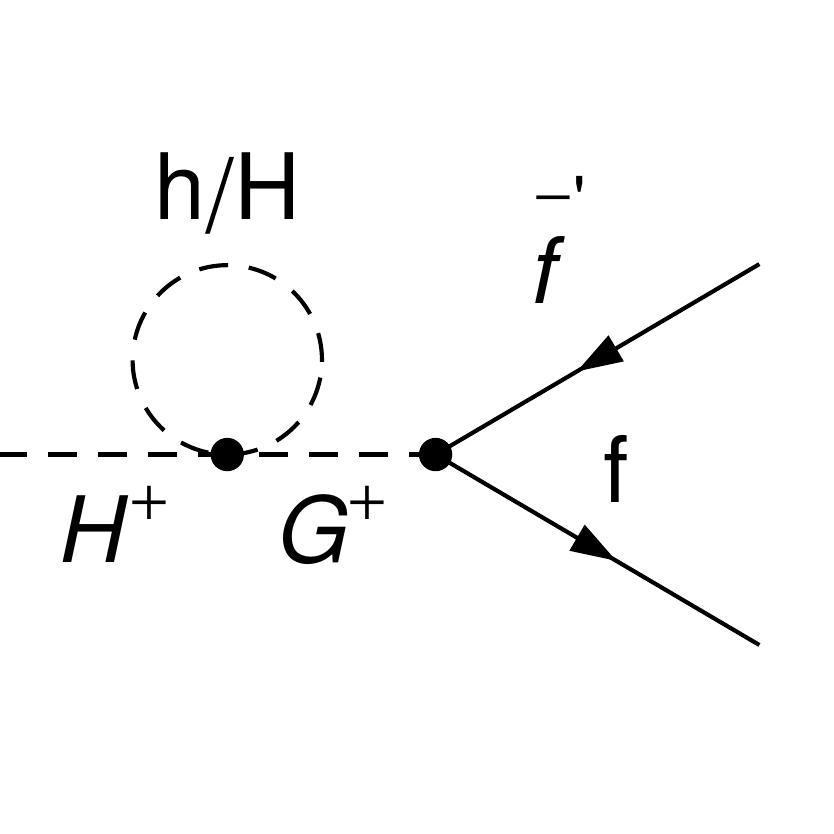} \\ \includegraphics[scale=0.4]{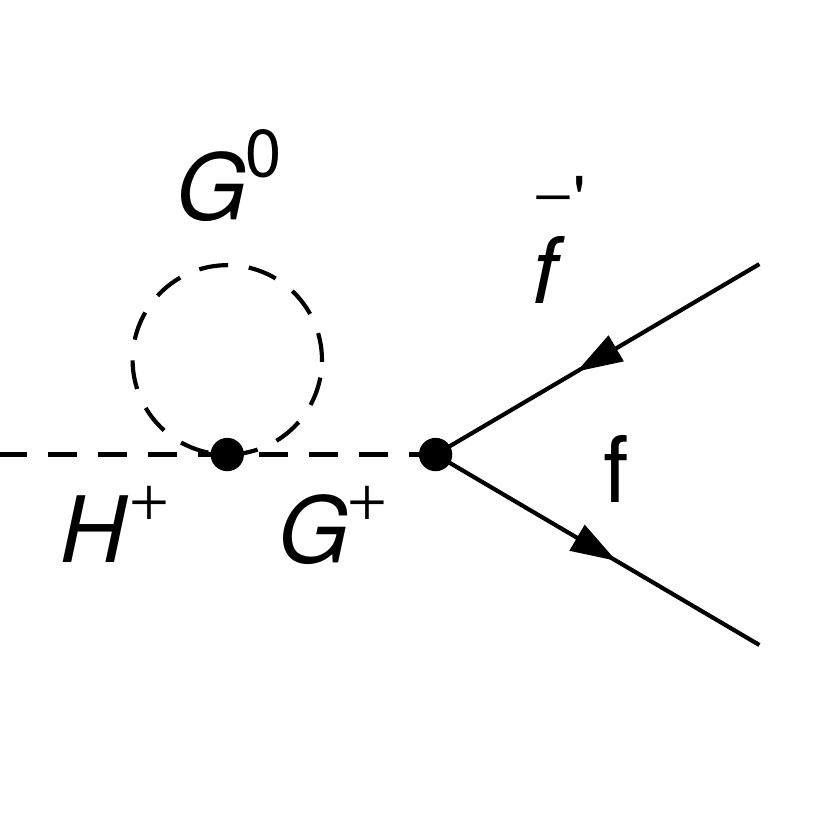} & \includegraphics[scale=0.4]{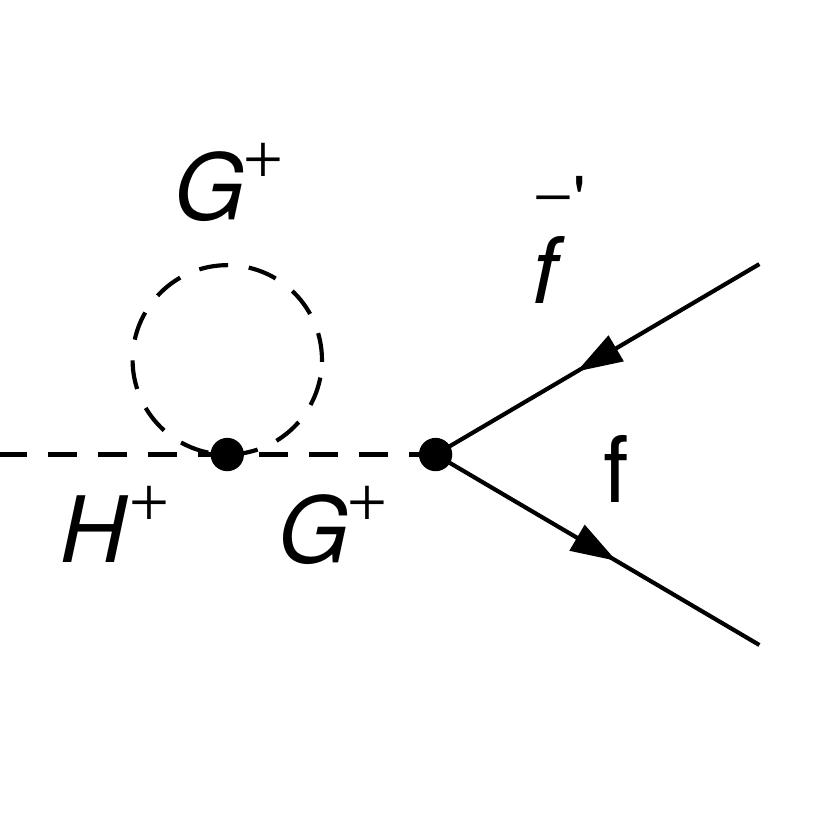} & 
\includegraphics[scale=0.4]{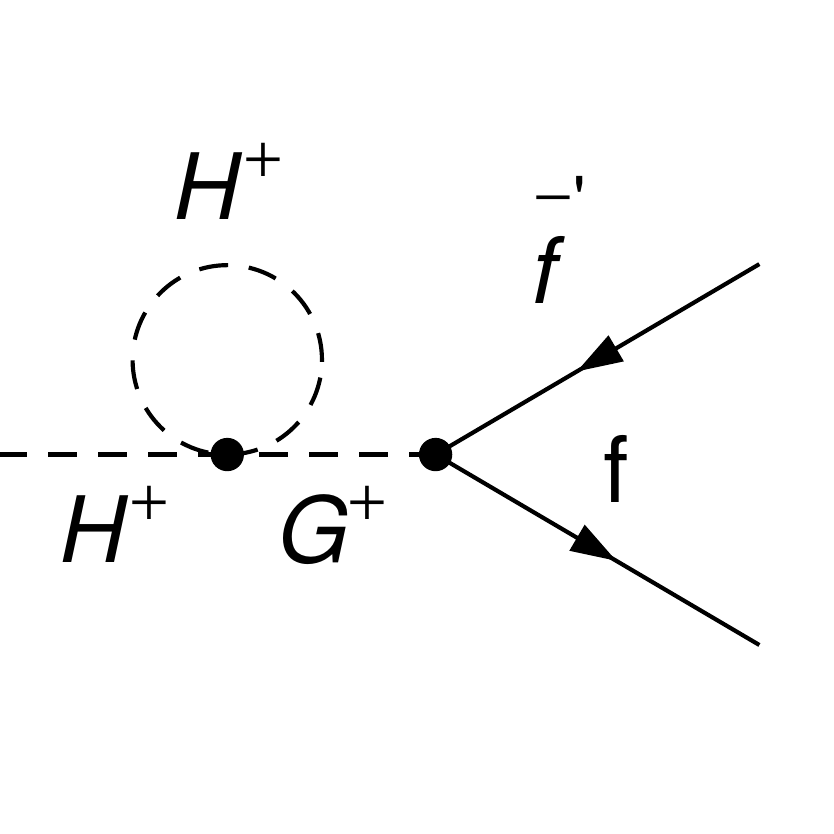}  & \includegraphics[scale=0.4]{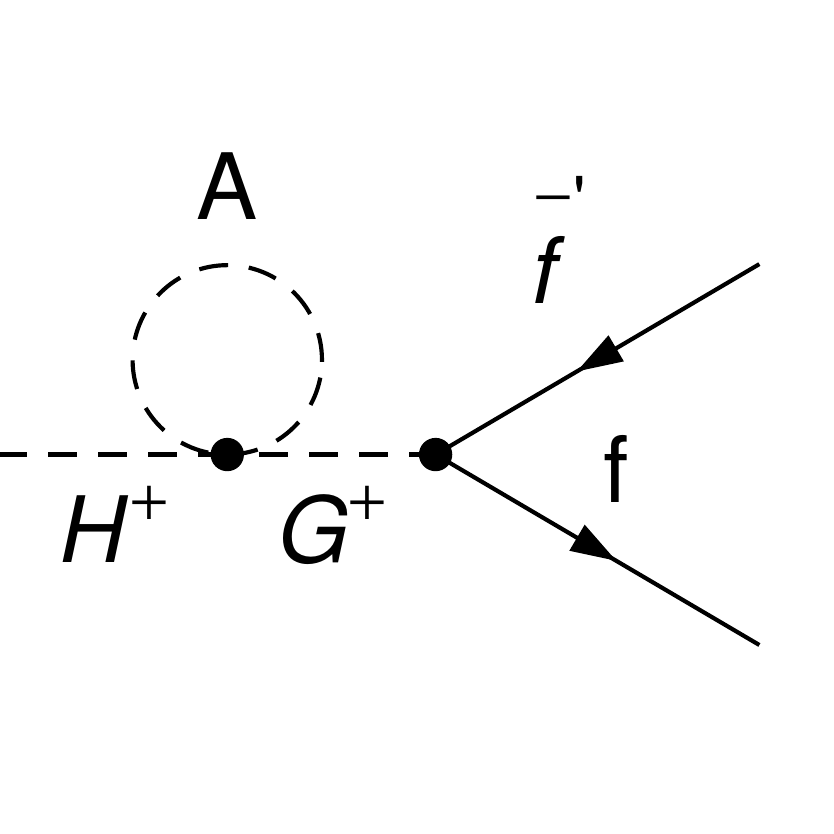} \\ 
\end{tabular}
\end{center}
\caption{$H^+G^-$ mixing contribution to $H^+ \to
f\bar{f}' $. The last five diagrams are purely real and vanish in on-shell renormalization schemes \cite{Arhrib:2006wd}. 
The same set of diagrams exists
for the $AG^0$ mixing contribution to $A \to f \bar{f}$ with the replacements $H^+ \to A$, 
$W^+  \to  Z$, $ G^+ \to G^0 $ and $\bar{f}^\prime \rightarrow \bar{f} $. }
\label{fig:HpmixFF1}
\end{figure}

Another contribution to the matrix element $\mathcal{M}_{H^\pm\to f\bar{f}^\prime}$ 
comes from mixing of the charged scalar
with the longitudinal component of the $W^\pm$
boson or the charged Goldstone boson $G^\pm$ since we are using $R_\xi$ gauge. 
This contribution also arises due to the
$ m^2_{12} \Phi_{1}^\dagger \Phi_{2} + \text{h.c.} $ term in the scalar potential. 
 Feynman diagrams for the $H^\pm W^\mp $ and
$H^\pm G^\mp$ mixing contribution to $H^\pm \to f\bar{f}^\prime$ are shown in figures
\ref{fig:HpmixFF2} and \ref{fig:HpmixFF1}.

In the present work, we follow the procedure for renormalization described in \cite{Arhrib:2006wd}, which means that no
tadpole diagrams contribute and the real parts of the $H^\pm W^\mp$ and $H^\pm G^\mp$ mixings are absent for on-shell charged
scalars. Again we refer to Appendix \ref{appendix-renormalization} for details.
Below the $hW^\pm$ threshold, only the vertex-diagrams in
\reffig{fig:HpLnuHpUD1} contribute to $ \Gamma_{H^\pm \to f\bar{f}^\prime} $ in the present renormalization scheme. 
As a consequence, for charged scalar masses below $m_h + m_W$, the width for $H^\pm \to f\bar{f}^\prime$ is proportional to the fermion mass $m_f$ and vanishes when $m_f\to 0$. Above the $m_h + m_W$ threshold, where the $\hp W^\mp$-mixing diagrams develops a non-zero imaginary part (which is unaffected by the renormalization scheme, see \reffig{fig:selfenergies}), the width will not vanish in the limit $m_f\to 0$. 
We have also verified, with our \FA{} and \FC{} implementation, that the final expression for the partial width $\Gamma_{H^\pm\to f\bar{f}^\prime}$, including all contributions, is indeed UV~finite.

Finally we want to emphasize that the $\hp \to f \bar{f}^\prime $ partial width is proportional to $ \sin^2 2\alpha $ and does not depend on the parameters $\lambda_2, \lambda_3$ or $ \lambda_7$. In our numerical calculations
 we include QCD radiative corrections for final state quarks up to order $\alpha_s^2$, according to~eq.~(14) in \cite{Eriksson:2009ws}, which is based on \cite{Braaten:1980yq,Drees:1990dq,Gorishnii:1990zu}. We will also in the following discussion set $V_{\text{CKM}}$ equal to the unit matrix.

\begin{figure}[t]
\begin{center}
\includegraphics[scale=1.05]{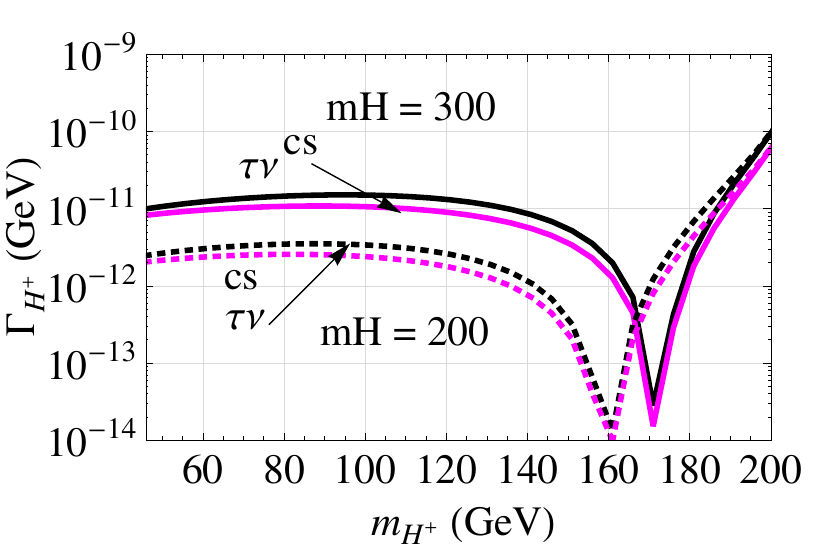} 
\end{center}
\caption{The partial widths $\Gamma_{H^\pm \to \tau \nu} $ (black) and $\Gamma_{H^\pm \to cs} $ (magenta) evaluated for $m_h = 125$~GeV, $\sin \alpha = 0.9$. For the solid lines we have $m_H = 300$~GeV, and for the dotted lines $m_H = 200$~GeV.}
\label{fig:Hpcstaunu}
\end{figure}

In \reffig{fig:Hpcstaunu}, the partial widths $\Gamma_{H^\pm \to \tau \nu} $ and $\Gamma_{H^\pm \to cs} $ are shown. The widths are very small, less than $\sim$ 1 eV. This is partially due to the small Yukawa couplings $m_s/v$, $m_c/v$ and $m_\tau /v$, on which all diagrams below the $hW^\pm$ threshold depend through the $H_{i}\bar f f$ vertex, $H_i = h,H$. Above the $hW^\pm$ threshold, the diagrams in \reffig{fig:HpmixFF2}, which are independent of the Yukawa couplings, start to contribute according to the chosen renormalization scheme. The smallness of the widths is also due to the loop suppression. In section \ref{Aff}, we compare the partial width for the process $A \to \tau^+ \tau^-$ (which is analogous to $H^\pm \to \tau \nu$) evaluated in our model and in a generic 2HDM in order to extract the size of the loop suppression.  We also note that the widths depend on $m_h$ and $m_H$ since diagrams with $h$ and $H$ propagators interfere destructively. Furthermore, the $\tau \nu$ and $cs$ widths are similar in size due to the scaling with the fermion masses in the $H_{i}\bar f f$ vertex.

\subsubsection{$H^\pm \to W^\pm Z/\gamma$ }
\label{HpwgammaZ}

\begin{figure}[t]
\begin{center}
\begin{tabular}{cccc}
\includegraphics[scale=0.4]{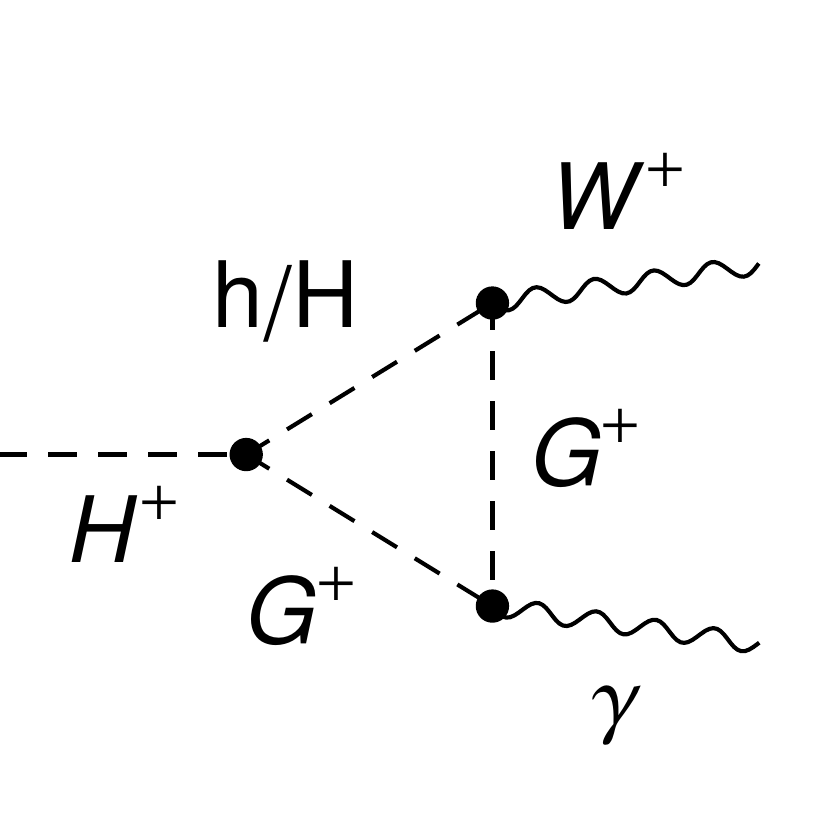} & \includegraphics[scale=0.4]{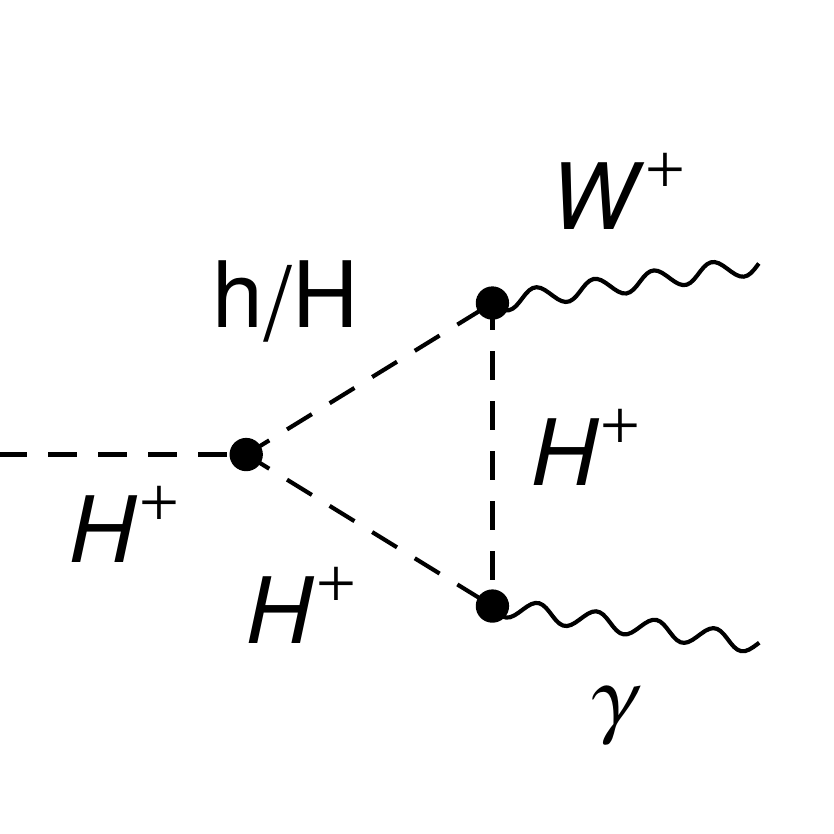} &\includegraphics[scale=0.4]{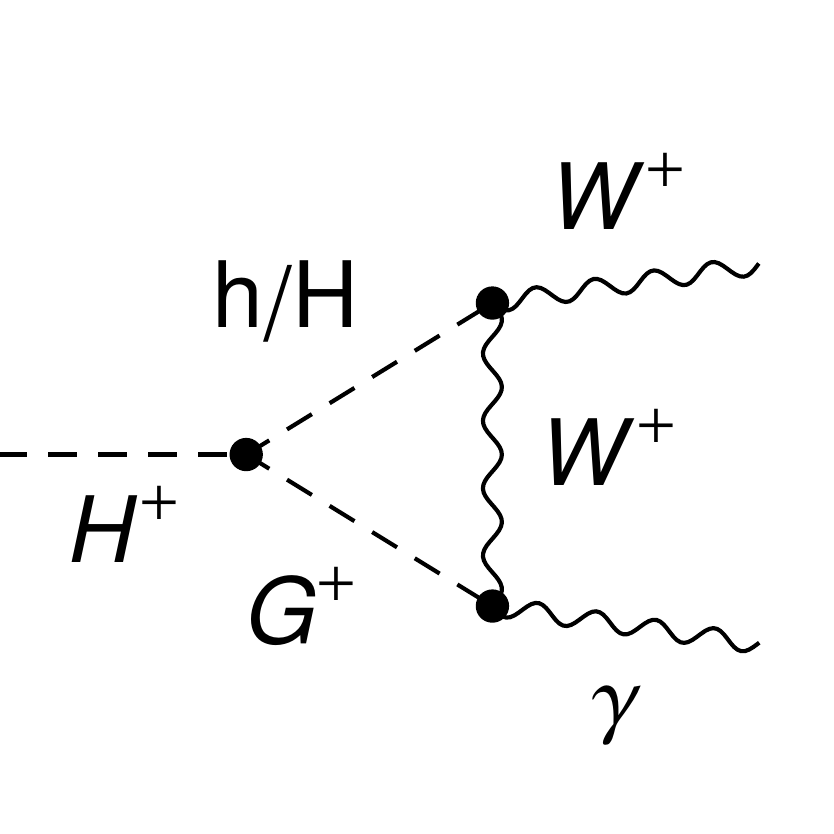} & 
\includegraphics[scale=0.4]{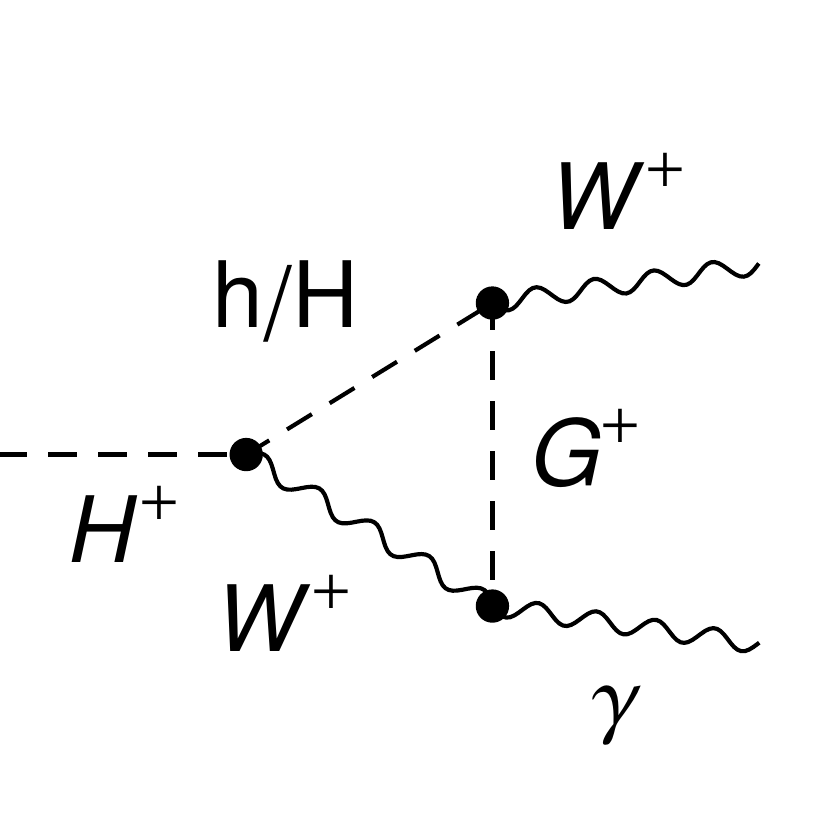} \\ \includegraphics[scale=0.4]{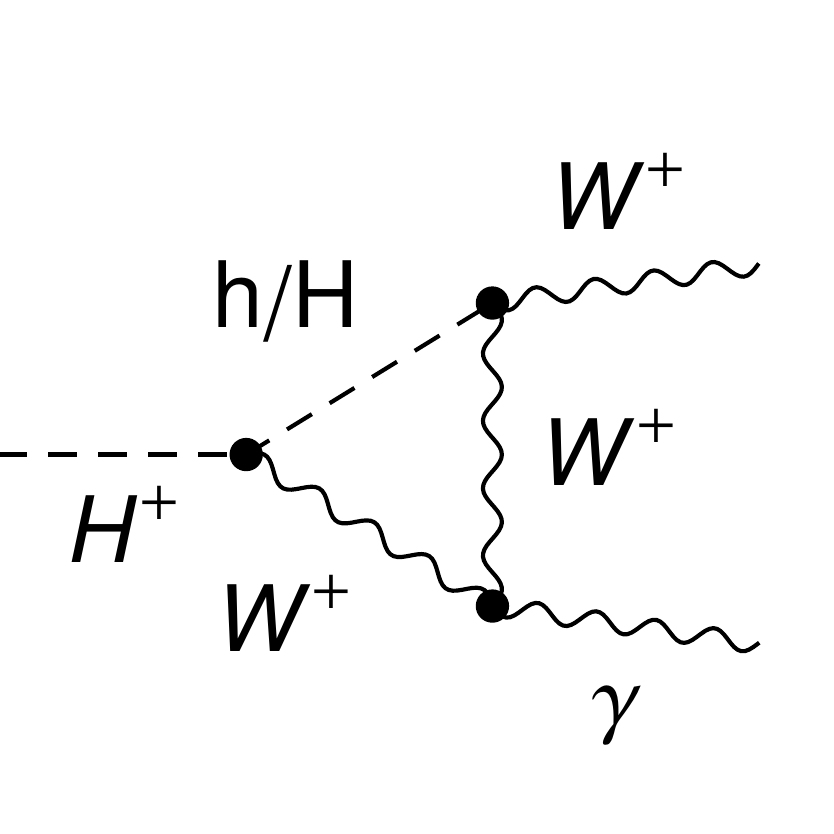} & \includegraphics[scale=0.4]{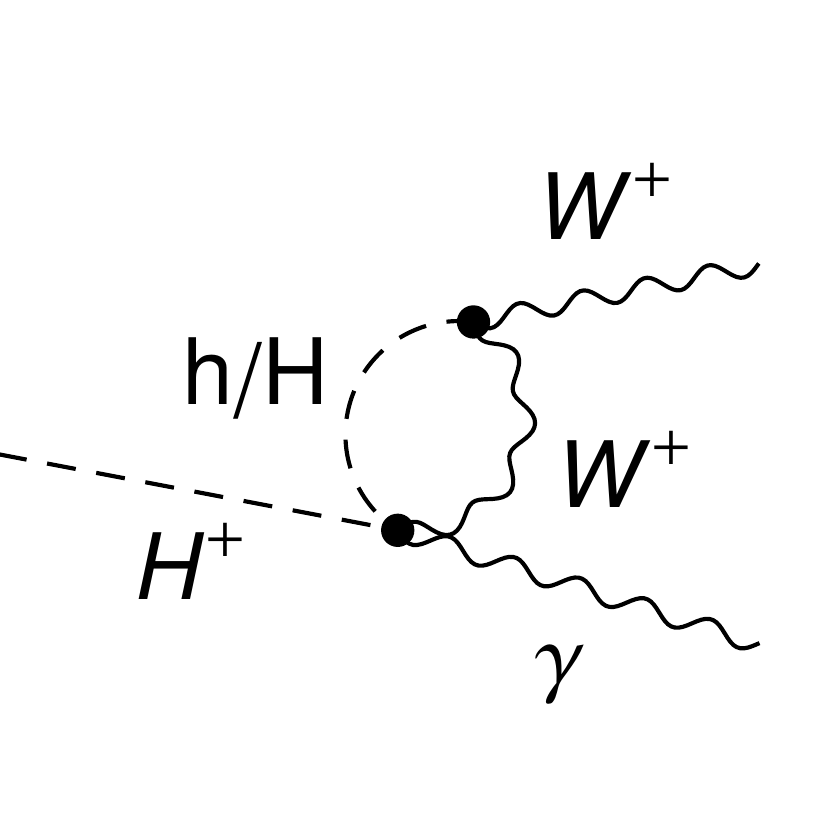} & 
\includegraphics[scale=0.4]{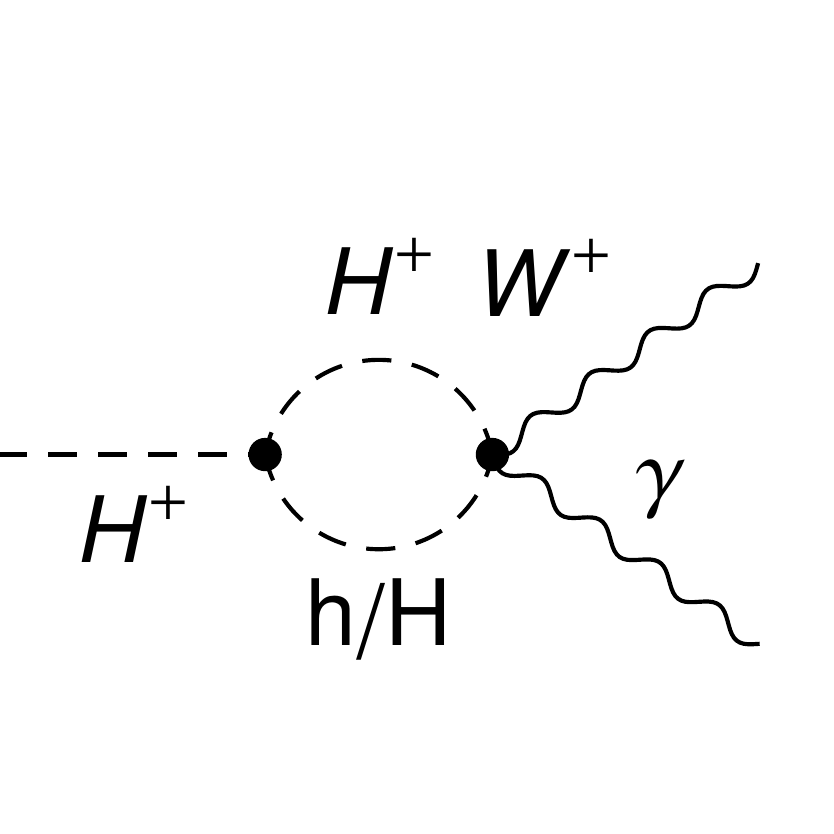}  & \includegraphics[scale=0.4]{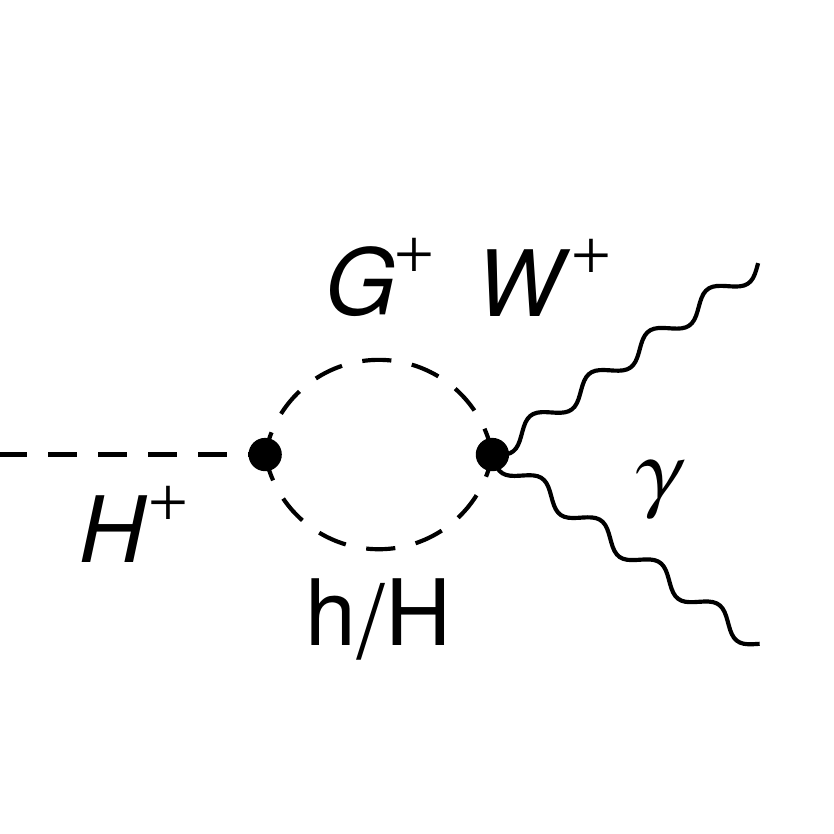} \\ 
\end{tabular}
\end{center}
\caption{Feynman diagrams in $R_\xi$ gauge for the $H^\pm W^\mp \gamma$ effective vertex at one-loop order. Diagrams that contain propagators
denoted by $h/H$ are to be counted as two diagrams:\ one with a $h$ boson running in the loop and one with
a $H$ boson instead.}
\label{fig:HpWgamma1}
\end{figure}

\begin{figure}[h]
\begin{center}
\begin{tabular}{ccc}
\includegraphics[scale=0.4]{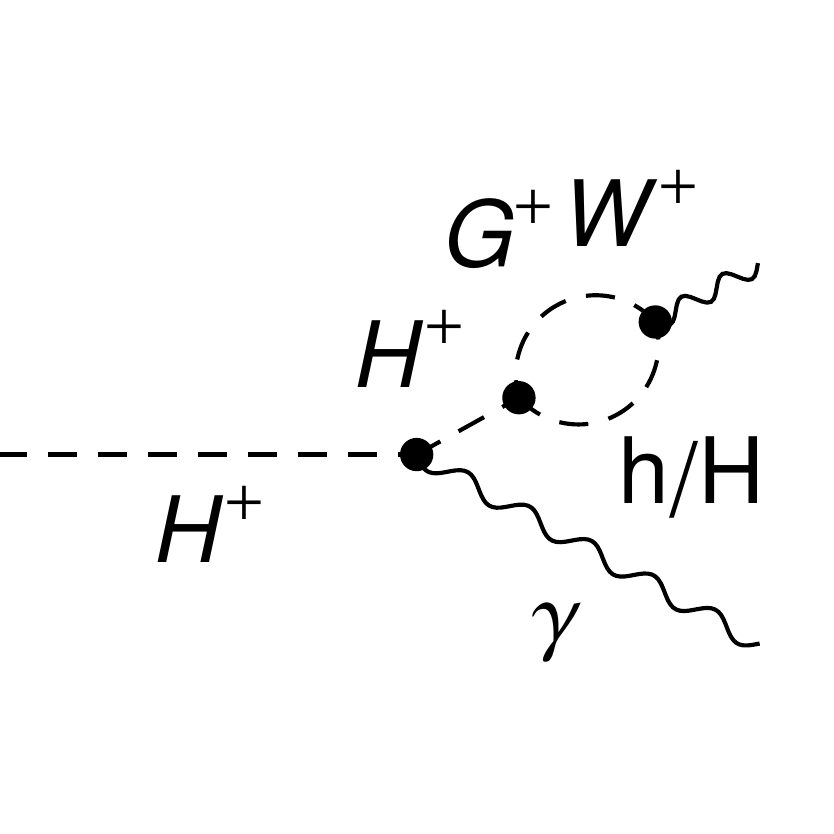} & \includegraphics[scale=0.4]{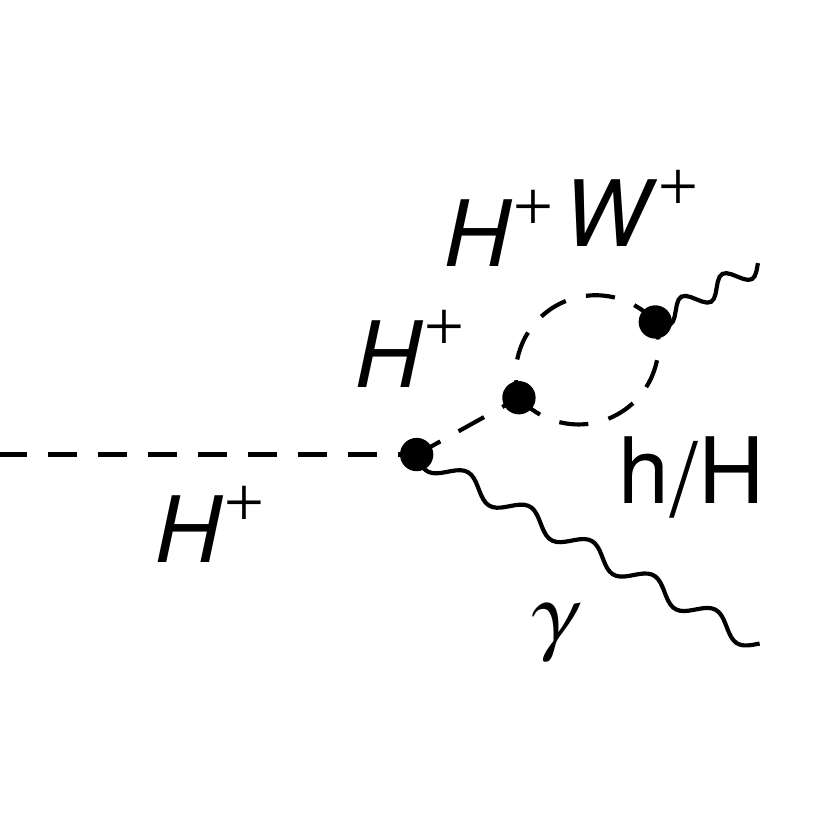} &\includegraphics[scale=0.4]{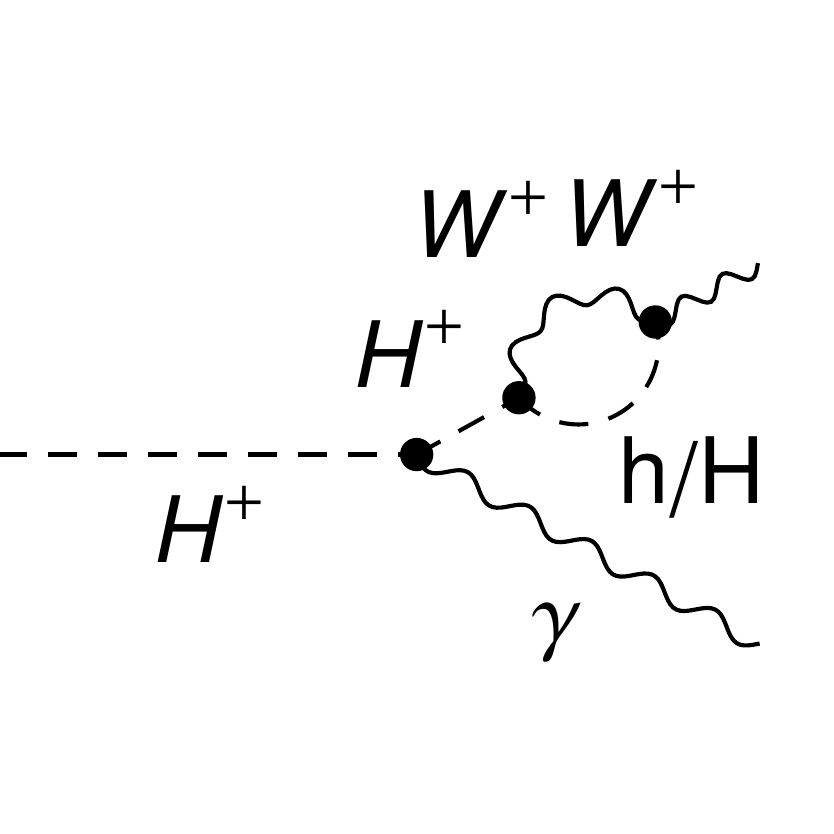} 
\end{tabular}
\end{center}
\caption{Feynman diagrams in $R_\xi$ gauge that contribute to the process $H^\pm \to W^\mp \gamma$ at one-loop level, where the external $W^\pm$ boson has longitudinal polarization, $W^\mp_L$. Diagrams that contain propagators denoted by $h/H$ are to be counted as two diagrams:\ one with a $h$ boson running in the loop and one with a $H$ boson instead. All of these diagrams vanish due to the form of the $H^+H^-\gamma$ vertex, as explained in the text.}
\label{fig:HpWgamma2}
\end{figure}

We now discuss the decay channels $H^\pm \to W^\pm Z/\gamma$, starting with $H^\pm \to W^\pm \gamma$. Because the electromagnetic current $j^\mu_{\text{EM}}$ must be conserved classically, only couplings between photons and particle--antiparticle pairs exist at tree level. This means in particular that the coupling $H^\pm W^\mp\gamma$ is absent, irrespective of the underlying model giving rise to the charged scalar $\hp$ state. However, this coupling can in general be generated at higher orders. The Feynman diagrams that contribute to the amplitude at one-loop order in $R_\xi$ gauge are shown in figure~\ref{fig:HpWgamma1}. 

In principle, the diagrams in \reffig{fig:HpWgamma2} could also contribute to longitudinally polarized $W^\pm$ bosons, $W^ \pm_L$, but in fact all vanish. 
This can be understood by the form of the $H^+H^-\gamma$ coupling, for which the Feynman rule reads
\begin{equation}
H^+H^-\gamma \: : \quad \ii e \, \left[ p^\mu_{H^+} - p^\mu_{H^-}     \right],
\end{equation}
where the four momenta are taken to be incoming. Due to four momentum conservation at each vertex, we obtain $ p^\mu_{H^{+ \text{ext.}} } - p^\mu_{\gamma} = p^\mu_{H^{+ \text{int.}}} = p^\mu_{W} $ (at the $H^+H^- \gamma$ vertex in the diagrams in \reffig{fig:HpWgamma2}), which contracted with the final state polarization vector $\epsilon_\mu$ for the $W^\pm$ boson gives
\begin{equation}
p^\mu_{W}\,\epsilon_\mu(\sigma,p_{W}) = 0,
\label{pWepsW}
\end{equation}
according to the gauge condition for massive spin-1 bosons, for all polarizations $\sigma$. 
This demonstrates that all the diagrams in \reffig{fig:HpWgamma2} vanish and has also been verified with our \FC{} implementation.

Similarly, the diagrams in \reffig{fig:HpWZ2}, which are a subset of possible diagrams for the matrix~element of $\hp \to W^\pm Z_L $, vanish by the same argument applied to the $A \hp W^\mp$ coupling,
\begin{equation}
A H^\pm W^\mp \: : \quad g_{A H^\pm W^\mp} \, \left[ p^\mu_{H^\pm} - p^\mu_{A}  \right],
\end{equation}
where the four-momenta are taken to be incoming. It is important that the contributions from $AZ$-mixing vanish at one-loop level in $\hp$ decays. If they did not, then we would not have a consistent renormalization scheme (see Appendix \ref{appendix-renormalization}).

One should also add to the matrix~element $\mathcal{M}_{\hp \to W^\pm \gamma}$ all the diagrams from the $H^\pm W^\mp$ and $H^\pm G^\mp$ mixing previously discussed for the $H^\pm\to f\bar{f}^\prime$ processes, by substituting $W^+ \gamma$ for $f\bar{f}^\prime$ in the diagrams depicted in the figures~\ref{fig:HpmixFF2}~and~\ref{fig:HpmixFF1}. We do not include diagrams with external Goldstone bosons in the processes $\hp \to W^\pm Z/ \gamma  $ since we employ the standard unitary gauge prescription for summing over the physical polarization states of the final state $W^\pm $ and $Z$ bosons,
\begin{equation}
\sum_{\sigma} \epsilon^*_\mu (\sigma, p) \, \epsilon_\nu (\sigma, p) = - g_{\mu\nu} +  \frac{p_\mu p_\nu}{m^2_V}  \,    ,
\end{equation}
where $V = W^\pm$ or $ Z$. 

\begin{figure}[t]
\begin{center}
\includegraphics[scale=1.05]{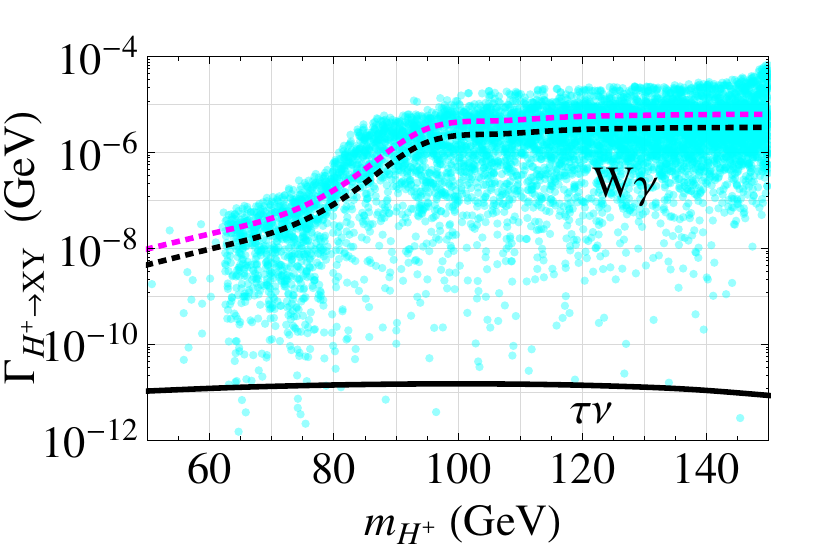} 
\end{center}
\caption{The cyan points shows the obtained $\Gamma_{H^\pm \to W \gamma }$ according to the scan described in the text. The dotted magenta line shows $\Gamma_{H^\pm \to W \gamma }$ and the solid black shows $\Gamma_{H^\pm \to \tau \nu }$, evaluated at $\lambda_3 = 2(m_{H^\pm}/v)^2$ and $\lambda_7 = \lambda_6$ respectively. The dotted black line shows $\Gamma_{H^\pm \to W \gamma }$ evaluated at $\lambda_3 = \lambda_7 = 0$ which makes the contribution from diagrams containing $H_i H^+ H^-$ vertices ($H_i = h,H$) vanish according to \eref{hphmhco}.  } 
\label{fig:HpWgammaL3}
\end{figure}

Before continuing with further $\hp$ decays, we now want to briefly compare the decay modes calculated so far. Above the on-shell threshold $m_\hp > m_W$ we find that $ \hp \to W^\pm \gamma $ dominates over $H^+ \to  \tau^+ \nu_\tau \,/ \, c\bar{s} $ by several orders of magnitude, as illustrated in \reffig{fig:HpWgammaL3}. However, as will be discussed below, it is possible to tune the parameters to make $\hp \to W^\pm \gamma$ become very small.

As was discussed in the previous section, all the diagrams that contribute to $\Gamma_{H^\pm \to f \bar{f}^\prime  } $  are proportional to the small Yukawa couplings for $ m_\hp < m_h + m_W $. Above the threshold this partial amplitude is more or less unchanged.
In contrast, the leading order diagrams that contribute to $ \hp \to W^\pm \gamma $ do not depend on the Yukawa couplings, and thus $ \hp \to W^\pm \gamma $ dominates over $H^+ \to  \tau^+ \nu_\tau \,/ \, c\bar{s} $. 
The situation is similar to the case in the Standard Model where $H_\text{SM} \to W^+ W^- $ dominates over the $b\bar{b}$ channel if it is open. It is well known that by including the width of the $W^\pm$ bosons, i.e.\ $H_\text{SM} \to W^{+*} W^{-*} \to $ fermions, the  $W^{+*} W^{-*} $  decay mode of the $H_\text{SM}$ dominates over
 $b\bar{b}$   far below the threshold; $ m_{H_\text{SM}} < 2m_W $. As we will now show, the situation is similar in our model, i.e.\ the $H^\pm \to W^{\pm*} \gamma$ mode dominates over the $\hp \to f\bar{f}^\prime $ modes even for charged scalar masses $m_\hp < m_W$.

To investigate this, we include the effect of subsequent decays of the $\wpm$ boson, by considering the process $\hp \rightarrow \wpm^*  \gamma $, using the method of ``smeared mass unstable particles''~\cite{Kuksa:2009de,Pasechnik:2010yu} described in Appendix~\ref{sect:smup}. 
Formally, one should consider all contributions to the process $\hp \to f \bar{f}^\prime \gamma $, with a photon energetic enough to be detected. 
The diagrams contributing to this process would be the same as those for $\hp \to  f \bar{f}^\prime $ with an external photon radiated off any charged particle. 
We do not do this here, since to be consistent, we would then also have to include all other $\mathcal{O}(\alpha_{\text{EM}} )$ corrections to those widths, which are needed to cancel IR~divergences. This procedure will then require two-loop calculations, a cumbersome task that should not alter the overall result regarding our $\hp \to W^{\pm *} \gamma  \to f \bar{f}^\prime \gamma $ calculation.

The result of the inclusion of the width of the $W^\pm$ boson is that, due to its broadness and the smallness of $\Gamma_{\hp
\rightarrow \tau \nu } $ and $ \Gamma_{\hp \rightarrow cs }$, the process $\hp \rightarrow \wpm^* \gamma $ clearly dominates the spectrum even below the threshold for $ \hp \to \wpm \gamma $, as shown in \reffig{fig:HpWgammaL3} above and in \reffig{fig:HpBR1} below.

The $\hp \to W^\pm Z/\gamma$ widths are proportional to $ \sin^ 2 2\alpha $ and are independent of $\lambda_2$. They do however depend on the $\lambda_3$ and $\lambda_7$ parameters through the $H^+ H^- h$ and $H^+ H^- H$ vertices present in the second and seventh diagram in \reffig{fig:HpWgamma1}. In \reffig{fig:HpWgammaL3} we give the partial decay width $\hp \to W^\pm \gamma$ for the canonical choice of $\lambda_3= 2 (m_\hp/v)^2 $ with $\lambda_7 = \lambda_6$ as well as when scanning over $\lambda_2$, $\lambda_3$ and $\lambda_7$ according to~\refeq{Lrandomscan} with $\sin \alpha = 0.9$, $m_h = 125$~GeV, $m_H = 300$~GeV and $m_A = m_\hp + 50 $~GeV. The conclusion is that for the vast majority of the scanned parameter points, the $W \gamma$ mode dominates over the $\tau \nu$ and $cs$ modes. We note that it seems possible to tune the parameters $\lambda_3$ and $\lambda_7$ for a given $m_{\hp}$ to give a very small $\Gamma_{\hp \to W^\pm \gamma}$. This is most likely due to cancellations between the diagrams containing $H_i H^+ H^-$ vertices, $H_i = h,H$ (which depend on $\lambda_3$ and $\lambda_7$, see eq.~\eref{hphmhco}) with diagrams containing $H_i \hp W^\mp$ vertices (which depend on gauge couplings, see appendix~\ref{sect:couplings}). However, we do not analyze this further here. 

\begin{figure}[t]
\begin{center}
\begin{tabular}{cccc}
\includegraphics[scale=0.4]{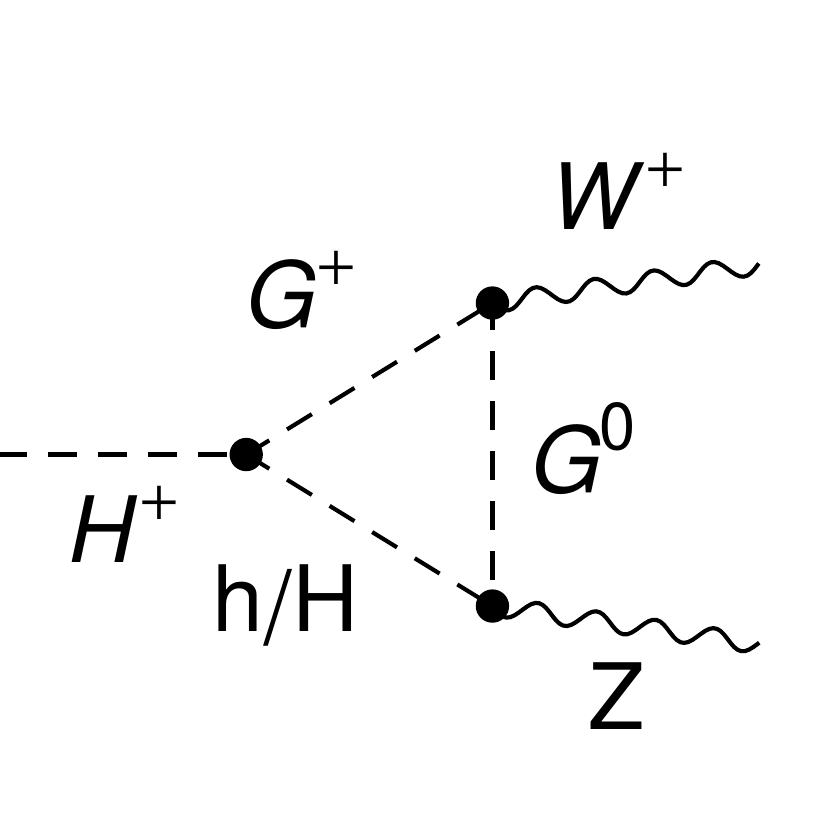} & \includegraphics[scale=0.4]{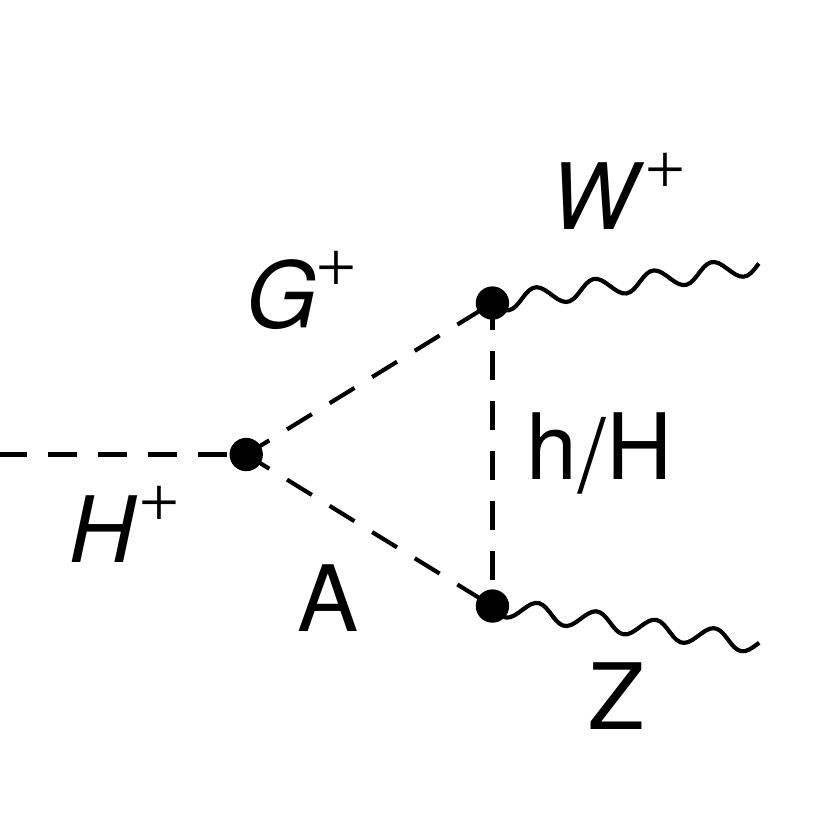} &\includegraphics[scale=0.4]{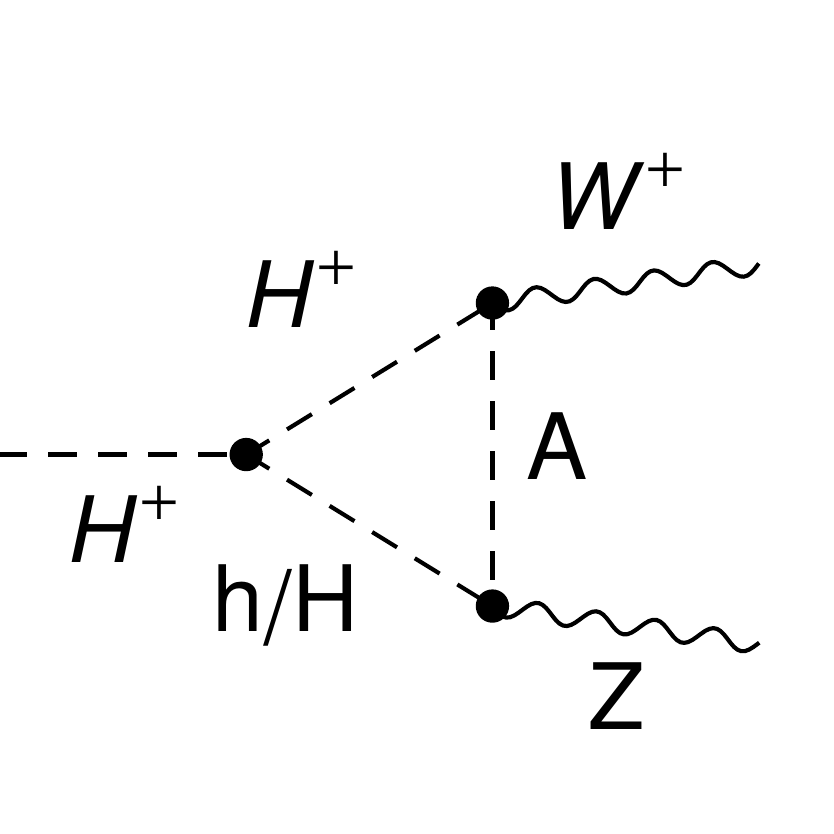}  & 
\includegraphics[scale=0.4]{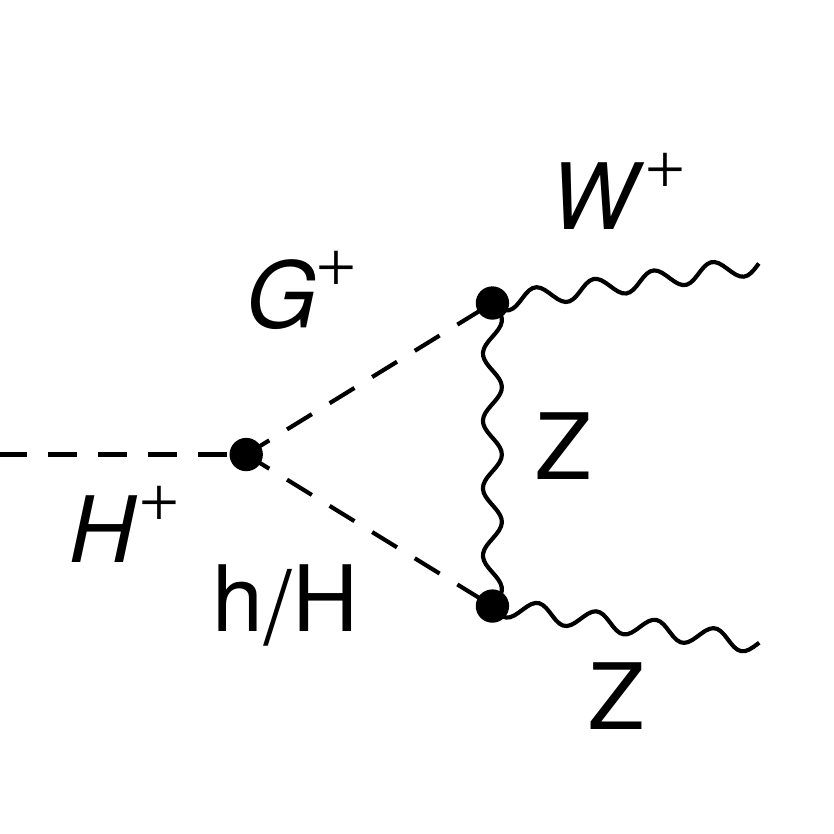} \\ \includegraphics[scale=0.4]{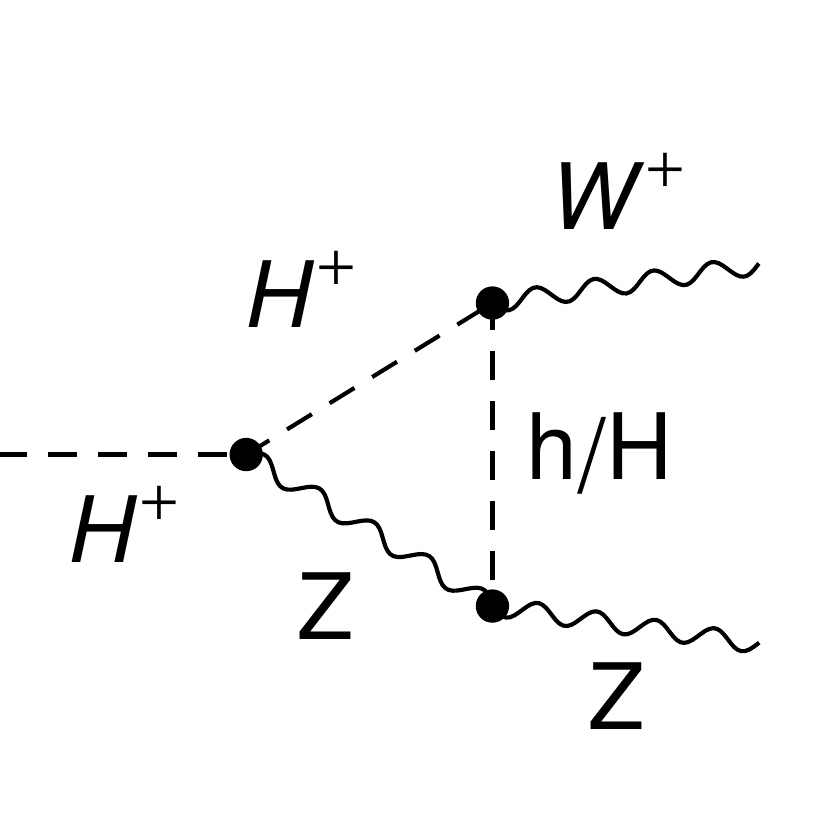} & \includegraphics[scale=0.4]{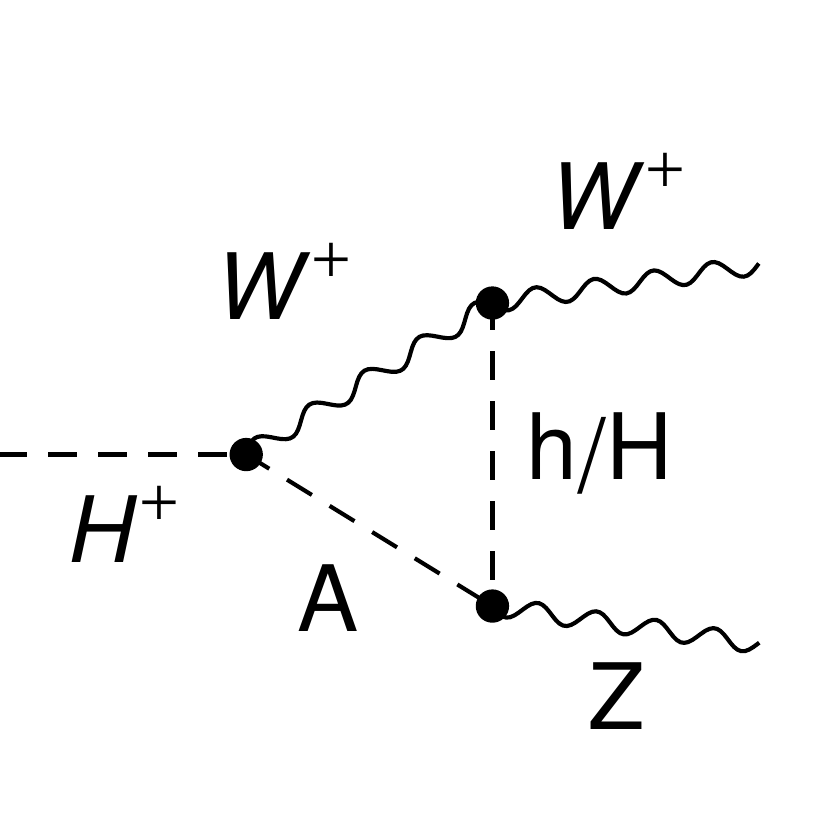} & 
\includegraphics[scale=0.4]{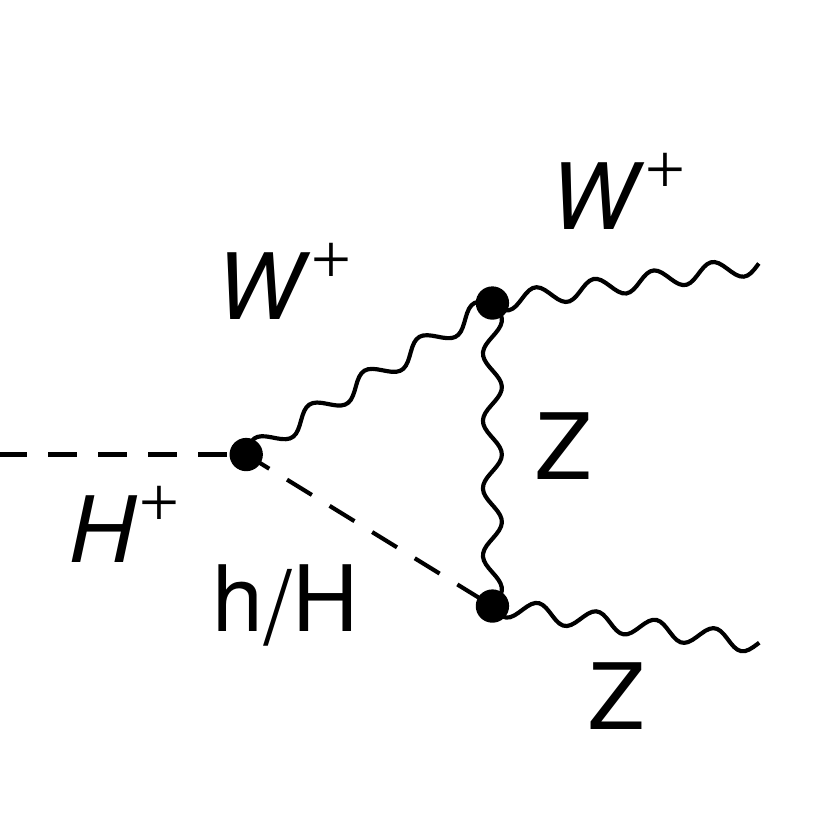}  & \includegraphics[scale=0.4]{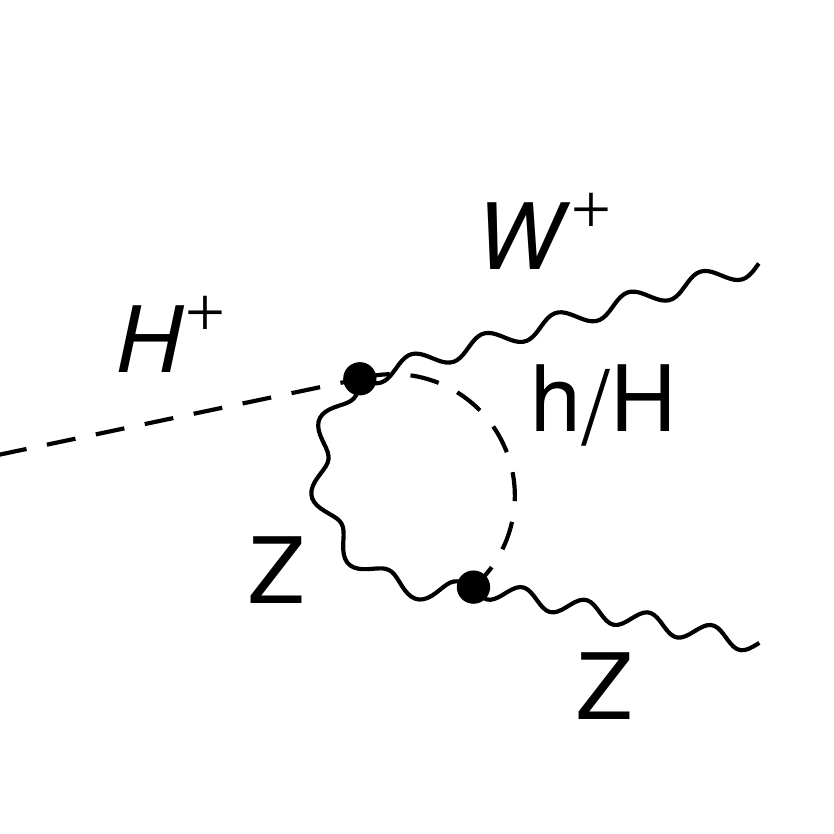} \\
\end{tabular}
\end{center}
\caption{Feynman diagrams in $R_\xi$ gauge that contribute to the process $H^\pm \to W^\mp Z$ at one-loop level. Diagrams that contain propagators denoted by $h/H$ are to be counted as two diagrams:\ one diagram with a $h$ boson running in the loop and one with a $H$ boson instead.}
\label{fig:HpWZ1}
\end{figure}

\begin{figure}[tb]
\begin{center}
\begin{tabular}{ccc}
\includegraphics[scale=0.4]{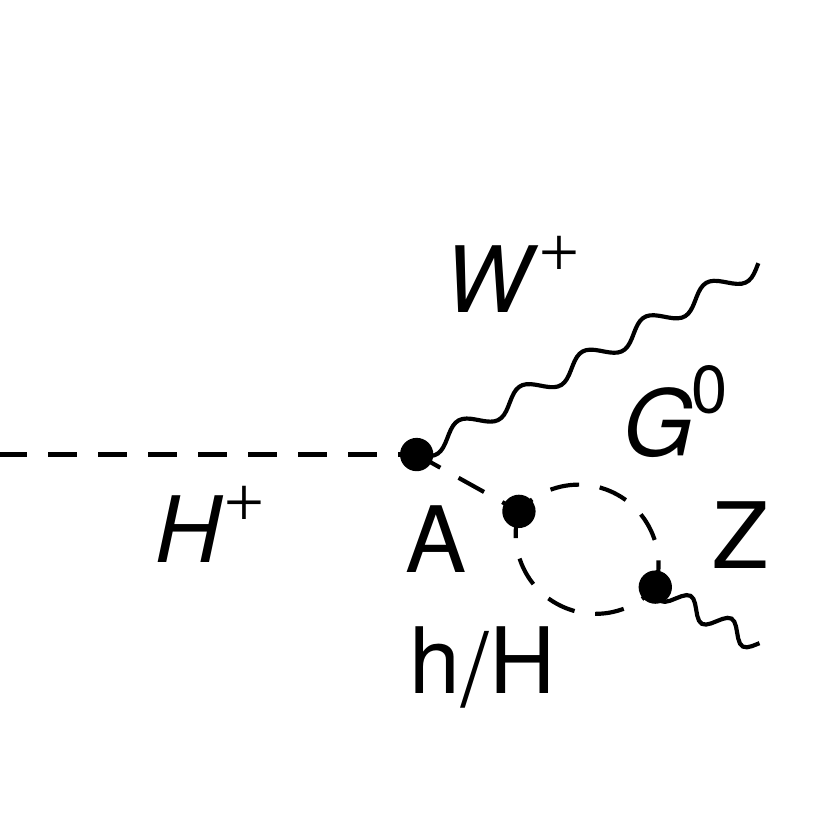} & \includegraphics[scale=0.4]{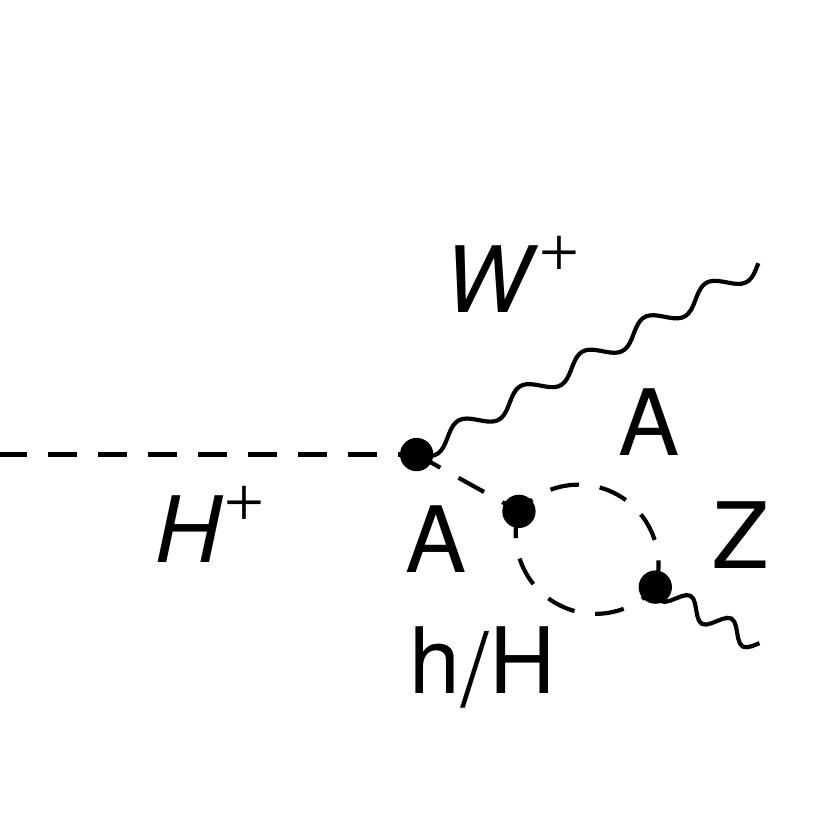} &\includegraphics[scale=0.4]{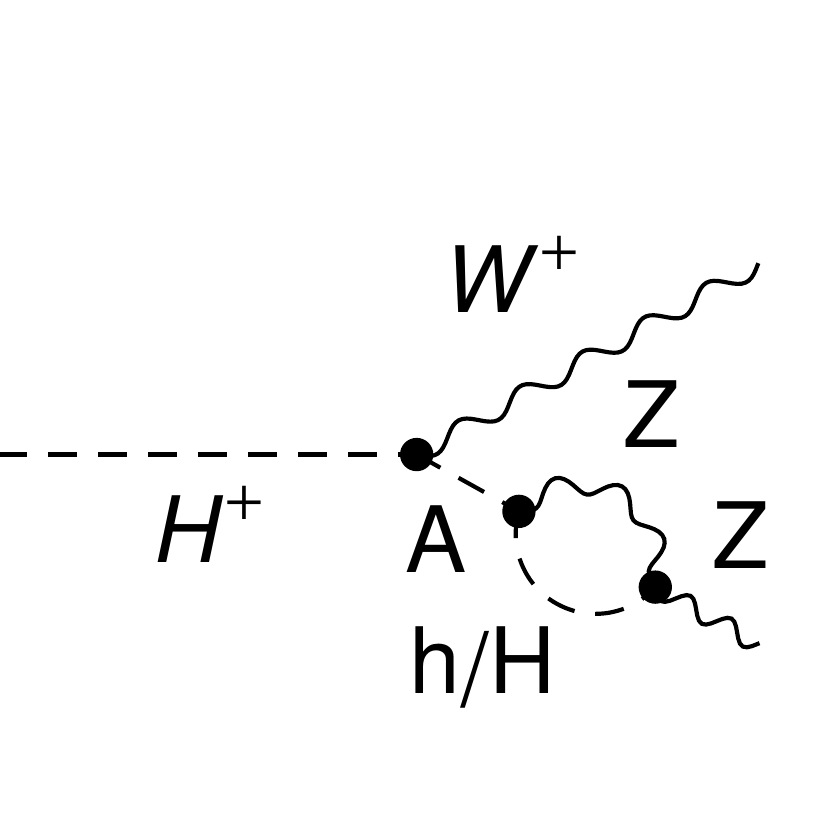} 
\end{tabular}
\end{center}
\caption{Feynman diagrams in $R_\xi$ gauge that contribute to the process $H^\pm \to W^\mp Z$ at one-loop level, where the external $Z$ boson has longitudinal polarization, $Z_L$. There is also the possibility to draw diagrams, with a $\hp W^\mp h/H$ vertex, where $h/H$ goes into an external $Z$ boson. Those diagrams vanish due to the different quantum numbers of $h/H$ and $Z$. Diagrams that contain propagators denoted by $h/H$ are to be counted as two diagrams:\ one diagram with a $h$ boson running in the loop and one with a $H$ boson instead. All of the diagrams in this figure vanish due to the form of the $AH^\pm W^\mp$ vertex as discussed in the text. }
\label{fig:HpWZ2}
\end{figure}

We now turn to the process $H^\pm \to W^\pm Z$. The tree-level coupling $g_{H^\pm W^\mp Z}$ depends on the $\grp{SU(2)}_L$ and $Y$ representations of the different scalar multiplets in a given model, and their vevs. In models where only $\grp{SU(2)}_L$ doublet representations are present, the coupling $g_{H^\pm W^\pm Z}$ vanishes at tree level. This coupling can in general be generated at higher orders. The diagrams for the process $H^\pm \to W^\pm Z$ at one-loop order are the same diagrams as for $H^\pm \to W^\mp \gamma $ (replace $\gamma \to Z$) plus the diagrams in \reffig{fig:HpWZ1}.

At this stage, we do not include off-shell effects in the $\hp \to W^\pm Z $ decays. 
The reason will become clear below in section \ref{Hp-results} where we will see 
 that since $m_h = 125$~GeV or lighter, the tree level decay 
$\hp \to W^{\pm (*)} h^{(*)}$ will dominate over $ \hp \to  W^\pm Z$ as soon as $h$ can be produced on-shell in $\hp \to W^{\pm *} h$. Now, since  $m_h = 125$~GeV is below the $W^\pm Z$ threshold, this will always be true. 
The inclusion of $\hp \to W^{\pm *} Z^* $ does not alter this result. However, it can in principle influence the importance of the $\hp \to W^\pm \gamma$ mode below the $WZ$ threshold, as indicated in \reffig{fig:HpBR1}. In addition, the inclusion of off-shell top quarks could also be important when we consider which decay mode is sub-dominant (at the percentage level). We leave these questions for future studies.

Finally, we have checked, using the \FA{} and \FC{} implementation of our model, that the calculated partial widths of $ \Gamma_{ \hp \to W^\pm Z/\gamma }$ are UV finite. For completeness we also note that the processes $H^\pm \to W^\pm Z/\gamma$  have been considered for the MSSM, as well as type I and II 2HDMs, in \cite{Arhrib:2006wd}.

\subsubsection{$H^\pm \to W^\pm h/H/A \to $ multiple fermions}
\label{Hphw}
In addition to the loop-decays already discussed, the $\hp$ can also decay into fermions via, possibly off-shell $W^\pm, h,H$ and $A$ bosons. 
Here we limit the discussion to decays into 4 or 6  fermions, $ \Gamma_{\hp \to 4f/6f } $. For 4 fermion decays the only relevant channel is
\begin{align}
\Gamma_{\hp \to 4f}= \Gamma (\hp \to [W^{\pm *}  \to 2f]+[h^*/H^* \to b \bar{b} ]).
\end{align}
For 6 fermion final states there are several different amplitudes that contribute.
In principle the partial width should be calculated from the sum of all of all these. In line with this we add the contributions from (possibly) virtual $h,H$ on the amplitude level. However, we do not consider possible interference terms between diagrams with different vector boson propagators. In other words, we approximate 
\begin{align}
\Gamma_{\hp \to 6f} & \approx \Gamma (\hp \to [W^{\pm *}  \to 2f] + [h^*/H^* \to W^* W^* \to 4f  ] )  \\ &
+ \Gamma( \hp \to [W^{\pm *} \to 2f]+ [h^*/H^* \to Z^* Z^* \to 4f  ])   ,
\end{align}
as is standard practice. We also define 
\begin{equation}
\Gamma_{\hp \to Wh/H} \equiv  \Gamma_{\hp \to 4f}   +   \Gamma_{\hp \to 6f}  .
\end{equation}
We calculate these widths using the \THDMC{} implementation of our model interfaced with the tree-level matrix-element and Monte Carlo phase-space generator \MadG{} \cite{Alwall:2007st,Alwall:2011uj,Alwall:2014hca}, with non-zero widths included for the internal propagators using the prescription  in \Eref{propagators}.

As we will see in section \ref{Hp-results},  $\Gamma_{\hp \to Wh/H } $ is negligible in comparison to the partial widths $\Gamma_{\hp \to f\bar{f}^\prime} $ and $ \Gamma_{\hp \to W^\pm \gamma}$ for $m_\hp < m_S \lesssim 2 m_W $, where $S$ is the lightest of $A$ and $h$, even after the inclusion of off-shell $h/H$ and $A$ bosons.
This is due to the smallness of the widths of the $h,H$ and $A$ bosons below the $h/H \to WW/ZZ$ and $ A \to Zh/H $ or $W^\pm H^\mp$ thresholds. The effects of off-shell $h,H$ and $A$ bosons can become sizable when we consider larger $m_{h,H,A}$, i.e.\ when the sub-channels $h/H \to VV$ or $A \to Z h/H$ are kinematically open, so that $ \Gamma_{h,H,A} = \mathcal{O}(1\text{ GeV})$. One should also remember that the $A H^\pm W^\mp$ coupling is independent of the mixing angle $\alpha$.

\begin{figure}[t]
\begin{center}
\begin{tabular}{cc}
\includegraphics[width=0.48\textwidth]{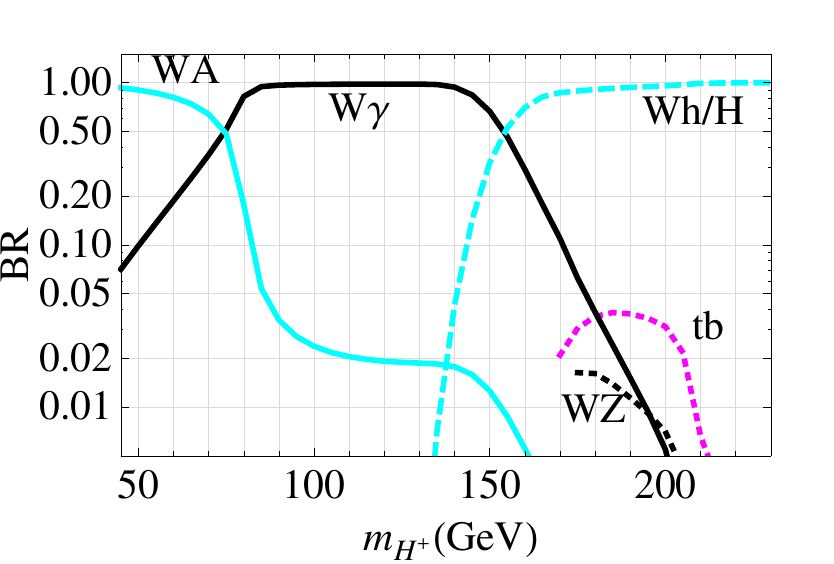} &  \includegraphics[width=0.48\textwidth]{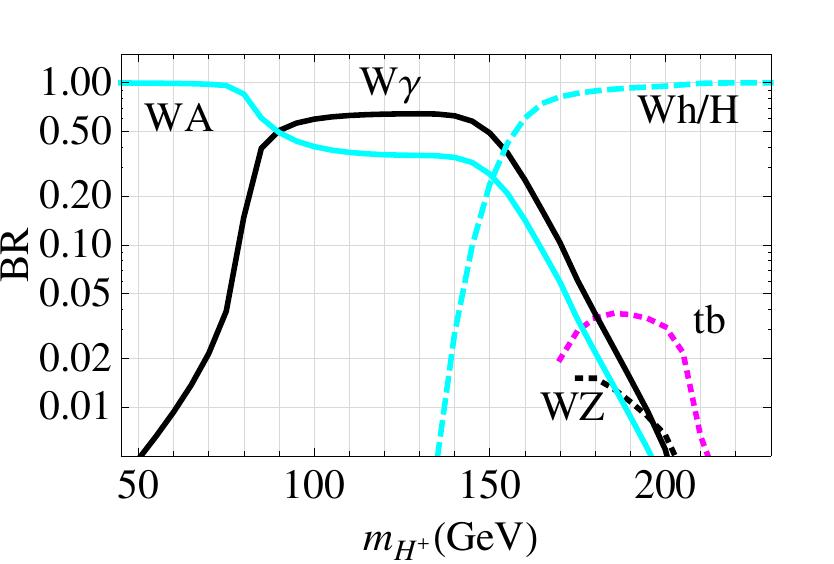}  \\
(a) $m_A = m_\hp - 10$ GeV, $m_h = 125$ GeV,  & (b) $m_A = m_\hp-20$ GeV, $m_h = 125$ GeV,  \\
$m_H = 300$ GeV, $\sina = 0.9$. & $m_H = 300$ GeV, $\sina = 0.9$.\\
\includegraphics[width=0.48\textwidth]{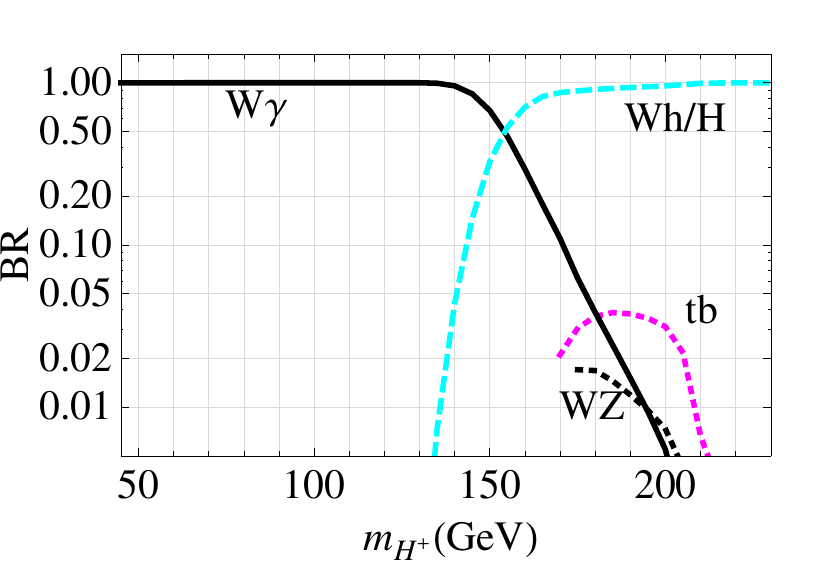} &  \includegraphics[width=0.48\textwidth]{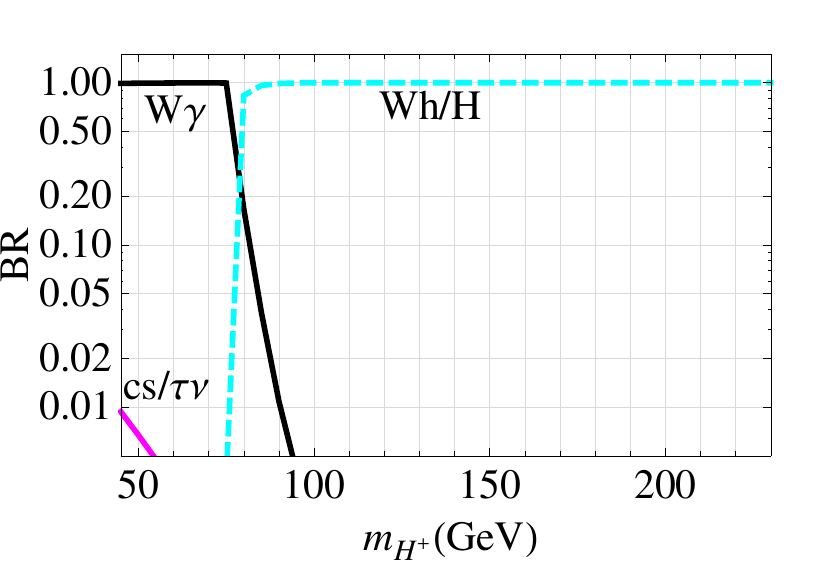}  \\
(c) $m_A = m_\hp $, $m_h = 125$ GeV, & (d) $m_A = m_\hp$, $m_h = 75$ GeV, \\
$m_H = 300$ GeV, $\sina = 0.9$. & $m_H = 125$ GeV, $\sina = 0.1$.\\
\end{tabular}
\end{center}
\caption{The branching ratios of the charged scalar $\hp$ as a function of $\mhp$. The solid black line shows the $W^\pm \gamma$ mode, dotted black $W^\pm Z$, solid cyan $W^\pm A$, dashed cyan $W^\pm h/H$ and dotted magenta $tb$. In this figure, we have $\lambda_3 = 2 (m_\hp/v)^2 $, $\lambda_2 = \lambda_1$ and $\lambda_7 = \lambda_6$. The scenarios in (a) and (b) are phenomenologically disfavored since EWPT require $m_A > m_\hp$ for these values of $m_\hp$ (see figure~\ref{fig:constraints} and the related discussion). }
\label{fig:HpBR1}
\end{figure}

\subsubsection{Decay widths and branching ratios for $\hp$} 
\label{Hp-results}

We have now come to the point where we can compare the magnitudes of the different decay modes under consideration in our standard cases with $\lambda_3 = 2 (m_\hp/v)^2 $, $\lambda_2 = \lambda_1$ and $\lambda_7 = \lambda_6$ as is illustrated in \reffig{fig:HpBR1}.
Here we have calculated the partial width of the decay mode $\hp \to W^{\pm *} A  $ using the results of \cite{Djouadi:1995gv} as implemented in \THDMC{}.

First of all it should be noted that the contribution to the decay modes of $\hp$ from the processes $\hp \to \tau \nu $ and $\hp \to cs $ is very small: BR$({\hp \to \tau \nu})$ + BR$({\hp \to cs})$ $< \mathcal{O}$(1\%). As mentioned, due to the broadness of the $W^\pm$ boson, the $\hp \to W^{\pm *} \gamma  $ mode dominates over $\hp \to \tau \nu $ and $\hp \to cs $ even below the threshold, $m_\hp < m_W$. If one considers $m_A < m_\hp$, the decay mode $\hp \to W^{\pm *} A$ can start to make a significant contribution and will dominate the branching ratios for the charged scalar below the $W^\pm \gamma$ threshold, $m_\hp < m_W$. If we consider Case~1, then the mass of the $A$ boson has to be heavier than $\hp$ for $m_\hp \lesssim M $ (according to the limits from EWPT illustrated in \reffig{fig:constraints}a with $M$ given by \refeq{eq:ms2}) and the decay mode $\hp \to W^\pm A$ is therefore not possible for light $\hp$. For charged scalar masses larger than $m_W$, the decay mode $\hp \to W^\pm \gamma$ will dominate, provided that $\hp$ is the lightest scalar in our model. The $tb$ and $WZ$ modes will contribute to the branching ratios of the order a few percent. 

A consequence is that the charged scalar in our model is \emph{not} in general constrained by the
LEP result $m_\hp \gtrsim 80$ GeV, valid for BR$({\hp \to cs})$ + BR$({\hp \to \tau \nu})$ = 1~\cite{Abdallah:2003wd,Heister:2002ev,Abbiendi:2013hk}. Moreover, the $W\gamma$ channel can be dominant, and $WZ$ of order 1\%. This is to be compared to the case of type-I or II  2HDMs and MSSM where the maximal branching ratios for the $W\gamma$ mode are $\sim \mathcal{O}(10^{-5}) $ and $WZ\,\sim \mathcal{O}(10^{-3}) $ \cite{Arhrib:2006wd}. 

Note that, as shown in figure~\ref{fig:HpWgammaL3}, the width of $\hp$ can become very small in some regions of parameter space. For example, if the width would be 1~eV, then the proper decay length is $c\tau\sim 0.2\,\mu$m, and if the width is as small as 1~meV, then  $c\tau\sim 0.2$ mm. It would therefore be interesting to study whether this could lead to tracks or displaced vertices in the detector. Such signatures have been studied by the CMS collaboration in \cite{Chatrchyan:2013oca}.

\subsection{Decays of the CP-odd scalar $A$}
\label{sect:cpodddecays}
We end this section on scalar decays by considering the decays of the $A$ boson. As mentioned for the $\hp$ bosons, we do not know \textit{a priori} if the decay modes of the $A$ boson into 4 or 6 fermions, via possible off-shell bosons, dominates over $A \to f \bar{f}$, which proceeds at one-loop at the lowest order in our model\footnote{Note that due to the quantum numbers of the $A$ boson, the amplitudes for $A \rightarrow VV$, where $ VV =
W^+W^-, ZZ, \gamma  \gamma, Z \gamma $ or $gg $, are zero at tree-level. 
In general 2HDMs, the $A$ boson can couple to a pair of gauge bosons at
one-loop order through a loop of fermions \cite{Arhrib:2006rx}. This is not the case in our model
due to the vanishing of the tree-level couplings between $A$ and a fermion pair, $C_{Af\bar{f} }
\sim \rho^F = 0$. This means that in our model, $A \to VV$ is a two-loop process. We will not consider these decay modes in this paper.}. The decay modes of the $A$ boson into 4 or 6 fermions through possible off-shell $h,H,\hp,Z$ and $W^\pm$ bosons are calculated in a very similar way as the decay of the charged scalar, in section \ref{Hphw}.

\subsubsection{$A \to f\bar{f}  $}
\label{Aff}
The situation here is similar to the situation for the charged scalar: the CP-odd scalar $A$
couples to a pair of fermions with the same diagrams as the charged scalar, but with the $W^\pm$
or $G^\pm$ bosons replaced with $Z$ or $G^0$ in the loop.
The $A$ bosons will mix with longitudinally polarized $Z$ bosons (and with $G^0$ bosons in $R_\xi $
gauge), which in turn go into a pair of fermions. We will renormalize the $AZ$ and $AG^0$ mixing in the
same way as for $H^\pm W^\pm$ and $H^\pm G^\pm$, i.e., the real part of the mixing vanishes for an on-shell $A$ boson.

\begin{figure}
\begin{center}
\includegraphics[scale=0.98]{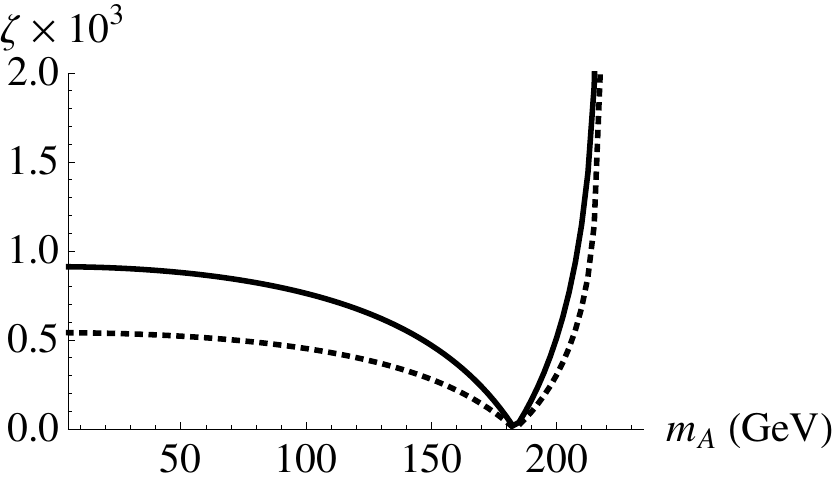}
\end{center}
\caption{The ratio  $  \zeta = \left[ \rho^{L} \right]_{33}/\left( m_\tau / v \right)  $ as a function of $m_A$. The
solid line is for $\sina = 0.7$ and the dotted for $\sina = 0.95$. The other parameters of the
model are taken to be $m_h = 125$ GeV, $m_H = 300$ GeV, $\mhp = m_A $.}
\label{fig:AtautauSDMvsTypeI}
\end{figure}

One way to give a measure of the magnitude of the loop-generated $\rho^F$ elements in our model is by comparing
e.g.\ $\Gamma_{A\to \tau^+\tau^- }$ calculated in our model (at one-loop level) with the
tree-level result obtained in a generic model. Writing the effective interaction as 
$\ii  \bar{\Psi}_\tau \left[ \rho^{L} \right]_{33}
\gamma_5 \Psi_\tau \,A$, we can calculate the effective coupling $ \left[ \rho^{L} \right]_{33}$
from
\begin{equation}
\Gamma_{A\to \tau^+\tau^-} =  \; \left(\left[ \rho^{L} \right]_{33}\right)^2 \frac{m_A}{8\pi} \sqrt{1 -  \frac{4m_\tau^2}{m_A^2} \:  } \:  .
\end{equation} 
Defining the ratio
\begin{equation}
\zeta \, \equiv \; \frac{ \left[ \rho^{L} \right]_{33} }{  m_\tau / v  } \:  ,
\end{equation}
where $\zeta = 1$ is the value obtained in a Type-I 2HDM with $\tan \beta = 1$, we find that the magnitude of $ \zeta$ in our model is $ \mathcal{O}(10^{-3})  $ if $m_A \lesssim m_h + m_Z$, see  \reffig{fig:AtautauSDMvsTypeI}. Note that this effective coupling is independent of the values of $\mhp$, $\lambda_3$, $\lambda_2
$, and $\lambda_7$. 

Another property of the model is that at lowest order we have
\begin{equation}
 \frac{\Gamma_{A\to c\bar{c}} }{  \Gamma_{A\to s\bar{s}} }   =  \frac{m_c^2}{m_s^2}.
\end{equation}
In this sense, our model is therefore Type I-like. Furthermore, as already mentioned, the off-diagonal entries in the $\rho^F$
matrices are zero at one-loop level. This is due to the absence of a $W^\pm$ boson in the diagrams
for the process $A \to f\bar{f}$.
At two-loop order, off-diagonal $\rho^F$ matrix
elements are generated and will introduce new FCNC in our model.
	
\begin{figure}[bt]
\begin{center}
\begin{tabular}{cc}
\includegraphics[width=0.48\textwidth]{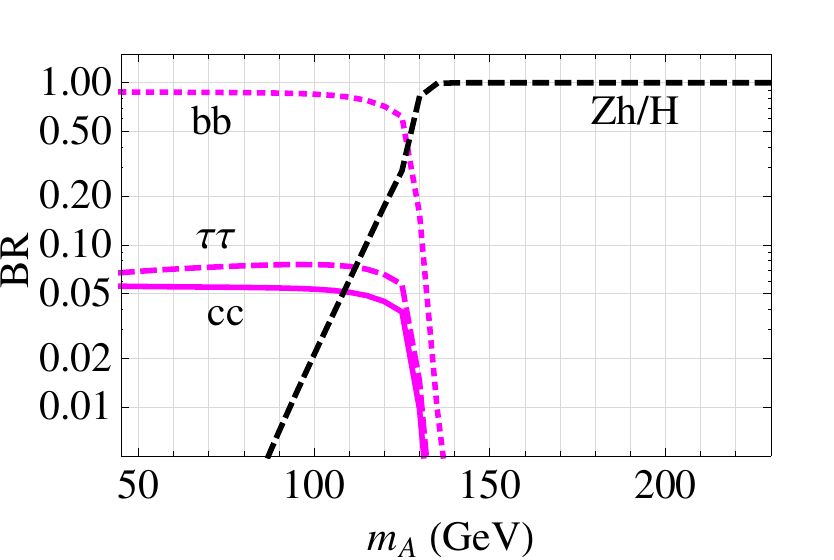} &  \includegraphics[width=0.48\textwidth]{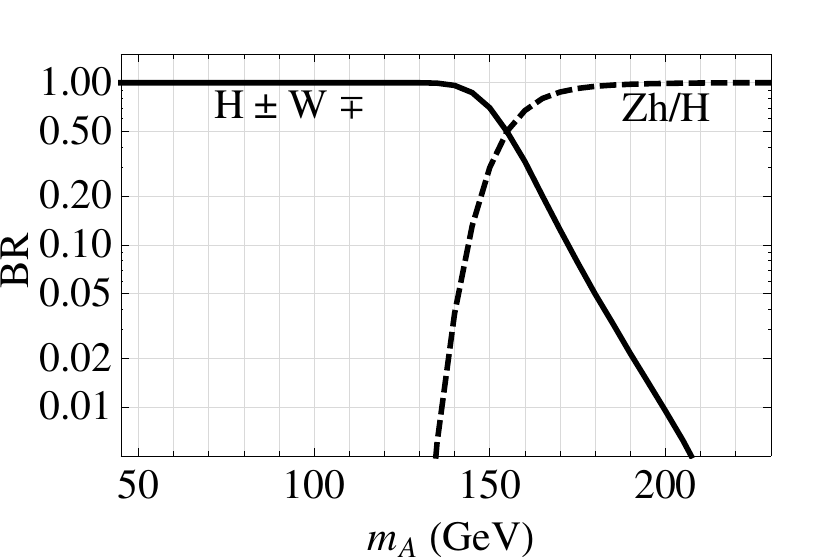}  \\
(a) $m_\hp = m_A $, $m_h = 125$ GeV, & (b) $m_\hp = m_A -  45$ GeV, $m_h = 125$ GeV, \\
$m_H = 300$ GeV, $\sina = 0.9$. & $m_H = 300$ GeV, $\sina = 0.9$.\\
\end{tabular}
\end{center}
\caption{The various branching ratios for the scalar $A$: dotted magenta $A \to b\bar{b}$, solid magenta $A \to c\bar{c}$, dashed magenta $A \to \tau \tau $, dashed black $A \to Zh/H$, solid black $A \to W^\pm H^\mp$. Here $\lambda_3 = 0$, $\lambda_2 = \lambda_1$ and $\lambda_7 = \lambda_6$. }
\label{fig:ABR1}
\end{figure}

\begin{figure}[bt]
\begin{center}
\includegraphics[width=0.6\textwidth]{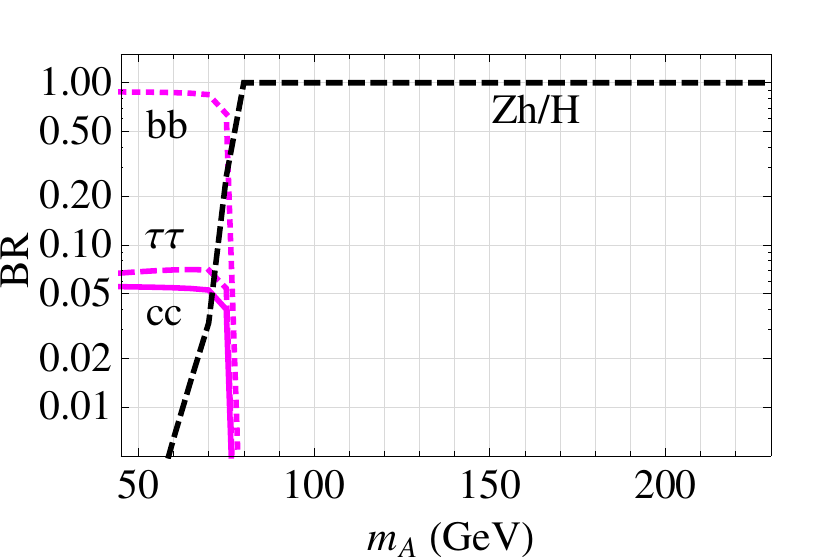} 
\end{center}
\caption{The branching ratios  for the scalar $A$:  dotted magenta is $b\bar{b}$, solid magenta $c\bar{c}$, dashed magenta $\tau \tau $, dashed black $Zh/H$. $m_\hp = m_A $, $m_h = 75$ GeV $m_H = 125$ GeV, $\sina = 0.1$. Here $\lambda_3 = 0$, $\lambda_2 = \lambda_1$ and $\lambda_7 = \lambda_6$.   }
\label{fig:ABR2}
\end{figure}

\begin{figure}[bt]
\begin{center}
\begin{tabular}{cc}
\includegraphics[width=0.48\textwidth]{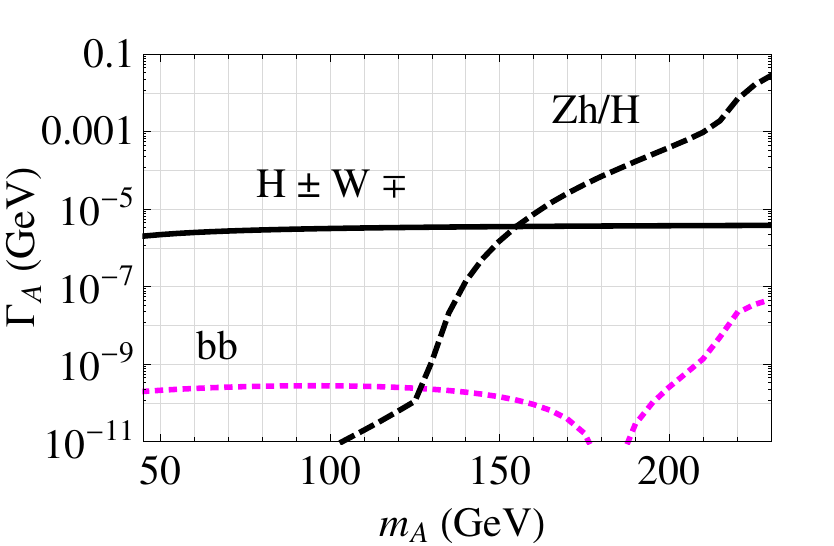} &  \includegraphics[width=0.48\textwidth]{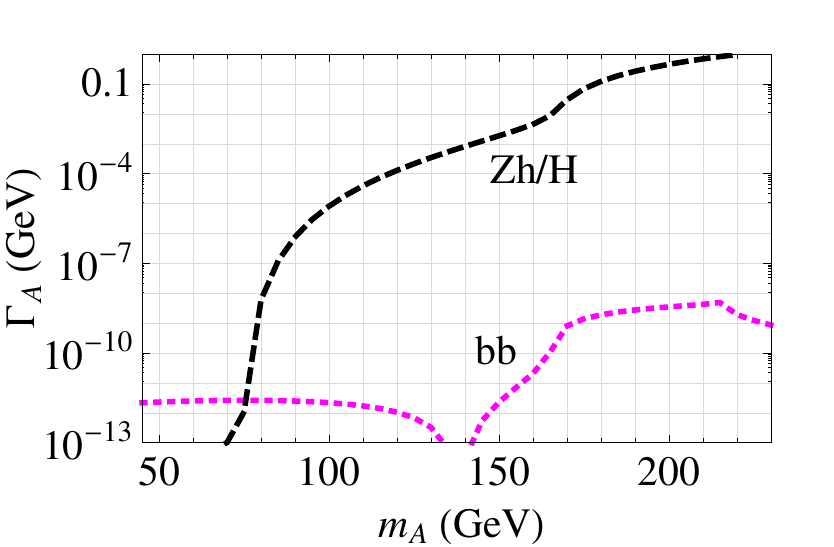}  \\
(a) $m_h = 125$ GeV, $m_H = 300$~GeV, & (b) $m_h = 75$ GeV, $m_H = 125$~GeV,  \\
$m_\hp = m_A - 45$ GeV, $\sina = 0.9$.  & $m_\hp = m_A $, $\sina = 0.1$.   \\
\end{tabular}
\end{center}
\caption{The partial decay widths $\Gamma_A$ as a function of $m_A$.  Dotted magenta is $b\bar{b}$, dashed black $Zh/H$, solid black $W^\pm H^\mp$. Here $\lambda_3 = 0$, $\lambda_2 = \lambda_1$ and $\lambda_7 = \lambda_6$.   }
\label{fig:AGamma1}
\end{figure}

\subsubsection{Decay widths and branching ratios for $A$}
\label{A-results}
The result of the calculations for the partial widths and branching ratios for the $A$ boson is similar to those of the charged scalar. If $A$ is not the lightest scalar in our model, the dominating decay mode is $A \to S V$, where $S$ is the lightest scalar and $V$ the associated vector boson. If $A$ is the lightest scalar, the $b \bar{b}$ mode dominates, see \reffig{fig:ABR1} and \reffig{fig:ABR2}. 
The partial decay widths $A \to f \bar{f} $ are proportional to $ \sin^2 2\alpha $ and can be very small, \reffig{fig:AGamma1}. 
In Case~1 there is no region in parameter space which allows the $A$ boson to be the lightest scalar. As was outlined in section \ref{sect:constraints}, in Case~1 one should have $m_A \gtrsim m_\hp + 50$ GeV in order to fulfill the constraints from EWPT for $m_\hp$ below $m_h = 125$ GeV. In Case~2 we have larger freedom to choose $m_A$ and $m_\hp$ according to \reffig{fig:constraints}b. But the recent LHC results restrict the possible $m_A$ and $m_\hp$ since e.g.\ the decay mode $H \to AZ^*$ (with $Z$ far off shell) should not be allowed.

\section{Possible signals of the SDM at collider experiments}
\label{SDMatColliders}
We have seen that the scalars in our model, and in particular $\hp$ and $A$, can have non-standard decay modes. In particular, if $\hp$ is the lightest scalar, its dominating decay mode will be $\hp\to W^\pm \gamma$ unless the parameters are fine tuned. In this section we now consider the production of the scalars. As mentioned in section \ref{sect:collconstraints}, the CP-even scalars $h$ and $H$ are produced in the same way as $H_{\text{SM}}$ in $gg$-fusion and VBF, but with modified couplings
\begin{equation}
\sigma_k(pp \to H_i) = \kappa_{H_i} \, \sigma_k(pp \to H_{\text{SM}}),
\label{eq:pphH}
\end{equation}
where $H_i = h,H$, $\kappa_h = \sin^2 \alpha, \kappa_H = \cos^2 \alpha$ and $\sigma_k$ are the production cross-sections through $gg$-fusion or VBF. The expression in \eref{eq:pphH} is valid up to electroweak corrections. The QCD corrections, which are the most important ones, are the same in our model as in the SM.

The discovery of a charged scalar $\hp$ has for long been considered a sure sign of physics beyond the SM. In the standard scenarios such as MSSM, NMSSM or 2HDMs, $\hp$ are produced primarily in top quark decays if they are light, or if they are heavy, in association with top and bottom quarks in $gg$ and $gb$ collisions. 

In our model, the $t b \hp$ coupling is zero at tree level and is instead generated by loops, and the same holds for the $At\bar t$ coupling. We have calculated the loop-generated decay width $\Gamma_{t\to H^+ \,b}$ in our model in the same way as we calculated $\Gamma_{H^+ \to t \, \bar{b}}$. The result is that the branching ratio $\text{BR}(t\to H^+ \,b) $ is less than $ 10^{-6} $ for allowed points in parameter space (the $\lambda$ parameters can not be arbitrarily large). So, due to the absence of tree-level fermion couplings for $\hp$ and $A$, the standard production mechanisms of the $\hp$ involving the $t \bar{b} H^+$ couplings and the $gg \to A$ channel for the $A$ are negligible. Other production mechanisms must therefore be considered. Our model thus leads to a novel phenomenology of the $\hp$ and $A$ bosons, with both production and decay modes being non-standard. More detailed phenomenological studies of $\hp$ and $A$ will be performed in future work, but in this section, we briefly outline some channels that will be important. Production cross sections for light charged Higgs bosons in general 2HDMs can be found in \cite{Aoki:2011wd}.

\begin{figure}[tb]
\begin{center}
\begin{tabular}{cc}
\includegraphics[scale=0.32]{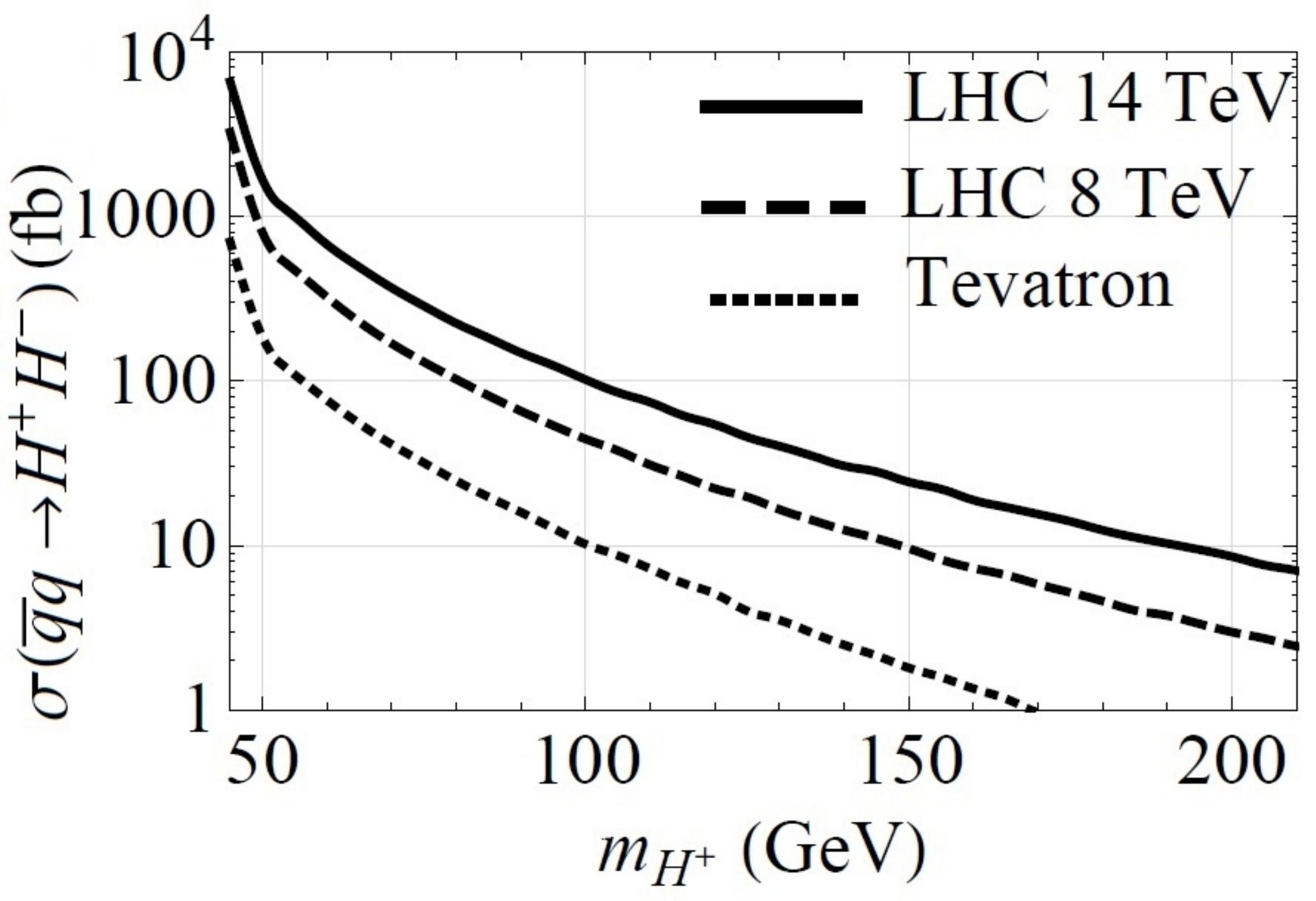} & \includegraphics[scale=0.33]{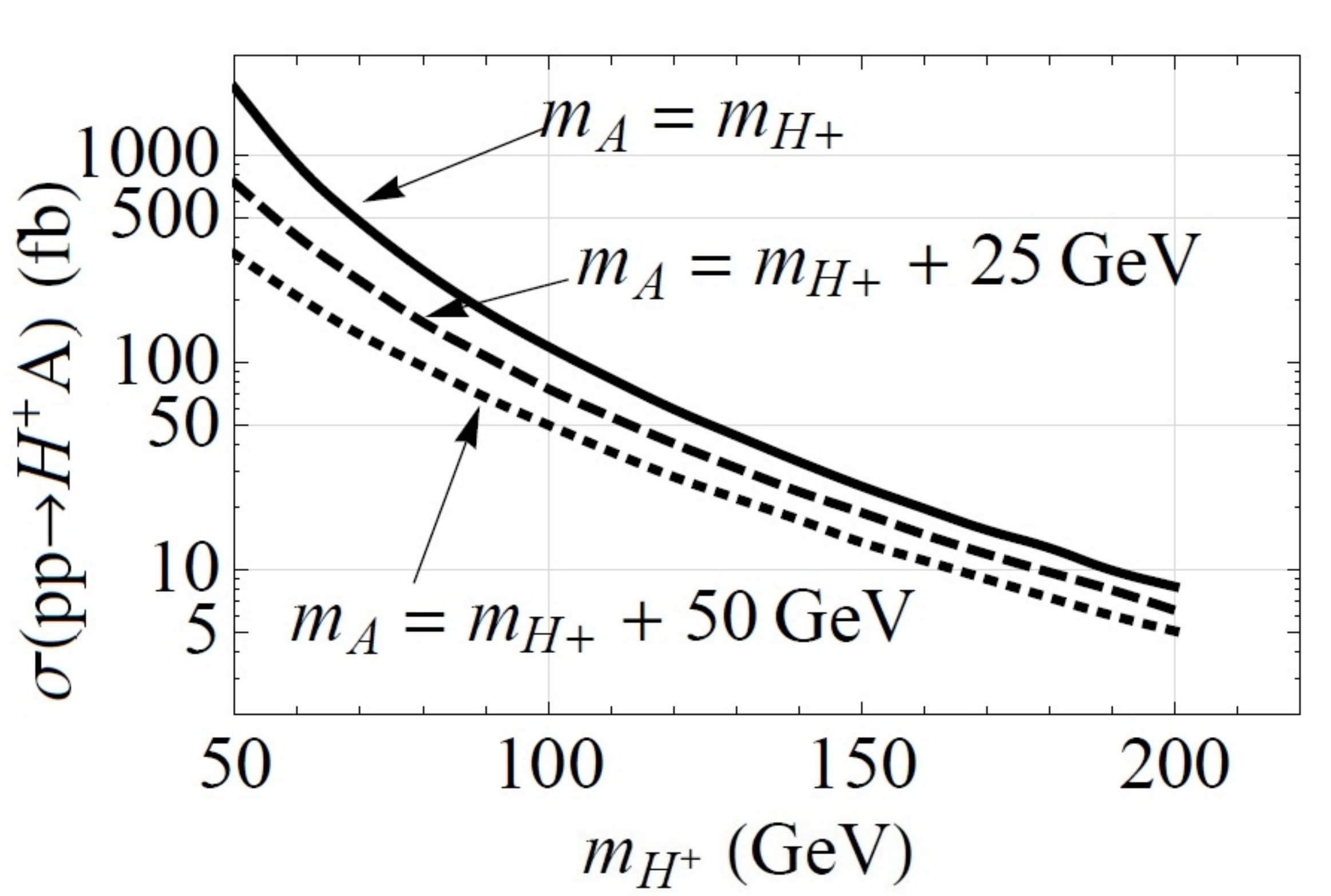} \\
(a) & (b) \\
\includegraphics[scale=0.33]{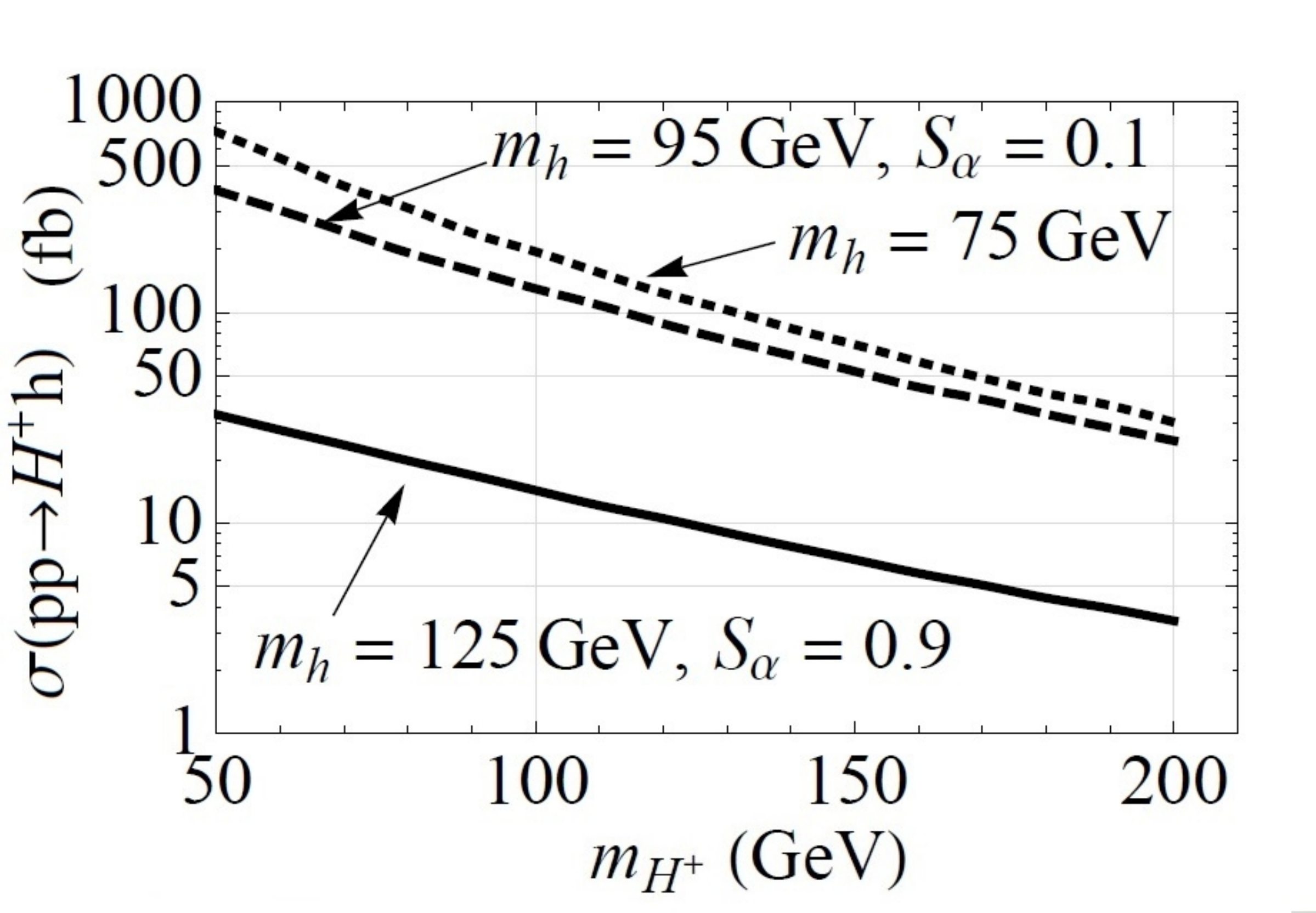} &   \includegraphics[scale=0.34]{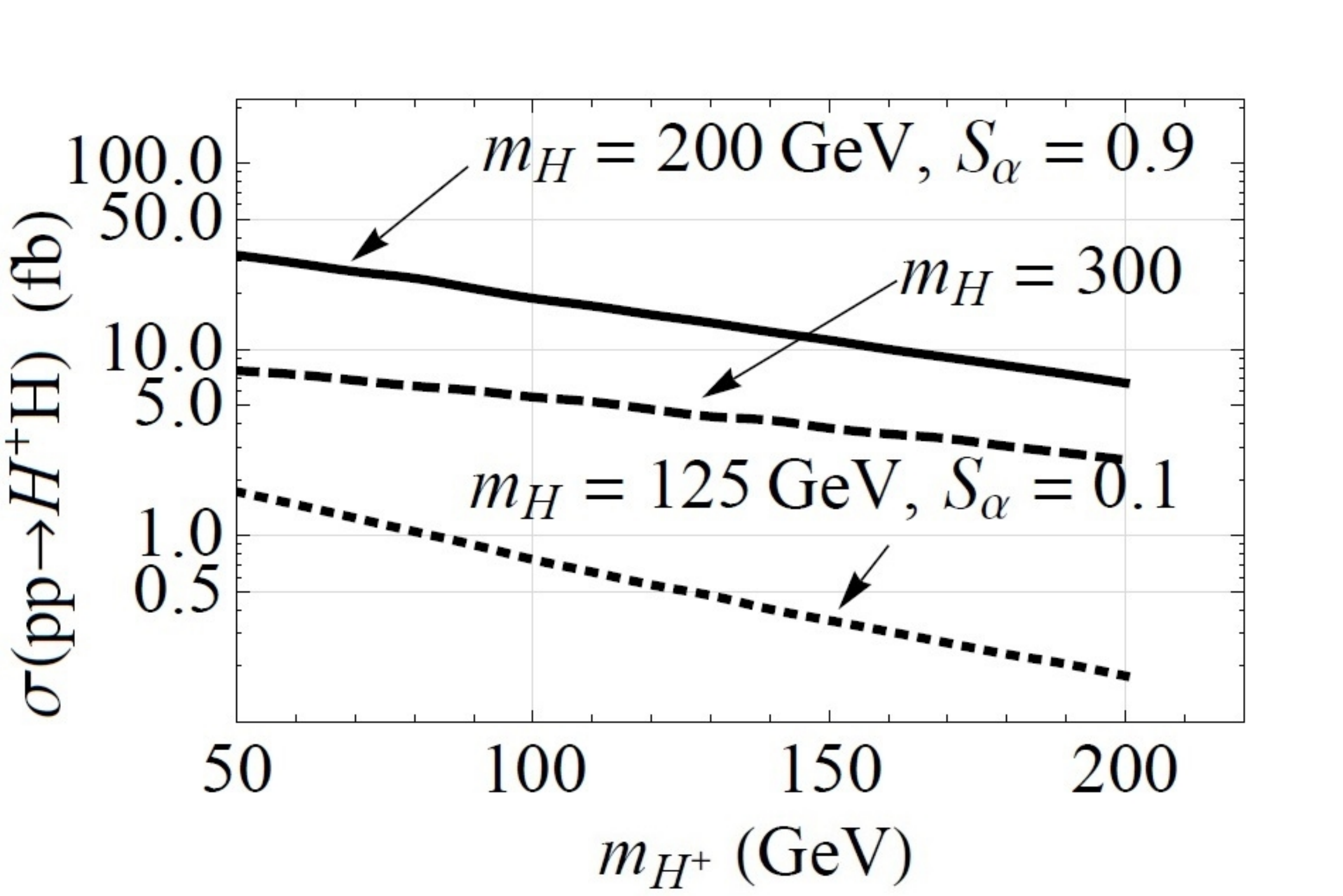}\\
(c) & (d) \\
\end{tabular}
\end{center}
\caption{
Hadronic cross-sections for various production mechanisms as functions of $m_\hp$:
(a)~$\sigma(pp \to H^+ H^-)$ at LHC 8 (14) TeV solid (dashed) and $\sigma(p\bar{p} \to H^+H^-)$ at the Tevatron (dotted),
(b) $\sigma(pp \to H^+ A)$ at LHC 8 TeV. For the solid/dashed/dotted lines, we have $m_A = m_\hp + 0/25/50$~GeV,
(c) $\sigma(pp \to H^+ h)$ at LHC 8 TeV. For the solid line, we have $m_h = 125$ GeV with $\sina = 0.9$, the dashed (dotted)  line $m_h = 95$ (75) GeV with $\sina = 0.1$, 
(d)~$\sigma(pp \to H^+ H)$ at LHC 8 TeV. For the solid (dashed) line, we have $m_H = 200$  (300) GeV with $\sina = 0.9$ and for the dotted line $m_H = 125$ GeV with $\sina = 0.1$. 
}
\label{fig:DY}
\end{figure}

\subsection{Production of $\hp$} 
\label{sec:HpProd}
The production of a pair of charged scalars in $q\bar{q}$ collisions through $s$-channel $\gamma^* / Z^*$ exchange depends on the electroweak couplings through the $ZH^+H^-$ and $\gamma H^+ H^-$ vertices.  Except for the dependence on $m_\hp$, the partonic cross section for $q\bar{q} \to H^+H^-$ does not depend on the parameters of the scalar potential, if one neglects the contribution from $s$-channel processes with $h$ and $H$ bosons, whose couplings to the quarks involved are very small. To get a first estimate of the hadronic production cross sections we have calculated  
 $\sigma({pp \to H^+H^-})$ at $\sqrt{s} = $ 8 TeV and 14 TeV and $\sigma(p\bar{p} \to H^+H^-)$ at $\sqrt{s} = $ 2 TeV using the LO Monte Carlo generator software \MadG{} \cite{Alwall:2007st,Alwall:2011uj,Alwall:2014hca} with CTEQ6L1 PDFs and using factorization and renormalization scales set to $\mu=M_Z$. The results are shown in \reffig{fig:DY}a as a function of $m_\hp$.

Another production process to consider is the associated production $q \bar{q}^\prime \to W^* \to \hp \, S $, where $S = h,H$ or $A$. This will give cross sections of similar magnitude as $q \bar{q} \to \gamma^* / Z^* \to H^+H^-$, provided that the sum of the final state rest masses are similar; $m_\hp + m_S \approx 2 m_\hp$. In figure \ref{fig:DY}b--d  
we show the leading order hadronic cross sections at the LHC with $\sqrt{s} = 8$ TeV, as calculated with \MadG, for $pp \to H^+ A$, $pp \to H^+ h$ and $pp \to H^+ H$ respectively. We note that the process $q \bar{q}^\prime \to W^* \to \hp \, A $ is independent of the mixing angle $\alpha$, whereas the $q \bar{q}^\prime \to W^* \to \hp \, H_i $ processes have a dependence on $\alpha$ through the $W^\pm H^\mp H_i$ coupling, where $H_i = h,H$. 

\begin{figure}[t]
\begin{center}
\begin{tabular}{cc}
\includegraphics[width=0.48\textwidth]{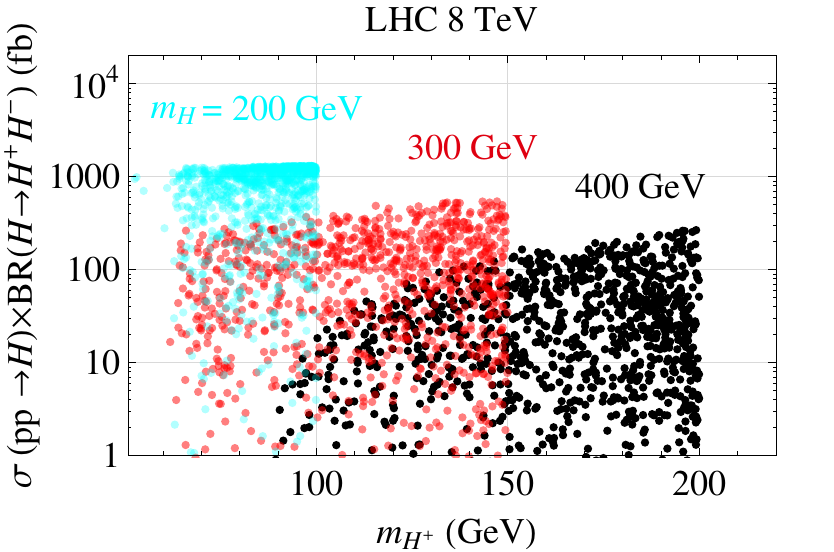} & \includegraphics[width=0.48\textwidth]{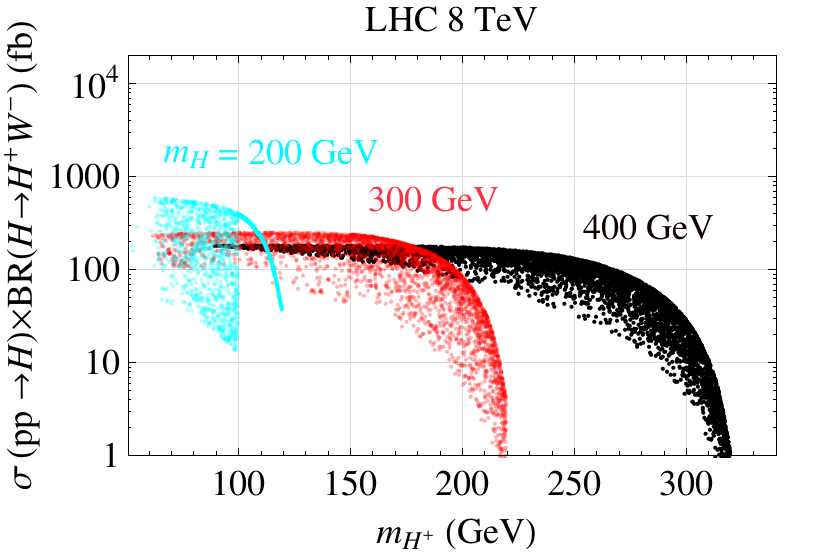} \\
\end{tabular}
\end{center}
\caption{The cross-section $\sigma(pp \to H)$ times branching ratio for $H\to H^+H^-$ (left), and $H\to H^+ W^-$~(right), at the LHC for $\sqrt{s} = 8$~TeV with $\sin \alpha = 0.9$ and different values of $m_H$. The values of the other parameters are described in the text. }
\label{fig:sigmaXBRH}
\end{figure}

The final production mechanisms of $\hp$ that we consider in this paper are via the $H$~boson from $pp\to H$, with subsequent decays $H\to H^+H^-$ and $H\to H^+W^-$. We have employed the same scan and constraints as in section \ref{sect:cpeven}, and the expression \eref{eq:pphH} together with the $H_\text{SM}$ cross sections from~\cite{Dittmaier:2011ti}. The results for $\sigma(pp\to H) \times \text{BR}(H\to H^+ H^-) $ and $\sigma(pp\to H) \times \text{BR}(H\to H^+ W^-) $ are shown in \reffig{fig:sigmaXBRH}, for $\sin \alpha = 0.9$ and $\sqrt{s} = 8$~TeV. The $m_\hp$ dependence of the cross section lies solely in the branching ratios of $H$ discussed in section~\ref{sect:cpeven}. We note that in Ref.~\cite{Ilisie:2014hea}, the off-shell contribution $\sigma(pp\to h^*/H^* \to H^\pm W^\mp)$ is calculated for the fermiophobic A2HDM. These results can be carried over to our model. We also note that in our model, given the cross sections in \reffig{fig:sigmaXBRH}, there could already exist a significant amount of events with charged scalars $H^\pm$ originated from $H$ decays. For $m_\hp \lesssim m_h $, the $H^\pm$ decays more or less exclusively into $W^\pm \gamma$, see section \ref{Hp-results}. Finally, the production of a heavy resonance $H$ in $gg$-fusion, with the decay chain $H \to (H^\pm \to (\hcal \to b\bar{b})W^\pm )W^\mp $, was studied for the LHC 8 TeV run in Ref.~\cite{Rompotis:2014qra}.

\subsection{Production of $A$ }
The process $gg \to A $ occurs at two-loop level in our model. We will instead consider those regions in parameter space where it can be produced in decays of the other scalars or in association with those. If $H$ is heavy enough, its decays $H \to A Z, AA $ could contribute a significant amount of the total production cross section of $A$ bosons at the LHC. One could also consider the process
$q \bar{q}^\prime  \to Z^* \to A h/H $ for which the cross section  is similar to the previously discussed $q \bar{q} \to H^\pm A/h/H$.  One would then have to consider the subsequent decay of the $A$ boson into $hZ$ or $\hp W^\mp$. If $A$ is the lightest scalar of the model one has to instead consider $A \to b \bar{b}$.

We also note that if $A$ is the lightest scalar, the decay width can be very small, $\Gamma_A < 1$~meV as shown in \reffig{fig:AGamma1}b, for $\sin \alpha \sim 0.1 $.
This feature might open for signatures with displaced vertices in the detectors, provided that the $A$ bosons are produced with sufficient  $p_T$ relative to its mass to that the $\gamma$-factor from the boost is large enough.

\section{Conclusions and summary}\label{sect:conclusions}
In this paper, we have discussed a novel type of 2HDMs, first introduced by us in \cite{Enberg:2013ara}, where the $\ztwo$ symmetry is only broken in the potential and only one of the doublets has tree-level fermion couplings such that new FCNCs occur first at the two-loop level. Since the $H^\pm$ and the $A$ bosons reside solely in the fermiophobic doublet, indirect constraints from flavor observables do not apply. We also demonstrated that there are substantial regions of the parameter space of the model which satisfy theoretical constraints, are compatible with EWPT and earlier Higgs searches, and with the new LHC results. In particular, we have considered the $\hcal \to  \gamma \gamma$ and $\hcal \to  ZZ$ signal strengths, where $\hcal$ denotes the observed Higgs boson. 

We have calculated the decay rates of all scalars, and in particular the decays of the $\hp$ and $A$ bosons that occur through one-loop processes at lowest order. Decay modes involving off-shell final state particles have also been considered in detail. These calculations show that if the $\hp$ boson is the lightest scalar of the model, the non-standard decay mode $\hp\to W^\pm \gamma$ will typically dominate. Otherwise, decays of $\hp$ into on-shell scalars and off-shell vector bosons will dominate. The decay modes of the $A$ boson show a similar behavior as for the $\hp$ boson. If $A$ is not the lightest scalar, then $A$ will decay into on-shell scalars and off-shell vector bosons. If $A$ is the lightest scalar, $A \to b \bar{b}$ is the dominating decay channel.

Since the $\hp$ and $A$ bosons of this model are fermiophobic at tree level, they have loop-suppressed standard production channels at hadron colliders. Therefore, we consider production of these scalars in pairs, and in association with vector bosons and other scalars. These production channels could originate from $ q \bar{q}^\prime$ collisions or $H$ decays. We estimate that, if light enough, $\hp$ and $A$ could already had been produced in considerable amounts at the LHC. Therefore, more detailed investigations of such scenarios should be considered, in particular the case where $\hp \to W^\pm \gamma$ is the dominating decay mode.

\subsection*{Acknowledgments}
We thank T.\ Hahn for being very helpful in answering questions about \FA{} and \FC{} and  A.\ Arhrib 
and R.\ Pasechnik for helpful discussions regarding renormalization and the smeared mass unstable particle model. We also thank M.\ Krawczyk and R.~Santos for helpful comments and discussions.
R.E.\ and J.R.\ are supported by the Swedish
Research Council under contracts 2007-4071 and 621-2011-5333.

\section*{Appendices}
\appendix

\section{Couplings}
\label{sect:couplings}
In this Appendix we give the three-particle couplings of scalars and gauge bosons. We do not list all
the Goldstone boson couplings or the four-particle couplings, but these can be easily obtained
using the \FRU{} implementation of the model. As before, we define $s_\alpha=\sin\alpha,
c_\alpha=\cos\alpha$, and $s_W=\sin\theta_W$.
The triple scalar couplings of the model are then given by $g_{ijk} = -\ii  v c_{ijk}$ for $i,j,k =
h,H,A,H^{\pm}$, where
\begin{align}
c_{hhh} &= 3   \left(-s_\alpha^3 \lambda_1 + 3 c_\alpha s_\alpha^2 \lambda_6 + c_\alpha^3 \lambda_7 -c_\alpha^2 s_\alpha \lambda_{345} \right) ,\\
c_{HHH} &= 3   \left(c_\alpha^3 \lambda_1+3 c_\alpha^2 s_\alpha \lambda_6+s_\alpha^3 \lambda
  _7+c_\alpha s_\alpha^2 \lambda_{345}\right),\\
c_{hhH} &=   3 s_\alpha^3 \lambda_6-3c_\alpha^2 s_\alpha \left(2 \lambda_6- \lambda
  _7\right)+c_\alpha s_\alpha^2 \left(3 \lambda_1-2 \lambda_{345}\right)+c_\alpha^3 \lambda
  _{345},\\
c_{hHH} &=   3 c_\alpha^3 \lambda_6-3c_\alpha s_\alpha^2 \left(2 \lambda_6- \lambda
  _7\right)-c_\alpha^2 s_\alpha \left(3 \lambda_1-2 \lambda
  _{345}\right)-s_\alpha^3 \lambda_{345},\\
c_{hAA} &=   -s_\alpha \left(\lambda_3+\lambda_4-\lambda_5\right)+c_\alpha \lambda_7,\\
c_{HAA} &=   c_\alpha \left(\lambda_3+\lambda_4-\lambda_5\right)+s_\alpha \lambda_7,\\
c_{hH^+H^-} &=  -s_\alpha \lambda_3+c_\alpha \lambda_7,\\
c_{HH^+H^-} &=   c_\alpha \lambda_3+s_\alpha \lambda_7,\\
c_{hH^+G^-} & =\half\left(2 \sa \lambda_6 - \ca (\lambda_4 + \lambda_5 ) \right),\\
c_{HH^+G^-} & = -\half\left(2 \ca \lambda_6 + \sa (\lambda_4 + \lambda_5 ) \right).
\end{align}
Coming to the gauge--scalar couplings, we start with  the $SSV$ couplings. Writing the Feynman rules as
\begin{equation}
S_1S_2V \: : \: g_{S_1S_2V}(p^\mu_{S_1} - p^\mu_{S_2}  ),
\end{equation}
where the momenta are taken to be incoming, we have
\begin{align}
g_{hAZ} &= \frac{e c_{\alpha }}{2 c_W s_W} \, ,\ \qquad g_{HAZ} = \frac{e s_{\alpha }}{2 c_W s_W}\, , \\
g_{hH^\pm W^\mp} &= \mp \frac{\ii  e c_{\alpha }}{2 s_W}\, ,\ \qquad  g_{HH^\pm W^\mp} = \mp \frac{\ii  e s_{\alpha }}{2 s_W}\, , \\
g_{AH^\pm W^\mp} &= -\frac{e}{2 s_W} \, ,\\
g_{H^+H^-Z} &= \frac{\ii e \left(c_W^2-s_W^2\right)}{2 c_W s_W} \,,  \qquad g_{H^+H^-\gamma } = \ii e\, .
\end{align}
Finally we have the $SVV$ couplings, which we write as 
\begin{equation}
SVV \: : \: g_{SVV} g^{\mu\nu}
\end{equation}
with
\begin{align}
g_{hZZ} &= -\frac{\ii  e^2 v s_{\alpha }}{2 c_W^2 s_W^2} \, ,\qquad  g_{HZZ} = \frac{\ii  e^2 v c_{\alpha }}{2 c_W^2 s_W^2}\, , \\
g_{hW^+W^-} &= -\frac{\ii  e^2 v s_{\alpha }}{2 s_W^2} \, ,\ \qquad    g_{HW^+W^-} = \frac{\ii  e^2 v c_{\alpha }}{2 s_W^2} \, ,\\
g_{hG^\pm W^\mp} &= \pm\frac{\ii e \sa}{2 s_W}\,, \qquad g_{HG^\pm W^\mp} = \mp\frac{\ii e \ca}{2 s_W}.\end{align}
The gauge--scalar couplings are thus the same as in general 2HDMs with the replacement $\beta\to 0$. 

\section{Renormalization}\label{appendix-renormalization}
We here give a summary of the on-shell renormalization scheme used in
\cite{Arhrib:2006wd}. The on-shell renormalization scheme at one-loop order for 2HDMs and the MSSM is also discussed in e.g.\ Refs.~\cite{Dabelstein:1994hb,Santos:1996vt,Arhrib:1999rg,Logan:2002jh,LopezVal:2009qy}.
We renormalize the doublets and vevs according to:
\begin{equation}
\Phi_i \to \sqrt{Z_i}\,\hat{\Phi}_i\,,\qquad v_i \to \hat{v}_i - \delta_{v_i}\, ,
\end{equation}
where $v_2 = 0$ at tree level in our model, and the wavefunction renormalization constants $Z_i$ are
expanded as $Z_i = 1 + \delta_{Z_i}$ at one-loop order. These redefinitions are then inserted into the
kinetic Lagrangian for the doublets. After this insertion, we obtain the following
counterterms ($A_\mu$ is the photon field):
\begin{align}
\delta_{H^\pm W^\mp}\, & (\partial^\mu H^\pm) W^\mp_\mu ,\label{counterterm1} \\
\delta_{H^\pm W^\mp \gamma}\, H^\pm W^\mp_\mu A_\nu &= e  \delta_{H^\pm W^\mp} H^\pm  W^\mp_\mu A_\nu  ,\label{counterterm2}\\
\delta_{H^\pm W^\mp Z}\, H^\pm W^\mp_\mu Z_\nu &= e \frac{s_W}{c_W} \delta_{H^\pm W^\mp} H^\pm W^\mp_\mu Z_\nu ,
\label{counterterm3}
\end{align}
for the mixings and vertices respectively, where
\begin{equation}
\delta_{H^\pm W^\mp} =  \frac{m_W}{\hat{v}_1^2 + \hat{v}_2^2} \left[ \hat{v}_1 \delta_{v_2} - \hat{v}_2 \delta_{v_1}  + \hat{v}_1 \hat{v}_2 ( \delta_{Z_1}   -\delta_{Z_2}) \right].
\end{equation}
Hence, the renormalization of the $H^\pm W^\mp Z$
and $H^\pm W^\mp \gamma$ vertices depends on the $H^\pm W^\mp$ mixing renormalization.

In order for the  one-loop potential to be minimized by $\hat{v}_1$ and $\hat{v}_2$, we require that the renormalized tadpoles vanish:
\begin{equation}
T_{h/H} + \delta_{t_{h/H}} = 0,
\label{tadpolecond}
\end{equation}
where $T_{h/H}$ denotes the sum of all tadpole diagrams for the field $h/H$ and $\delta_{t_{h/H}}$
the tadpole counterterms at one-loop order. 

The on-shell renormalization scheme proceeds by requiring that the real
part\footnote{$\delta_{H^\pm W^\mp}$ is real since we consider a CP-conserving scalar sector.} of
the renormalized off-diagonal self-energy $\hat{\Sigma}_{H^\pm W^\mp}$ vanishes for an on-shell $H^\pm$:  
\begin{equation}
\text{Re} \left[ \hat{\Sigma}_{H^\pm W^\mp}(k^2 = m_{H^\pm}^2 ) \right] = 0 ,
\end{equation}
which then determines $\delta_{H^\pm W^\mp}$ according to
\begin{equation}
\text{Re} \left[\hat{\Sigma}_{H^\pm W^\mp}(k^2 = m_{H^\pm}^2 ) \right ] =  \text{Re} \left[\Sigma_{H^\pm W^\mp}(k^2 = m_{H^\pm}^2 ) \right ] +  \delta_{H^\pm W^\mp}   = 0 ,
\end{equation}
where the bare self-energy $\Sigma_{H^\pm W^\mp}$ is given by \Eref{eq:self}. Furthermore, the renormalization of the $H^\pm G^\mp $
mixing is also determined by $\delta_{H^\pm W^\mp}$ due to a Slavnov--Taylor identity that forces $\hat{\Sigma}_{H^\pm W^\mp}$ and $\hat{\Sigma}_{H^\pm G^\mp}$ to be proportional to each other \cite{LopezVal:2009qy,Coarasa:1996qa,Arhrib:2006wd}.

\begin{figure}[t]
\begin{center}
\begin{tabular}{cc}
\includegraphics[scale=0.85]{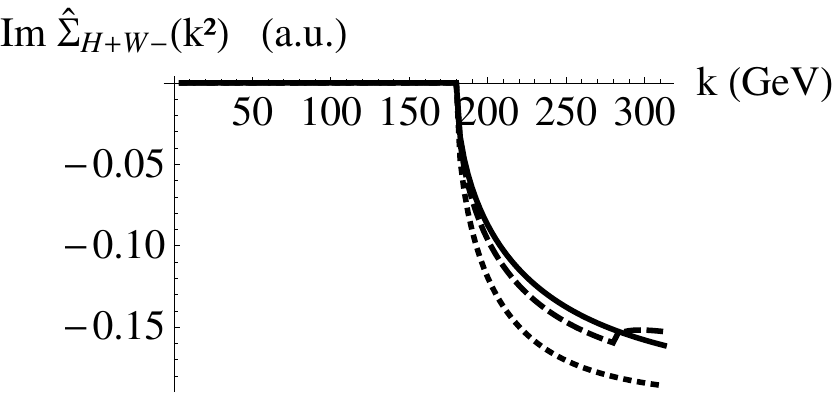} & \quad \includegraphics[scale=0.85]{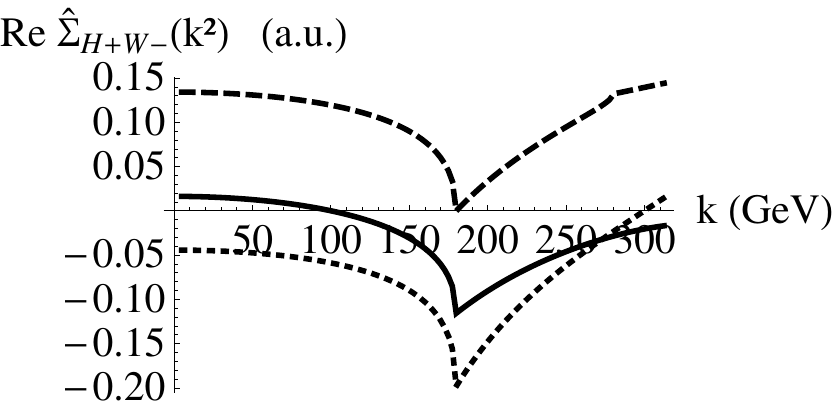} \\
(a) &$\quad$ (b)
\end{tabular}
\end{center}
\caption{The imaginary part (a) and the real part (b) of the on-shell renormalized off-diagonal self
energy $\hat{\Sigma}_{H^\pm W^\mp}$ as a function of the invariant mass $k$. In this figure we have
$\mhp = $ 100 GeV (solid), $\mhp = m_h+m_{W^\pm}$ (dashed) and $\mhp = $ 300 GeV (dotted). The other parameters in
our model are taken to be $m_h = 100$ GeV, $m_H = 300$ GeV, $m_A = \mhp $, $\sina = 0.9$,
$\lambda_3 = 0$, $\lambda_2 = \lambda_1$ and $\lambda_7 = \lambda_6$.}
\label{fig:selfenergies}
\end{figure}

For illustration we show  the real and imaginary parts of the renormalized self-energy $\hat{\Sigma}_{H^\pm W^\mp}$ in figure \ref{fig:selfenergies}. 
Note that the
real part vanishes for an on-shell $H^\pm$ as prescribed. Note also that the imaginary part is only non-zero when the internal particles in the loop ($W^\pm, h$ and $H $) can be produced on-shell, i.e.\ when $ k  > m_h + m_W $.  

By following the same prescription outlined here and in \cite{Arhrib:2006wd}, we find that the counterterm for $AZ$ mixing is proportional to the one obtained for $H^\pm W^\mp$ mixing, $\delta_{AZ} = \ii (m_Z/m_W)\,\delta_{H^\pm W^\mp}$. 
The $A Z$ mixing is also defined to vanish on-shell,
\begin{equation}
\text{Re}  \left[\hat{\Sigma}_{A Z}(k^2 = m_A^2) \right] = 0 \, ,
\end{equation}
and the $AG^0$ mixing is related to this by a similar Slavnov--Taylor identity as for the $H^\pm G^\mp $ mixing. All in all this means that the  $AZ$ and $H^\pm W^\mp$ mixing cannot vanish on-shell at the same time. At one-loop order this is not a problem since
the $AZ$ and $H^\pm W^\mp$ mixing cannot both be present in the same set of diagrams, and
we are free to choose whatever scheme (i.e.\ values of the counterterms) we want. 
However, if we include two-loop diagrams, then inconsistencies may
arise but this is not relevant for this study, so we leave aside
the issue of on-shell renormalization of 2HDMs, and in particular of our model, at arbitrary order in
perturbation theory.

In a perturbative expansion using $R_\xi$ gauge one must also include Faddeev--Popov ghosts. The ghosts
corresponding to $W^\pm$ and $Z$ couple only to $h/H$ in the scalar sector and only occur in
loop diagrams. For diagrams that contribute to the matrix elements for $\Gamma_{H^\pm \to
f \bar{f^\prime}}$, $\Gamma_{H^\pm \to W^\pm Z /\gamma} $ and $\Gamma_{A \to f \bar{f} }$ at
one loop order, the tadpole diagrams are the only ones that contain ghosts. 
Since we require the sum of the tadpole diagrams to vanish according to \Eref{tadpolecond} we do not need to include the ghost contributions explicitly in our calculations. 
It is however straightforward to include ghosts in our model. One just
makes the replacement $ H_{\text{SM}} \to  H \cosa - h \sina $ in 
$\mathcal{L}_{\text{ghost}}^{\text{SM}}$ \cite{Santos:1996vt}, which gives the following couplings
between ghosts ($\eta_V$) and $h, H$,
\begin{equation}
g_{h\eta_V\bar{\eta}_V} = \ii  \sa\, \xi m_V^2/v\,, \quad g_{H\eta_V\bar{\eta}_V} = -\ii  \ca\,\xi m_V^2/v\,,
\end{equation}
where $V = W^+, W^-$ or $Z$.

\section{Expressions for the vertices and mixing self-energies} \label{verticesandselfE}
In this appendix we give the expressions for the unrenormalized vertices and self-energies.
The vertex function $V_{H^+ L^-\nu}$ for $H^+ \to L^+\nu$ in Feynman--'t Hooft gauge is at leading order defined as
\begin{equation}
\mathcal{M}_{H^+\to L^+\nu} \equiv [\bar{u}_{L^+}\,P_R \,v_\nu ] V_{H^+L^-\nu}\, (m^2_\hp,m_L^2 ,0 \,) \,,
\end{equation}
where $\mathcal{M}_{H^+ \to L^+\nu}$ is the matrix element for the triangle loop contribution to $ H^+ \to L^+\nu$, see figure~\ref{fig:HpLnuHpUD1}b. The vertex function reads
\begin{align}
 16 \pi^2  \, V_{H^+ L^-\nu}\,(m^2_\hp,m_L^2 ,0 \,) \, &= \, g_{hL^+L^-}g_{hH^+W^-}\tilde{g}\,B_0(0,m_L^2,m^2_W) \nn \\
&- \, g_{hL^+L^-} \left[ \,g_{hH^+G^-} g_{G^+ L^- \nu} m_L - g_{hH^+W^-} \tilde{g}(m^2_\hp + m_h^2 - 4 m_L^2 )\, \right] \,   \nn \\
&\quad \times C_0( m^2_\hp , m_L^2, 0, m_W^2, m_h^2, m_L^2 ) \nn \\
&+ \, g_{hL^+L^-} \left[ g_{hH^+G^-} g_{G^+ L^- \nu} m_L - g_{hH^+W^-} \tilde{g}\,( m^2_\hp - 2 m_L^2 )\, \right] \, \nn \\
&\quad \times C_1( m^2_\hp , m_L^2, 0, m_W^2, m_h^2, m_L^2 )  \nn \\ 
&+ g_{hL^+L^-} g_{hH^+W^-} \tilde{g}\,(m^2_\hp - m_L^2) \, \nn \\ 
&\quad \times C_2( m^2_\hp , m_L^2, 0, m_W^2, m_h^2, m_L^2 ) \, +
 (h \to H) ,
\label{eq:vertex}
\end{align}
where $B_0$, $C_0$, $C_1$, $C_2$ are Passarino--Veltman integrals \cite{Passarino:1978jh}, $\tilde{g} = \ii e/\sqrt{2}s_W $, $g_{hL^+L^-} = \ii m_L/v$, $ g_{G^+L^-\nu} = -\ii \sqrt{2} m_L/v $, and the remaining $g_{ijk}$ are given in Appendix \ref{sect:couplings}.
The $ (h \to H)$ indicates the four terms that have a $H$ boson running
in the loop instead of $h$, which are obtained if one makes the replacement $h \to H $. The vertex function for $H^+ u_i\bar{d}_j$ is
analogous to $V_{H^+L^-\nu}$, but has more terms due to the non-vanishing quark masses.

The bare off-diagonal $H^+W^-$ self-energy in Feynman--'t Hooft-gauge reads
\begin{equation}
\begin{split}
16 \pi^2\,\Sigma_{H^+ W^-}(k^2) =&\, g_{hH^+W^-}\,g_{hH^+H^⁻}\left[ B_0(k^2,m_h^2,m^2_\hp) + 2\, B_1(k^2,m_h^2,m^2_\hp)  \right] \\
 & - \,g_{hH^+W^-}\,g_{hW^+W^-}\left[ 2\, B_0(k^2,m_h^2,m^2_{W}) + \, B_1(k^2,m_W^2, m^2_h)  \right]  \\
 & + \,g_{hG^+W^-}\,g_{hH^+G^-}\left[ B_0(k^2,m_h^2,m^2_{W}) + 2\, B_1(k^2,m_h^2,m^2_W)  \right]  \\
 &  +\,  (h \to H) \:,
\end{split}
\label{eq:self}
\end{equation}
where again $B_0$ and $B_1$ are Passarino--Veltman functions. 
The $ (h \to H)$ indicates the three terms that have a $H$ boson running in the loop instead of $h$ are obtained if one makes the replacement $h \to H $. 

One should notice that the matrix element $\mathcal{M}_{\hp \to W^\pm} $ for the transition $\hp \to W^\pm$ vanishes. This is because of the Feynman rules for the $h \hp W^\mp $ and $H \hp W^\mp $ vertices, which are present in the diagrams in \reffig{fig:HpmixFF2}:
\begin{equation}
S \hp W^\mp \: : \quad g_{S \hp W^\mp} \, \left[ p^\mu_{H^\pm} - p^\mu_{S}  \right] ,
\end{equation}
where $S = h,H$ and the four-momenta are taken to be incoming. This means that the mixing diagrams are all proportional to $p^\mu_{H^\pm} = p^\mu_{W}$, which, combined with \Eref{pWepsW} for a final state $W^\pm$ boson results in $\mathcal{M}_{\hp \to W^\pm}  = 0$. A $\hp$ boson can therefore not fluctuate into an (on-shell) $W^\pm$ boson, which is a renormalization-scheme independent statement.

The vertex functions for $A\to 2f$ are obtained similarly. We do not give the expressions for the vertex functions for $H^\pm \to W^\pm V $ here, but they can be found in Ref.~\cite{Arhrib:2006wd}.

\section{The smeared mass unstable particle model}\label{appendix-smup}
\label{sect:smup}
The smeared mass unstable particle (SMUP) model is based on the time--energy uncertainty
relation and the
K\"all\'{e}n--Lehmann form of the exact propagator where finite width effects are taken into
account in the spectral density, see \cite{Kuksa:2009de,Pasechnik:2010yu} and references therein. 
The reason we use this model is that it requires only the use and knowledge of
$\Gamma^*_{\hp \rightarrow \wpm^* \gamma } (\mhp,q)$ defined below.

To evaluate the decay width for $\hp \rightarrow  \wpm^* \gamma \ $ for a given mass of the charged
scalar, $\mhp$, one considers the invariant mass of the virtual $W$, $m_{ \wpm^*}\equiv q$, as a free parameter and defines
\begin{equation}
\Gamma_{\hp \rightarrow  \wpm^* \gamma} (\mhp) = \int_{0}^{m^2_{\hp}} \Gamma^*_{\hp \rightarrow
\wpm^* \gamma } (\mhp,q)\,\rho( q )\, \mathrm{d}q^2 ,
\label{eq:SMUP1}
\end{equation}
where $\Gamma^*_{\hp \rightarrow \wpm^* \gamma } (\mhp,q)$ is the decay width for $\hp
\rightarrow  \wpm^*\gamma$ with the off-shell $W^\pm$ having a specific invariant mass $q$. 
This is folded with the spectral density $\rho( q )$, defined as
\begin{equation}
\rho( q ) = \frac{1}{\pi} \frac{q \Gamma_{W}(q) }{[q^2  - m^2_{W} ]^2 + [q \Gamma_{W}(q)]^2}
\label{eq:SMUP2}
\end{equation}
where we have used $m_{W} = 80.4$ GeV and
\begin{equation}
\Gamma_{W}(q) = \frac{9\, g^2}{48\pi}q   .
\label{eq:gammaW}
\end{equation}
We evaluate \Eref{eq:SMUP1} by using our code for $\hp \rightarrow   \wpm \gamma $ with on-shell $W^\pm$ but allowing the
$\wpm$-mass to vary. We then integrate numerically over the spectral
density. 

\begin{figure}[tb]
\begin{tabular}{cc}
\includegraphics[scale=0.90]{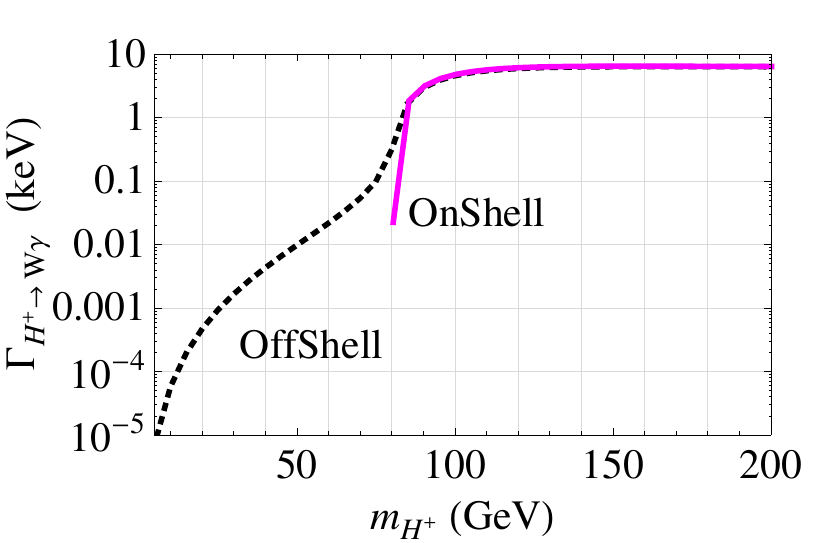}\qquad & \includegraphics[scale=0.85]{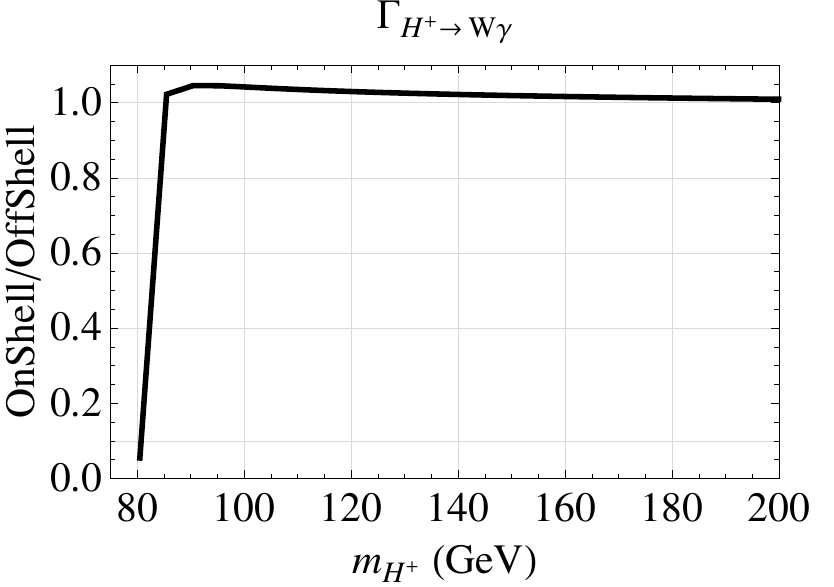} \\
(a) & (b) 
\end{tabular}
\caption{(a) Comparison of off-shell (dashed blue line) and on-shell (solid red line) decay widths for
$\hp \rightarrow \gamma \wpm$. (b) The ratio of the on-shell and the off-shell decay widths. The
parameters of the model take the values $m_h = 125, m_H = 300$ GeV, $m_A = \mhp$,
$\sina = 0.9$ and $\lambda_3 = 2(m_\hp/v)^2$. }
\label{fig:hpwgammaoff1}
\end{figure}

As a check of the formalism, we also applied the SMUP model to the well-known SM process $H_\text{SM} \rightarrow W^{-*}W^{+*}$. Comparison
with known ``standard'' formulas \cite{Romao:1998sr,Djouadi:2005gi} show excellent agreement with a
difference of less than 2\%. The standard formula for $ H_\text{SM} \rightarrow W^{-*}W^{+*}$ with a fixed width reads
\cite{Djouadi:2005gi}
\begin{equation}
\Gamma _{H_\text{SM} \rightarrow W^{-*}W^{+*} }(m_{H_\text{SM}}) =  \int_0^{m_{H_\text{SM}}^2} \frac{\mathrm{d} q_1^2\,m_W\Gamma_W / \pi }{[q_1^2 - m^2_W]^2 + m_W^2\Gamma_W^ 2  } \int_0^{k^2} \frac{\mathrm{d}q_2^2\,m_W\Gamma_W/ \pi}{[q_2^2 - m^2_W]^2 + m_W^2\Gamma_W^2  } \Gamma_0
\label{eq:djou235}
\end{equation}
where $ k= m_{H_\text{SM}}-q_1$, and
\begin{equation}
\Gamma_0 = \frac{m_{H_\text{SM}}^3}{16\pi v^2}\sqrt{\left( 1-\frac{q_1^2}{m_{H_\text{SM}}^2}-\frac{q_2^2}{m_{H_\text{SM}}^2} \right)^2 - 4\frac{q_1^2q_2^2}{m_{H_\text{SM}}^4}  }\: \left[ \left( 1-\frac{q_1^2}{m_{H_\text{SM}}^2}-\frac{q_2^2}{m_{H_\text{SM}}^2} \right)^2 +8\frac{q_1^2q_2^2}{m_{H_\text{SM}}^4}   \right] \, .
\label{eq:djou236}
\end{equation}
This formula is obtained by denoting the denominator of the respective $W^\pm$-propagators as 
\begin{equation}
q_i^2 - m^2_W + \ii m_W \Gamma_W ,
\label{propagators}
\end{equation}
where $ q_i^2  $ is the invariant mass squared of the $i$'th off-shell $W^\pm$-boson.
We stress that, differently from the SMUP method,
the quantity $\Gamma_0$ in \eqref{eq:djou235} 
should not be literally interpreted as neither the
decay width of the Higgs boson to a pair of virtual 
bosons with invariant masses $q_1, q_2$
nor as the matrix element squared.

As a further check, we evaluate $\Gamma_{\hp \rightarrow  \wpm^* \gamma}$ for $\mhp$ far above the threshold, with the
result that the off-shell calculation coincides with the on-shell result, as shown in~\reffig{fig:hpwgammaoff1}b.

\bibliographystyle{apsrev4-1}
\bibliography{lopsided}

\end{document}